\documentclass[preprint]{aastex631}

\received{}
\revised{}
\accepted{}

\shorttitle{Two Types of Kuiper Belt Object}
\shortauthors{Fraser et al.}

\begin{document}

\title{Col-OSSOS: The Two Types of Kuiper Belt Surfaces}

\author[0000-0001-6680-6558]{Wesley C. Fraser}
\affiliation{Herzberg Astronomy and Astrophysics Research Centre, National Research Council, 5071 W. Saanich Rd. Victoria, BC, V9E 2E7, Canada}
\affiliation{Department of Physics and Astronomy, University of Victoria, Elliott Building, 3800 Finnerty Road, Victoria, BC V8P 5C2, Canada}

\author[0000-0003-4797-5262]{Rosemary~E. Pike}
\affiliation{Center for Astrophysics | Harvard \& Smithsonian; 60 Garden Street, Cambridge, MA, 02138, USA}

\author[0000-0001-8617-2425]{Micha\"el Marsset}
\affiliation{European Southern Observatory, Alonso de C\'ordova 3107, Santiago, Chile}
\affiliation{Department of Earth, Atmospheric and Planetary Sciences, MIT, 77 Massachusetts Avenue, Cambridge, MA 02139, USA}

\author[0000-0003-4365-1455]{Megan E. Schwamb}
\affiliation{Astrophysics Research Centre, Queen’s University Belfast, Belfast, County Antrim BT7 1NN, UK}

\author[0000-0003-3257-4490]{Michele T. Bannister}
\affiliation{School of Physical and Chemical Sciences | Te Kura Mat\=u University of Canterbury, Private Bag 4800, Christchurch 8140, New Zealand}

\author[0000-0002-8032-4528]{Laura Buchanan}
\affiliation{Astrophysics Research Centre, Queen’s University Belfast, Belfast, County Antrim BT7 1NN, UK}

\author[0000-0001-7032-5255]{JJ Kavelaars}
\affiliation{Herzberg Astronomy and Astrophysics Research Centre, National Research Council, 5071 W. Saanich Rd. Victoria, BC, V9E 2E7, Canada}
\affiliation{Department of Physics and Astronomy, University of Victoria, Elliott Building, 3800 Finnerty Road, Victoria, BC V8P 5C2, Canada}

\author[0000-0001-8821-5927]{Susan D. Benecchi}
\affiliation{Planetary Science Institute, 1700 East Fort Lowell, Suite 106, Tucson, AZ 85719, USA}

\author[0000-0001-6541-8887]{Nicole J. Tan}
\affiliation{School of Physical and Chemical Sciences | Te Kura Mat\=u University of Canterbury, Private Bag 4800, Christchurch 8140, New Zealand}

\author[0000-0002-6830-476X]{Nuno Peixinho}
\affiliation{Instituto de Astrof\'{\i}sica e Ci\^{e}ncias do Espa\c{c}o, Universidade de Coimbra, PT3040-004 Coimbra, Portugal}

\author[0000-0001-8221-8406]{Stephen D. J. Gwyn}
\affiliation{Herzberg Astronomy and Astrophysics Research Centre, National Research Council, 5071 W. Saanich Rd. Victoria, BC, V9E 2E7, Canada}

\author[0000-0003-4143-8589 ]{Mike Alexandersen}
\affiliation{Center for Astrophysics | Harvard \& Smithsonian; 60 Garden Street, Cambridge, MA, 02138, USA}

\author[0000-0001-7244-6069]{Ying-Tung Chen}
\affiliation{Institute of Astronomy and Astrophysics, Academia Sinica; 11F of AS/NTU Astronomy-Mathematics Building, No. 1, Sec. 4, Roosevelt Road, Taipei 10617, Taiwan, R.O.C}

\author[0000-0002-0283-2260]{Brett Gladman}
\affiliation{Department of Physics and Astronomy, University of British Columbia, Vancouver, BC, Canada}

%\author[0000-0003-0407-2266]{Jean-Marc Petit}
%\affiliation{Institut UTINAM UMR6213, CNRS, Univ. Bourgogne Franche-Comté, OSU Theta F-25000 Besançon, France}

\author[0000-0001-8736-236X]{Kathryn Volk}
\affiliation{Lunar and Planetary Laboratory, The University of Arizona, 1629 E University Blvd, Tucson, AZ 85721}

\correspondingauthor{Wesley C. Fraser}
\email{wesley.fraser@nrc-cnrc.gc.ca}

\begin{abstract}

The Colours of the Outer Solar System Origins Survey (Col-OSSOS) has gathered high quality, near-simultaneous (g-r) and (r-J) colours of 92 Kuiper Belt Objects (KBOs) with (u-g) and (r-z) gathered for some. We present the current state of the survey and data analysis. Recognizing that the optical colours of most icy bodies broadly follow the reddening curve, we present a new projection of the optical-NIR colours, which rectifies the main non-linear features in the optical-NIR along the ordinates. We find evidence for a bifurcation in the projected colours which presents itself as a diagonal empty region in the optical-NIR. A reanalysis of past colour surveys reveals the same bifurcation. We interpret this as evidence for two separate surface classes: the BrightIR class spans the full range of optical colours and broadly follows the reddening curve, while the FaintIR objects are limited in optical colour, and are less bright in the NIR than the BrightIR objects. We present a two class model. Objects in each class consist of a mix of separate blue and red materials, and span a broad range in colour. Spectra are modelled as linear optical and NIR spectra with different slopes, that intersect at some transition wavelength. The underlying spectral properties of the two classes fully reproduce the observed structures in the UV-optical-NIR colour space ($0.4\lesssim\lambda\lesssim1.4 \mbox{ $\mu$m}$), including the bifurcation observed in the Col-OSSOS and H/WTSOSS datasets, the tendency for cold classical KBOs to have lower (r-z) colours than excited objects, and the well known bimodal optical colour distribution. 
\end{abstract}

\keywords{}

\section{Introduction} \label{sec:intro}

The colours of Kuiper Belt Objects (KBOs) have long been a topic of interest to the Solar System science community, having been used to reveal a range of compositional and cosmogonic properties about those bodies. Broadband colours have been used to provide compositional inference when targets are too faint for detailed spectroscopy. For KBOs, colours are often as informative as spectroscopy, as shortward of the $1.5\mbox{ $\mu$m}$ water-ice absorption feature, most small KBOs exhibit nearly featureless spectra, that are well described by a red optical slope \citep{Fornasier2009} and a less-red NIR slope \citep{Barucci2011}, with a smooth transition from one to the other \citep[e.g.,][]{Gourgeot2015}. Where known absorptions exist, specific compositional information for certain materials can be gleaned with well chosen filters \citep[e.g.,][]{Trujillo2011}. 

%-KBO colours have long been a topic of great interest to the Solar System science community. Most colours data reported from surveys dedicated to the purpose. Typically used to provide compositional inference when too faint for spectroscopy. Well chosen filters can provide information on surface content of specific materials. 

Arguably more useful than compositional information, large samples of broadband colours have also been used for taxonomic purposes. For example, optical colours combined with H-band photometry have been useful in identifying faint Haumea family members \citep{Snodgrass2010}. In recent years, multiple efforts have been made to generate a taxonomy of KBO surfaces from their broadband optical-NIR colours \citep{DalleOre2013,Fraser2012,Peixinho2015,Pike2017,Schwamb2019}. Thus far, no consensus has been found regarding the true number of main KBO compositional classes, with as few as three classes being proposed \citep[e.g.,][]{Fraser2012, Pike2017} and as many as 10 \citep{DalleOre2013}. The count of KBO taxons has important cosmogonic implications, as it is believed that taxon membership is a consequence of compositional structure that occurred while objects resided in the protoplanetesimal disk. It is believed that the taxons reflect the level of radial compositional stratification that occurred between formation and the dispersal of the protoplanetesimal disk. 

The Colours of the Outer Solar System Origins Survey (Col-OSSOS) was designed around the experiences of past optical and NIR colour surveys \citep{Schwamb2019}. The spectrophotometric observations of this program were designed to be sensitive to the optical and NIR spectral slopes, so as to detect any potential taxons that may reveal themselves in these wavelength regions \citep{Tegler2003b,Delsanti2006,Peixinho2012,Fraser2012}. Additional UV and optical bands were acquired for some targets as telescope time permitted. With a goal of homogenous signal-to-noise (SNR) in all bands, and for all targets, the Col-OSSOS sample was designed with a focus on providing confident taxonomic classification for KBOs. Here we report the current state of the Col-OSSOS survey, presenting the grJ colours of 92 targets acquired at the Gemini-North Telescope. We also present the first results of the u-band observations that were acquired by an accompanying program on the Canada-France-Hawaii Telescope.

We present a new colour projection scheme that is motivated by the spectral behaviour of the bulk of icy bodies to fall near the so-called reddening curve, or the curve of constant spectral slope through colour-space. This projection into axes with distance along, and perpendicularly away from the reddening curve, rectifies the non-linear structures seen in the optical-NIR colour space of KBOs, for the Col-OSSOS dataset and others. With this reprojection we find a bifurcation in the optical-NIR colours of KBOs, with significant overlap of the optical  and NIR colours of each class. The bifurcation is apparent for colours which include a NIR filter, with wavelength $\lambda\gtrsim0.8 \mbox{ $\mu$m}$. These findings are similar to those of the Hubble/WFC3 Test of Surfaces in the Outer Solar System \citep[H/WTSOSS,][]{Fraser2012}, and a reanalysis of those data shows the same bifurcation. These results argue for only two surface classes for small KBOs (absolute magnitudes $H_r\gtrsim5$). One class is dominated mainly by Cold Classical KBOs (CCKBOs) with very red optical, and less red NIR colours. The other class has optical colours that span the full range exhibited by KBOs, but for objects that are equally as red as the first class, have much higher NIR reflectivities. Further, we argue that the now well-known bifurcation in optical colours is not due to one spectral class being found with only neutral to moderately red optical colours, and the other with very red colours. Rather, the class that the majority of excited KBOs belongs to spans the full optical colour range. 

In Section~\ref{sec:observations} we present a summary of the Col-OSSOS survey design, target selection, observations, and data reductions. In Section~\ref{sec:colours} we present the colours of Col-OSSOS targets, along with a new projection technique that helps reveal structures in the multi-dimensional UV-optical-NIR colour space by rectifying those colours along and perpendicular to the reddening line. We use that projection to search for evidence of separate classes of KBOs in the optical and NIR range of KBO colours. In Section~\ref{sec:publishedcolours}, independent of the Col-OSSOS dataset, we re-consider other large optical and NIR colour datasets to increase the wavelength range and sample size in our analysis. Those results corroborate what we find with the Col-OSSOS dataset. In Section~\ref{sec:model} we present a simple spectral model that accounts for the bulk behaviours of the different colours datasets we have considered. In Section~\ref{sec:discussion} we present a synthesis or our results, and show how they fit within past compositional interpretations. We discuss the cosmogonic significance of our results, and finish with concluding remarks in Section~\ref{sec:conclusions}.

\section{Observational Data} \label{sec:observations}

We make use of the Col-OSSOS dataset. This consists mainly of observations in the Sloan u, g, and r, filters, and the Mauna Kea J filter, which were acquired with the Gemini Multi-Object Spectrograph \citep[GMOS;][]{Hook2004} and Near Infrared Imager \citep[NIRI;][]{Hodapp2003} at Gemini-North, and the MegaPrime imager \citep{Boulade2003} on the Canada-France-Hawaii Telescope. A small number of targets received supplemental observations with the Suprime-Cam imager \citep{Miyazaki2002} on the Subaru Telescope. The data we present here include measurements presented previously \citep{Fraser2017, Pike2017, Schwamb2019, Marsset2019, Marsset2020, Fraser2021}, but are updated to reflect improvements in the reduction pipelines. Full details of the observing program, data reductions, and colour analysis can be found in \citet{Schwamb2019}, though we summarize the procedures here. We also list specific changes and improvements to the pipelines and analysis procedures since last publication \citep{Schwamb2019}.

The nominal Col-OSSOS sample are those 98 objects found in the E, H, L, O, S, and T blocks of the Outer Solar System Origins Survey \citep[OSSOS,][]{Bannister2018}, with discovery triplet brightness r$<23.6$ mags. The E, S, and T blocks have 6, 1, and 2 targets that are yet to be observed at the time of writing. A summary of the field details is available in Table~\ref{tab:fields}.

The standard observing procedure for a given target was to image the target in a r-g-J-g-r pattern, with g, and r observations acquired with GMOS, and the J observations acquired with NIRI. Exposure times were 300~s exposure for GMOS images, and 120~s for individual NIRI frames, unless otherwise noted (see Table~\ref{tab:observations}). The number of exposures acquired in each filter was tuned to match the target's discovery brightness, and designed to achieve a signal to noise greater than 20 in J, and 25 in g, and r, assuming pessimistic KBO colours of (g-r)=1.1 and (r-J)=1.2. If circumstance allowed it, GMOS images in z-band were acquired in an ad-hoc approach, either before or after, or bracketing the nominal sequence. 

When available, observations in r and u where acquired at CFHT, simultaneously with the Gemini sequence. These observations usually used an r-u-r pattern, with $\sim14$ u-band images of 320~s, with two r-band bracketing exposures of 300~s. Though some sequences acquired in 2014 did not have bracketing r-band images. A couple sequences also included g-band observations. Here we present only measurements from the u-band data, and have not made use of the r-band imagery. A full list of all exposures considered here are listed in Table~\ref{tab:observations}.

\subsection{Reductions}

GMOS imagery was first flat fielded and de-biased in the standard way using flats produced as part of Gemini's day calibrations program. Rough photometric calibrations were made using in-image stellar sources, with magnitudes uncorrected for differences between the GMOS and catalog photometric systems. Linear colour terms were measured, and applied to correct the rough image zeropoints, to place them in the catalog calibration magnitude system. Source magnitudes were measured using the Trailed Image Photometry in Python package \citep[TRIPPy;][]{Fraser2017} with pill apertures accounting for the source's rate of motion, and radius of 1.2 times the Full Width at Half Maximum (FWHM) of nearby stellar sources. Point-spread functions (PSFs) were generated using in-image point sources, from which aperture corrections were measured, enabling correction to the circular apertures with radius $4\times$~FWHM used on the calibration stars.

After bad-pixel masking and cosmic ray rejection, NIRI images were flattened, and had dark-current removed using sky-frames created from a rolling 15-frame window for each target NIRI sequence. The images of a science sequence were aligned using in-field point-sources, at the sidereal rate, as well as the rate of motion of the science target. Two mean stacks consisting of the first and second halves of the images in the sequence were produced. PSFs were generated from the sidereal imagery, from which FWHMs were measured. TRIPPy pill apertures of radius $1.2\times$~FWHM were used, and aperture corrections applied to match the $4\times$~FWHM were used on the calibration stars. 

The Subaru photometry we report here is an updated reduction of the values previously reported in \citet{Pike2017}. We highlight a change to the reductions. Specifically, sources were more closely scrutinized to better exclude non-point-sources from the list of sources used to evaluate image PSFs. Unfortunately, in the original analysis some PSF reference sources were non-stellar (eg. galaxy cores), resulting in inflated PSFs, and therefore, inflated TRIPPy apertures. 

NIRI photometric calibrations were made from standard star sequences acquired before and after the standard r-g-J-g-r sequence. Each standard sequence consisted of 9 frames each consisting of 6 1.5~s coadds, and were reduced in the same way as the science frames, but utilizing a single sky frame for the full sequence. During analysis, an inconsistency was discovered in measured calibration magnitudes from night to night. The source was traced to non-symmetric variations in the PSF, presumably due to the primary mirror aberrations not settling during the short calibration sequences. The result was an asymmetric PSF whose profile was still well described by a Moffat profile along any given radial line (from centre outwards) but whose width varied with angle around source peak. The inconsistent calibration photometry was measured using a circular aperture scaled off the FWHM measured from the average radial profile of the star, resulting in large portions of the calibrator source flux falling outside the aperture in regions where the PSF was particularly wide. The solution was to measure an azimuthally varying Moffat profile, in 8 equal pie segments around the source. The adopted FWHM was then twice the widest measured Half-Width at Half Maximum from the separate segments. Reported zeropoints are the median values extracted from all nine images of a sequence. By scaling apertures in this way, the image-to-image zeropoints were reflected correctly, enabling calibrations accurate to $\sim3\%$. See Appendix~\ref{sec:Jcals} for a discussion of the precision of this technique. 

The MegaPrime u-band images were debiased, flattened, and photometrically calibrated to the u$_{\scriptscriptstyle CFHT}$ bandpass following the MegaPipe procedures \citep{Gwyn2008}. All calibrated u-band images of a science sequence were then stacked at sidereal and non-sidereal rates using the same procedures as used for the J-band data. PSFs and aperture radii were evaluated from the sidereal stacks, and photometry of the science target was measured from the non-sidereal stack. We report photometry of only cases of clear source detection. No effort was made to measure upper limits of ambiguous- or non-detections. Analysis of u-band data was completed only for some data taken in semesters 2014B through 2016A. A more robust analysis including upper limits and all outstanding u-band Col-OSSOS data -- the majority of available u-band data -- will be the focus of a future work. 

With photometry in hand, colours were measured in the same way as previously reported. That is, each object was assumed to exhibit a linear lightcurve during the 1.5-4 hour Gemini sequences. A linear lightcurve, and colours with respect to r was fit to the optical data reported in the GMOS passbands. The fitted line was then extrapolated to the J and u-band epochs to measure (u-g) and (r-J) colour values. We point the reader to \citet{Schwamb2019} for further details of this fitting procedure.

Finally, using the colour terms evaluated during our optical data calibrations (see Appendix~\ref{sec:c-terms}) and from \citet{Gwyn2008}, colours were evaluated in the 
Sloan Digital Sky Survey \citep[SDSS][]{Fukugita1996, Padmanabhan2008} (for the u-band photometry) and the
Panoramic Survey Telescope and Rapid Response System Pan-STARRS  PS1 colour bands \citep{Tonry2012}, with appropriate propagation of uncertainties. Colours of our targets in the Gemini/CFHT, SDSS, and PS1 colour systems are reported in Table~\ref{tab:colours}. The updated colours derived from Subaru observations are presented in the SDSS colour system in Table~\ref{tab:subaru_colours}.

A number of notable changes and improvements have been made to the Col-OSSOS pipeline. These changes are:

\begin{enumerate}
\item All optical photometric calibrations are against the Pan-STARRS1 archive only. We no longer make use of calibrations against the SDSS archive.
\item J-band calibration source PSF profiles were measured using the azimuthally varying moffat profile technique described above.
\item Reported J-band uncertainties now include a 0.03 magnitude (0.02 in prior reductions) contribution due to the precision by which we can measure zeropoints, added in quadrature with contributions from other noise sources.
\item A bug was fixed in the exposure times of NIRI imagery that was caused by alteration of the \emph{EXPTIME} header keyword as part of the Gemini \emph{nirlin} procedure.
\item We utilized the background measurement routine in \emph{SExtractor in Python} package \citep{Barbary2016}, with a 128x128 box, to remove the background in each quadrant of each NIRI frame, in advance of the \emph{cleanir}\footnote{\url{https://www.gemini.edu/instrumentation/niri/data-reduction}} artifact removal and cosmic ray rejection routines. This resulted in a drastic improvement in residual background structure compared to using the default \emph{SExtractor} background estimation for the whole image.
\item The linear lightcurve fitting routing was altered to weight each measurement's contributions to the overall fit likelihood according to the uncertainty of each measurement. 
\end{enumerate}

Special consideration was made for observations in the 2016B semester. The observing plan was to make use of stars GD246 and G-129 as NIR calibration sources. It was readily apparent from our calibration data that GD246 was variable. Fortunately, two other stars happened to fall in the field of view, providing an extra calibration source. A sequence of observations were acquired in 8 different epochs in 2019B at twilight during which GD246, and calibration sources  F108, and SA114-750 were acquired, enabling brightness calibration; the two stars near GD246 have $J=11.879\pm0.005$ and $J=15.03\pm0.03$, respectively. Across 12 separate epochs, there was clear variability for the fainter source, but not for the brighter.

In addition to the unexpected variability of GD246, an error was made in the observing plan, causing observations to be acquired of a similarly bright star at 30' higher RA than for G-129. This affected the Col-OSSOS sequences of 7 targets from the S and T blocks. Unfortunately this star has a faint secondary within the wings of the primary, and about 5\% the brightness of the primary.  Removal of the secondary was a relatively simple task by making a PSF from the primary, and manually removing the secondary from the look up table. Subtraction residuals from the secondary were roughly 0.1\% of the primary's flux. At five separate epochs, GD246 and this incorrect star were observed in the same Col-OSSOS sequence, enabling a flux measurement of $J=14.56\pm0.03$. No variability was detected for the primary within the precision of the brightness measurement. With brightness measurements of the two alternate calibration sources -- the incorrectly pointed star, and the star below GD246 -- in hand, and confirmation of non-variability for both sources, photometric calibration of the 2016B data was performed as normal, and accounted for the precision of the brightness measurements of the two alternate calibrations sources. 

A full list of observations utilized in this paper are reported in Table~\ref{tab:observations}. The raw Gemini data files are available through the Gemini Observatory Archive (\url{https://archive.gemini.edu}). The reduced imagery are available for download at the Canadian Astronomy Data Centre: doi: \emph{to be updated at publication.}

\section{Colours of Col-OSSOS} \label{sec:colours}

\begin{figure*}[h]
\epsscale{1.16}
\plottwo{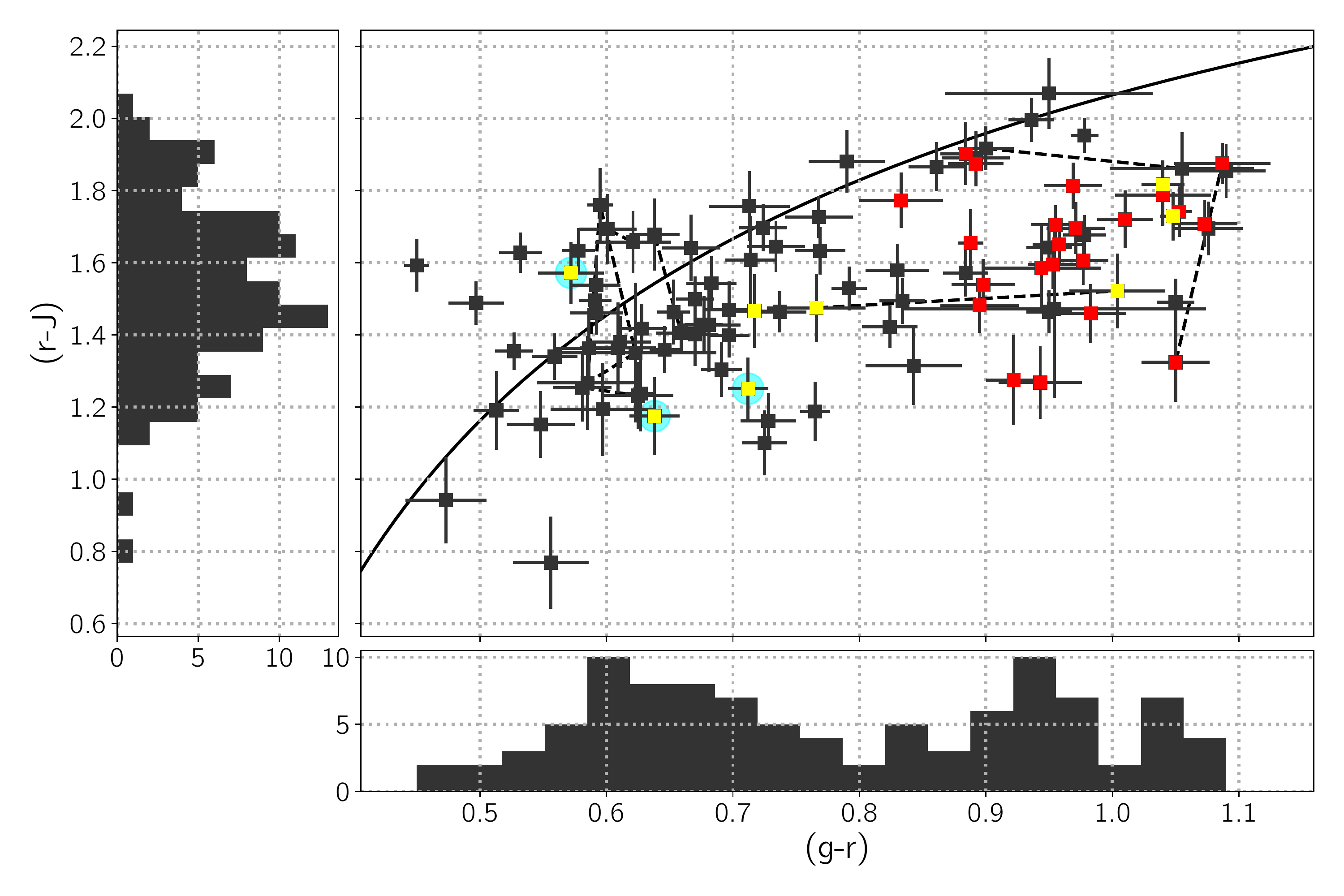}{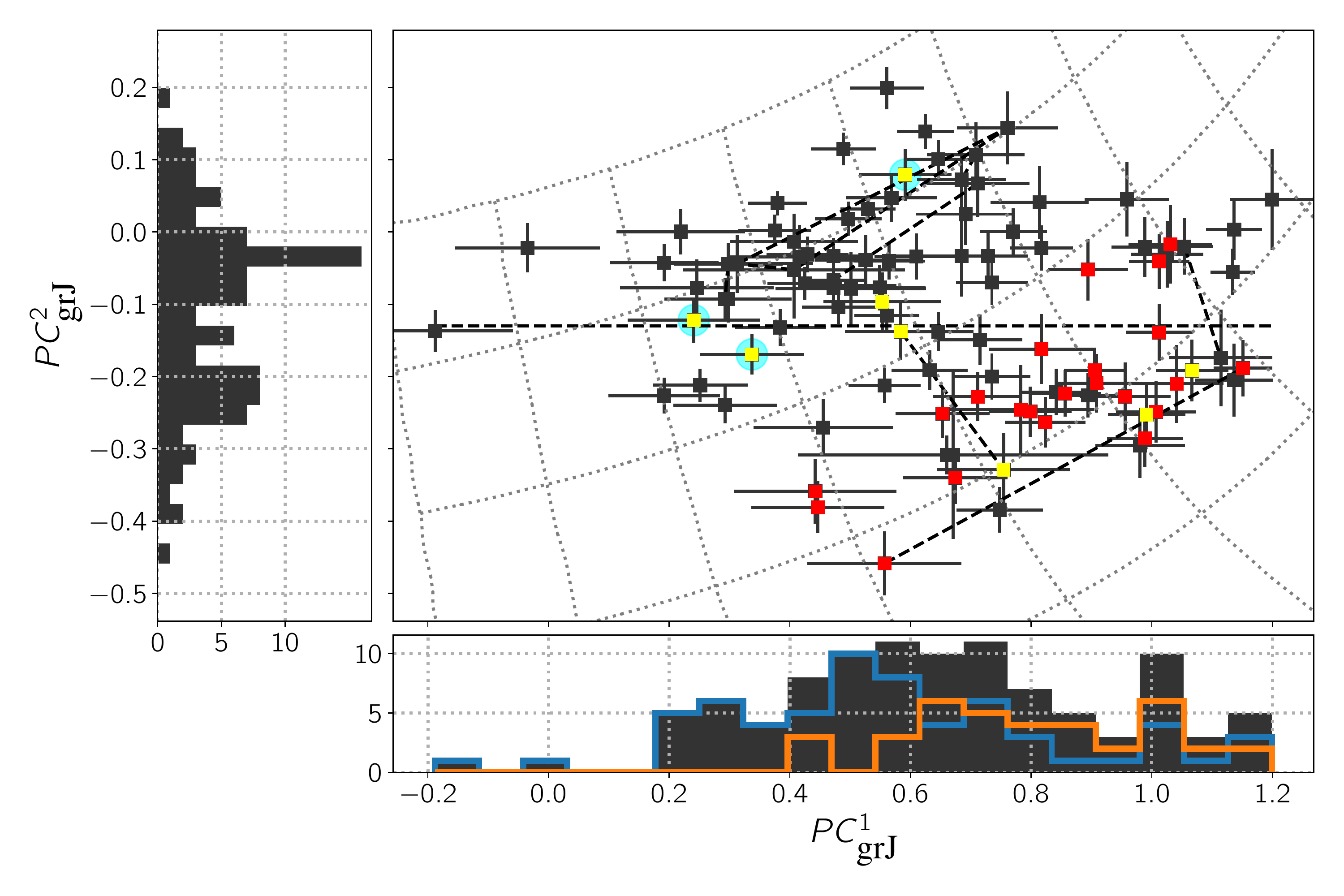}
\caption{ \textbf{Left:} The (g-r) and (r-J) colours of Col-OSSOS targets. Repeat measurements for targets (6 objects) have been connected by dashed lines. The reddening curve is shown by the solid black line. \textbf{Right:} The projection values $PC^1$ and $PC^2$ of the (g-r) and (r-J) colours. In this projection, the reddening curve is a horizontal line with $PC^2=0$. Grey hashes show lines of constant (g-r) or (r-J) colour. The horizontal line at $PC^2=-0.13$ marks the value below which all FaintIR objects are found. These mostly consist of CCKBOs. The bottom and side panels show histograms of the full dataset (black) and of the objects in the BrightIR (blue, bottom panel) and FaintIR classes (orange, bottom panel).  In both panels, red points are CCKBOs defined by their $i_{\textrm{free}}<4^\circ$, yellow and red points together are CCKBOs by the historical definition based on $i$ (see Section~2), and black points are all those excited objects outside the CC region. Objects identified as belonging to the class of \emph{blue binary} KBOs \citep{Fraser2021} are outlined with cyan halos. \label{fig:grJ}}
\end{figure*}

In Figure~\ref{fig:grJ} we show the full (g-r) and (r-J) measurements of the Col-OSSOS sample, which consists of 110 individual measurements and 92 unique objects. In this figure, we highlight two broad dynamical groups: the cold classical Kuiper Belt objects (CCKBOs), and the dynamically excited objects. For the former, we consider two different definitions. Classically, CCKBOs have been defined as those with low inclinations, $i<6^\circ$ and high perihelia $q>37$~AU, and semi-major axes bounded by the 3:2 and 2:1 mean-motion resonances, $40<a<48$~AU. We note that the exact maximum inclination has varied slightly \citep[eg.][]{Brown2001,Elliot2005,Fraser2012}, though slight variations in the adopted value have little bearing on our results. Like in past papers, we also include in this group objects in the 7:4 and 5:3 resonances with $i<6^\circ$. For a more modern definition, we consider objects' free inclinations -- the inclination vector after the forced component is removed \citep{Volk2017, vanLaerhoven2019,Huang2022preprint}, $i_{\textrm{free}}<4^\circ$, and a more restrictive semi-major axis range $42<a<48$ to select the least disturbed sample of classical KBOs, which are those most likely to have formed in-situ. Through out this manuscript, we discuss each subset separately where the distinction between these two definitions of cold classical object is important. Finally, we take as the dynamically excited sample, all those objects that are not considered CCKBOs by either definition.

Figure~\ref{fig:grJ} presents some of the features of the optical-NIR KBO colour space that past surveys have shown, including the tendency for CCKBOs to be optically red, and for most objects to fall to the lower-right of the reddening curve, or curve of constant spectral slope in colour-space. This sample also shows a bimodality in the optical colour distribution as has been reported numerous times in the past \citep[for recent examples, see][]{Tegler2016,Fraser2012,Peixinho2015,Marsset2019}. Hartigan's DIP test of bimodality \citep{Hartigan1985} implies that the full sample has 3.5\% probability of being drawn from a unimodal distribution. That probability of unimodality changes to near 100\% when excluding the CCKBOs. Interestingly, the optical plane clustering test \citep[FOP test;][]{Fraser2012} does not reveal any statistically significant bifurcation in the (g-r) and (r-J) colour space. We note, however, an intriguing region with a dearth of sources. This region can be roughly described as a diagonal region redward in both the optical and NIR colours of $(g-r)\sim0.75$ and $(r-J)\sim1.5$. %We will discuss this empty region further in this section. 

\begin{figure}[h]
\plotone{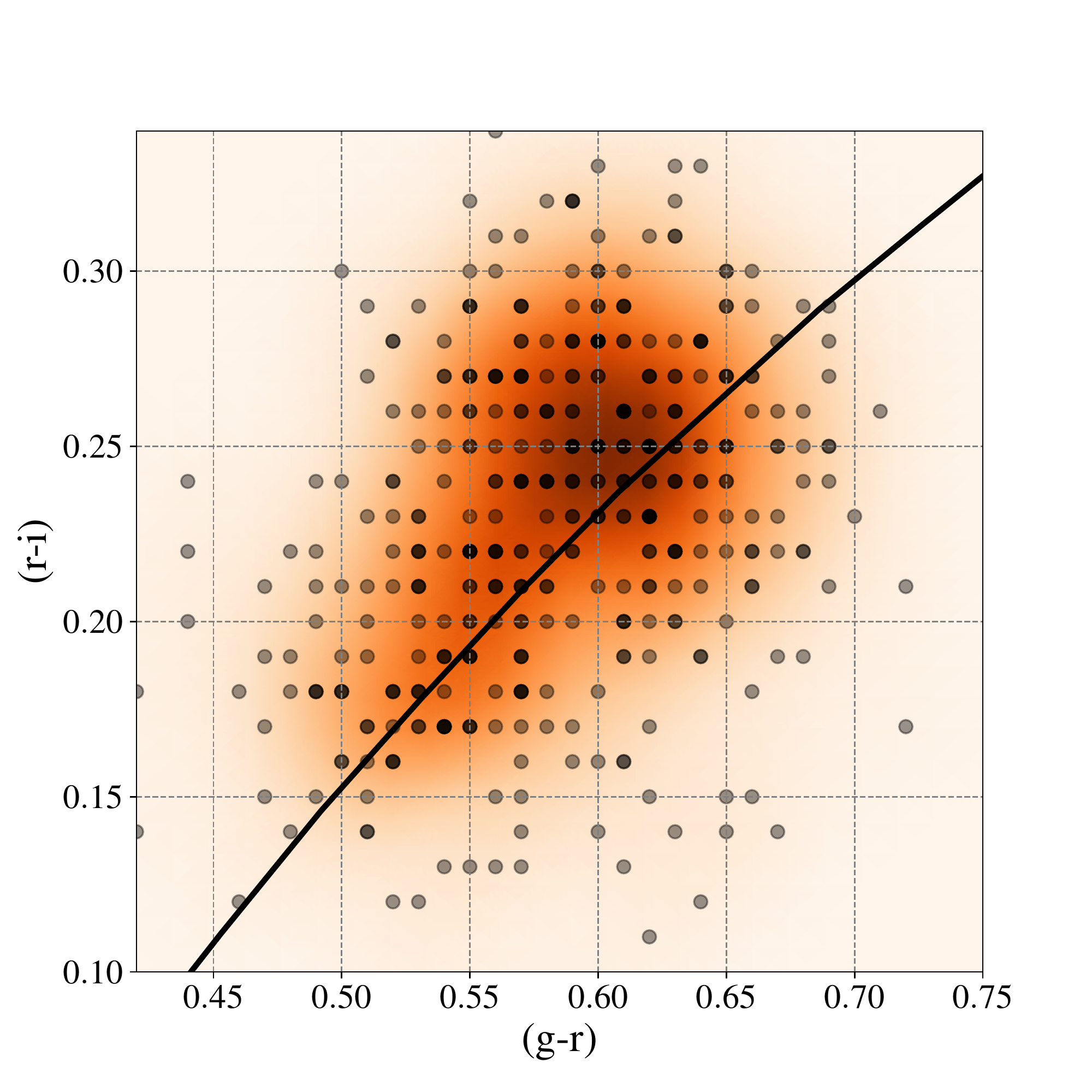}
\caption{ The optical colours of Jupiter Trojans reported in the Sloan Digital Sky Survey Moving Object Catalog v4 \citep{Ivezic2001} with $\Delta(g-r)<0.04$ and $\Delta(r-i)<0.04$. The datapoints generally scatter around and follow the reddening curve which is shown by the solid line. Errorbars are omitted for clarity. Semi-transparent points are used to show overlapping table values.  The shading is a kernel density estimate to highlight the densest regions of the data. \label{fig:Trojans}}
\end{figure}

\begin{figure*}[h]
\epsscale{1.16}
\plottwo{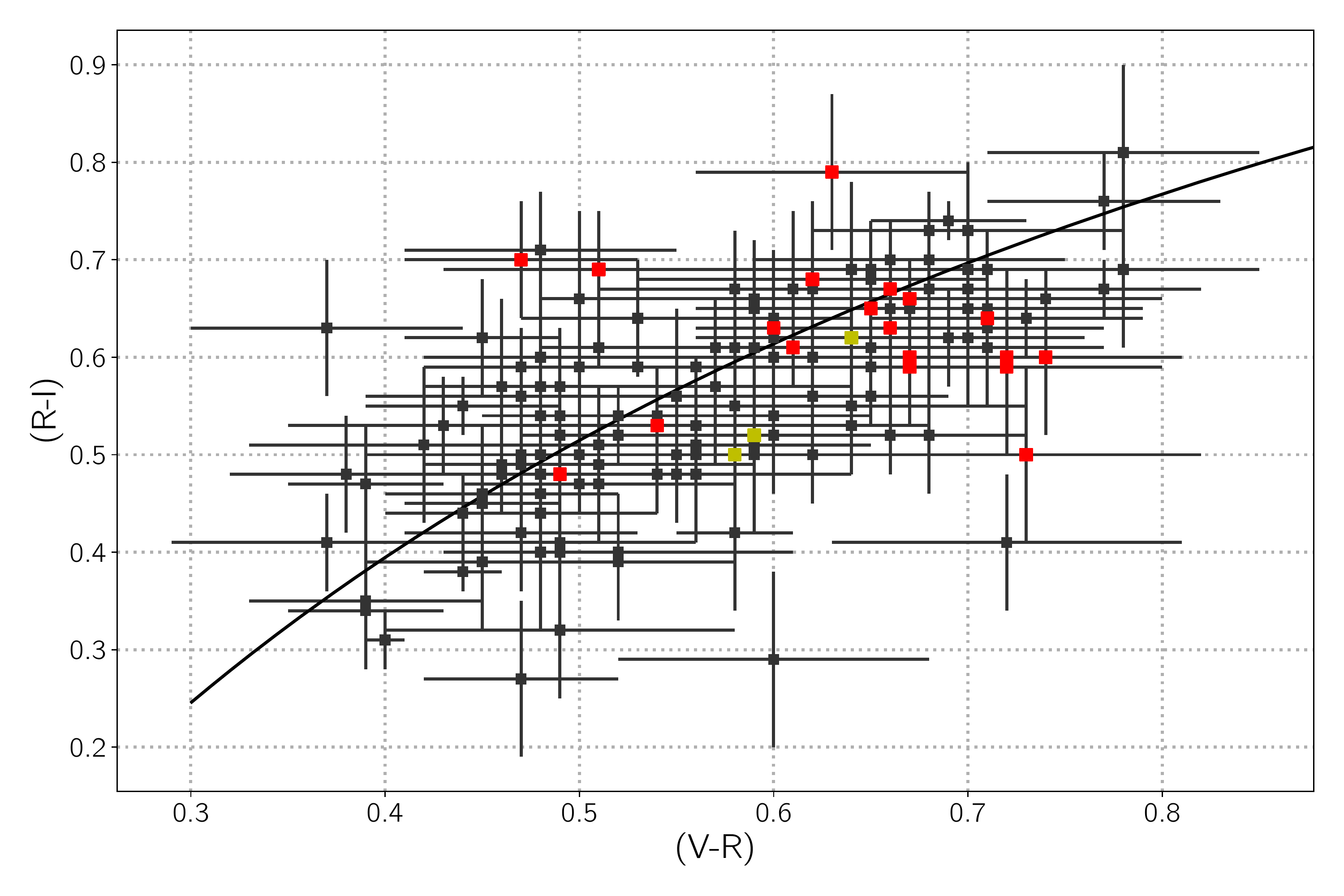}{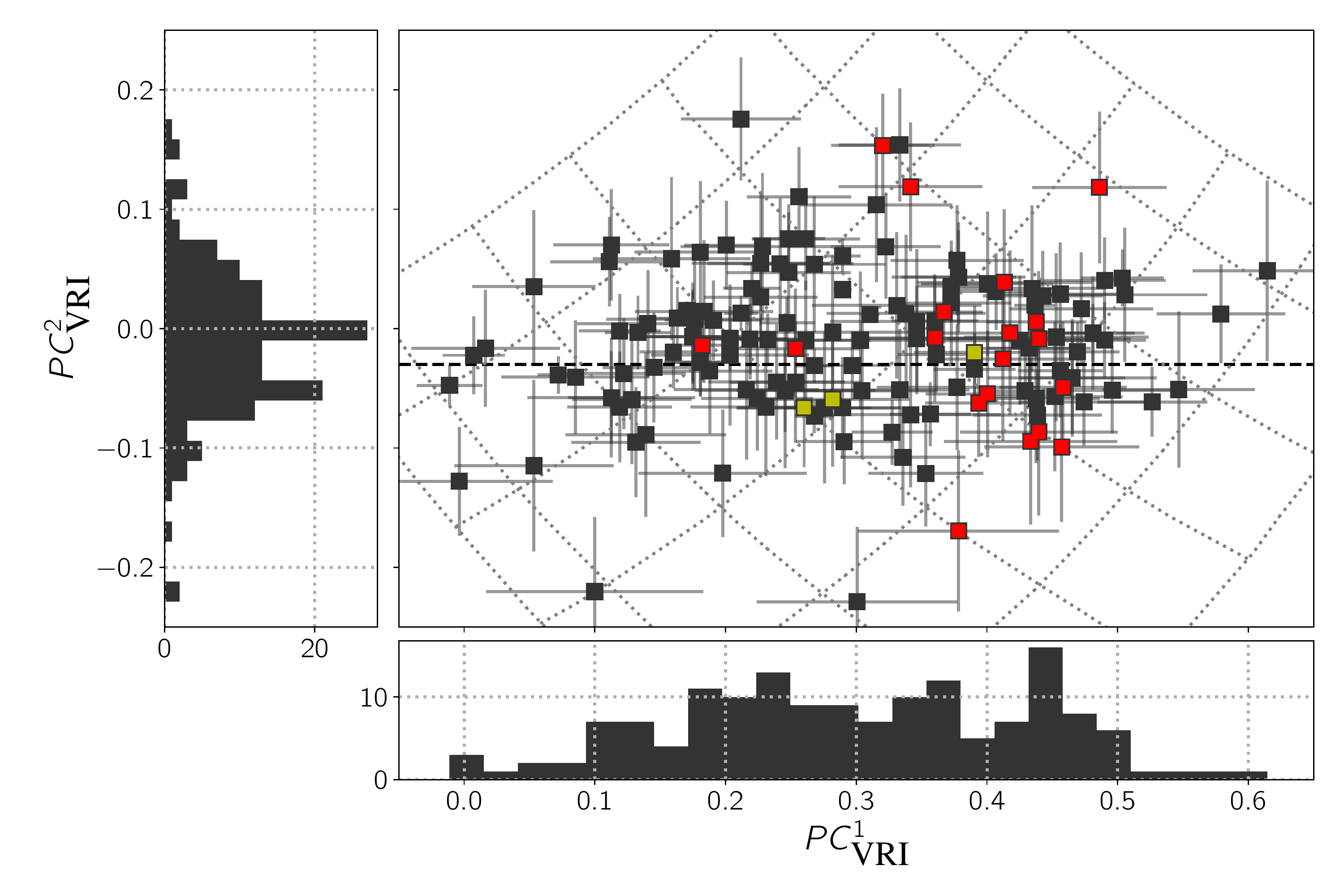}
\caption{\textbf{Left:} The optical colours of Kuiper Belt Objects and Centaurs reported in \citet{Peixinho2015}.  The datapoints generally scatter around and follow the reddening curve which is shown by the solid line. \textbf{Right:} The projection values $PC^1$ and $PC^2$ of the (V-R) and (R-I) colours shown in the left panel. Only the colours of objects with colour uncertainty $\Delta(V-R)<0.1$, $\Delta(R-I)$, and $H_\textrm{R}>4$ are plotted. The limit on absolute magnitude is made to avoid the few dwarf planets who have surfaces dominated by their own rheologies, and therefore have colours that are outliers on this plot. Red points are CCKBOs defined by their $i_{\textrm{free}}$, yellow and red points together are CCKBOs by the historical definition based on $i$, and black points are all those excited KBOs outside the CC region.  \label{fig:Peixinho_VRI}}
\end{figure*}

To analyze the grJ colour space, we introduce a new approach inspired by the recognition that the colours of icy bodies tend to roughly follow the reddening curve -- the curve of equal spectral slope in all colours (see Figure~\ref{fig:grJ} for an example). This is true of Jupiter Trojans \citep[see][for example]{Szabo2007}, KBOs, and Centaurs \citep{Peixinho2015}, the bulk of which have optical colours usually within a few tenths of a magnitude of the reddening curve (we present the optical colours of Trojans, KBOs, and Centaurs in Figures~\ref{fig:Trojans}, \ref{fig:Peixinho_VRI}). The majority of small bodies in each of these populations exhibit spectra dominated by the so-called \emph{optical gap} feature associated with C-H bonds in the organic materials  \citep[see][for recent discussions]{Grundy2020, Seccull2021}. The tendency for the spectra of these bodies to follow the reddening curve is true to a lesser extent in the near-UV, and NIR where albedos tend to deviate away from the reddening curve. This deviation is most dramatic in the NIR, though these populations still exhibit correlated optical and NIR colours \citep{Emery2011,Fraser2012}. The non-linear shape of the reddening curve is a common feature of colour-space that can be utilized in interpretation of observed colours.

As such, we adopt a projection of a 2-dimensional colour space onto a new set of basis vectors that removes the reddening curve trends seen in the colours samples. Specifically, we adopt  two basis vectors that are the instantaneously parallel and perpendicular vectors along the reddening curve, with projection values that we label $PC^1$ and $PC^2$, respectively. We use subscripts to denote which filters are considered. $PC^1$ and $PC^2$  are respectively, the line integral along the reddening curve from some predefined reference value, and the distance from the reddening curve along a vector perpendicular to the reddening curve. In this way, negative values of $PC^2_{\textrm{grJ}}$ reflect objects with concave spectra, or lower spectral slope in (r-J) than in (g-r), and positive values of  $PC^2_{\textrm{grJ}}$ reflect convex spectra. We choose as the origin the Solar (g-r) and (r-J) colours (see Appendix C) for $PC^1_{\textrm{grJ}} =0$, or when put in terms of spectral slope, a value $s=0 \mbox{ \%/100 $n$m}$\footnote{Spectral slope, $s$, is defined as increase in reddening per 100~nm, with respect to the V-band ($\sim550$~$n$m). eg., a spectral slope of 50 \%/100~$n$m is 50\% more reflective at 650~$n$m than at 550~$n$m.}. For a visual demonstration of this projection, see Appendix Figure~\ref{fig:projection_instruction}. The projection values are tabulated in Table~\ref{tab:projextras}.

This new projection is similar in concept to projections derived from Principle Component Analysis. Though, rather than adopting an algorithm that finds the  \emph{linear} projection that maximizes the data variance along a single basis vector, our approach is agnostic of the dataset in question. Rather this projection is physically driven by the observation that in general, bulk colour samples of icy bodies (Jupiter Trojans, comet nuclei, centaurs, KBOs, etc.) tend toward the reddening curve, with usually only moderate spectral deviations away from it.

We present the projection of the main Col-OSSOS data, (g-r) and (r-J), into the PC space in Figure~\ref{fig:grJ}. From this figure, the primary advantage of this projection is apparent - the non-linear structures in colour space that follow the reddening curve are nearly rectified parallel to the ordinates of the new basis. For example, the diagonal region that is absent of objects in grJ colour space now appears as an obvious bifurcation in $PC^2_{\textrm{grJ}}$.  Hartigan's DIP test of bimodality \citep{Hartigan1985} suggests that this apparent bifurcation has a 20\% chance of occurring randomly. The gap between the two peaks is only $\sim0.1$~mags, and so we consider smaller uncertainty subsets of the (g-r) and (r-J) sample. The probability that apparent bifurcation is random chance is $<2\%$ for uncertainty values in both colours of 0.06 to 0.1~mags. For example, with uncertainties of $\Delta(g-r)\leq0.1 \mbox{ and } \Delta(r-J)\leq0.1$ (79 of 110 g-r and r-J colour pairs), the probability is 2\%. The probability reduces even further still, to 0.0015\% for uncertainties in both colours $<0.09$~mags. The sample with uncertainty $<0.06$~mags is too small for meaningful inference.  
Considering a sample that rejects the most uncertain points provides a better representation of the significance of the bifurcation in $PC^2_{\textrm{grJ}}$ because the observed gap between the two inferred populations is only $\sim0.1$ magnitudes. Thus, including objects with larger uncertainties will only mask the bifurcation. We find no evidence for further bifurcations in the colour, or projected colour spaces. That is, we only have evidence for 2 separate groups in the (g-r) and (r-J) colour space. %Those two groups are relatively cleanly divided into two classes above and below $PC^2_{\textrm{grJ}}=-0.13$. 
A value of $PC^2_{\textrm{grJ}}=-0.13$ provides a relatively clean division into two classes.
We label these two classes, BrightIR and FaintIR, as one class is notably brighter in the NIR with respect to their optical colours than is the other class.

In the Col-OSSOS sample, the objects  with $PC^2_{\textrm{grJ}}>-0.13$ are almost entirely all dynamically excited objects. Of the 52 objects (57 unique measurements) in that region, only 6 are CCKBOs, and only 3 by the $i_\textrm{free}$ definition.  A higher concentration of CCKBOs are FaintIR objects ($PC^2_{\textrm{grJ}}<-0.13$).  Of the 40 with $PC^2_{\textrm{grJ}}<-0.13$ (excluding 2013 JK64, see below), 22 are dynamically excited, and the rest are CCKBOs; $PC^2_{\textrm{grJ}}<-0.13$ is where the majority of CCKBOs are found.

Intriguingly, the majority of objects with $PC^2_{\textrm{grJ}}<-0.13$ also have $PC^1_{\textrm{grJ}}>0.4$. Only 6 objects have values $PC^2_{\textrm{grJ}}<-0.13$ and $PC^1_{\textrm{grJ}}<0.4$. Five of those 6 objects have similar colours, with $(g-r)\sim0.75$ and $(r-J)\sim1.2$ ($PC^1_\textrm{grJ}\sim0.2$, $PC^2_{\textrm{grJ}}\sim-0.2$). Examination of their colours reveals that those 6 objects are closer to the objects with $PC^2_{\textrm{grJ}}>-0.13$, suggesting that they are a tail of the BrightIR class rather than belonging to the FaintIR class with lower $PC^2_{\textrm{grJ}}<-0.13$ and $PC^1_{\textrm{grJ}}>0.4$. This is true of those 6 objects in both the colour and the projection spaces. All six are also dynamically excited KBOS. We suggest that these 6 objects are members of  the BrightIR class, and that a two-dimensional boundary exists between the two classes of KBOs: FaintIR has $PC^1_{\textrm{grJ}}\gtrsim0.4$ and $PC^2_{\textrm{grJ}}<-0.13$, and BrightIR objects fall outside this region.

Using the above two dimensional boundary between  BrightIR and FaintIR objects, of the 6 BrightIR CCKBOs, three objects with $i_\textrm{free}$ too inflated to be considered CCKBOs by that definition are so-called \emph{blue binaries} \citep{Fraser2017}. These objects, identified by their less red optical colours, and 100\% binary rate, are theorized to be push-out survivors from more interior regions of the Solar System -- though some difficulties with this interpretation have been pointed out \citep{Nesvorny2022preprint}. We consider these interesting objects in the Discussion Section. %That these objects all have larger $i_\textrm{free}$ compared to the truly cold population demonstrates that these objects have also experienced more excitation than the rest of the cold population, that at least is consistent in principle, with the idea that the origins of the \emph{blue binaries} are found outside the cold classical region. 

Our results suggest that the dynamically excited KBOs occupy a different optical-NIR colour space than do the CCKBOs. To test the statistical significance of the apparent differences in distribution of $PC^2_{\textrm{grJ}}$ values between the CCKBOs and the dynamically excited KBOs, we turn to the non-parametric, 2-sample Anderson-Darling (AD) statistic, which we calibrate with bootstrap sampling with repeated samples allowed \citep{Fraser2012}. From this we find that the probability that the CCKBOs and excited KBOs share the same $PC^2_{\textrm{grJ}}$ distribution is essentially 0; in 10,000 evaluations of the bootstrapped AD statistic, not one evaluation produced a value as extremal as the observed sample. That is, the CCKBOs -- by either definition -- span an entirely different distribution of $PC^2_{\textrm{grJ}}$ values than does the full excited sample. We highlight that the application of the AD statistic directly to the (r-J) colour distributions of the two dynamical samples results in a probability of 3\%. Thus, by projecting into PC space, the apparent bifurcation of the colour distributions is maximally separate, and significantly more rectilinear.

\begin{figure*}[h]
\epsscale{1.16}
\plottwo{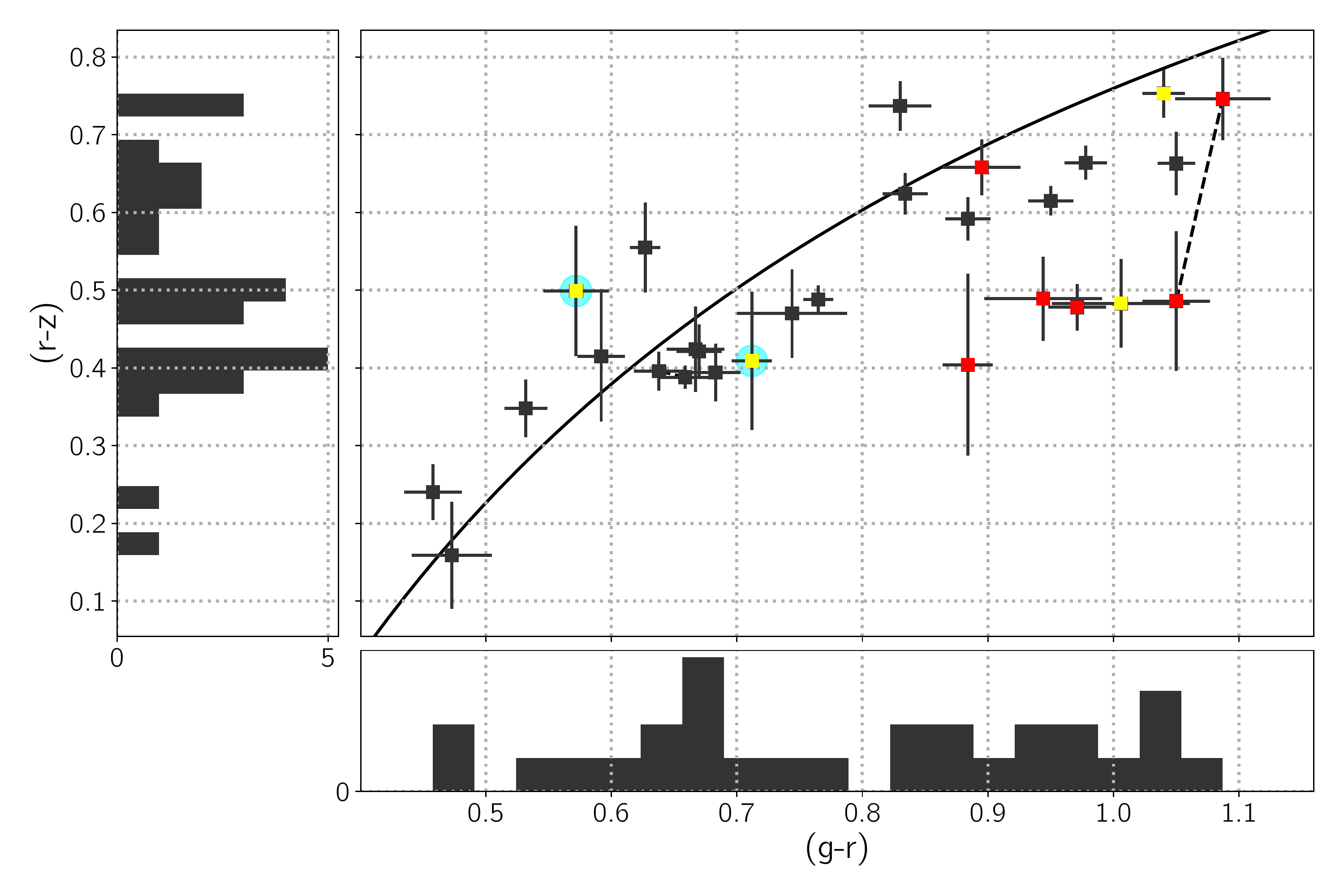}{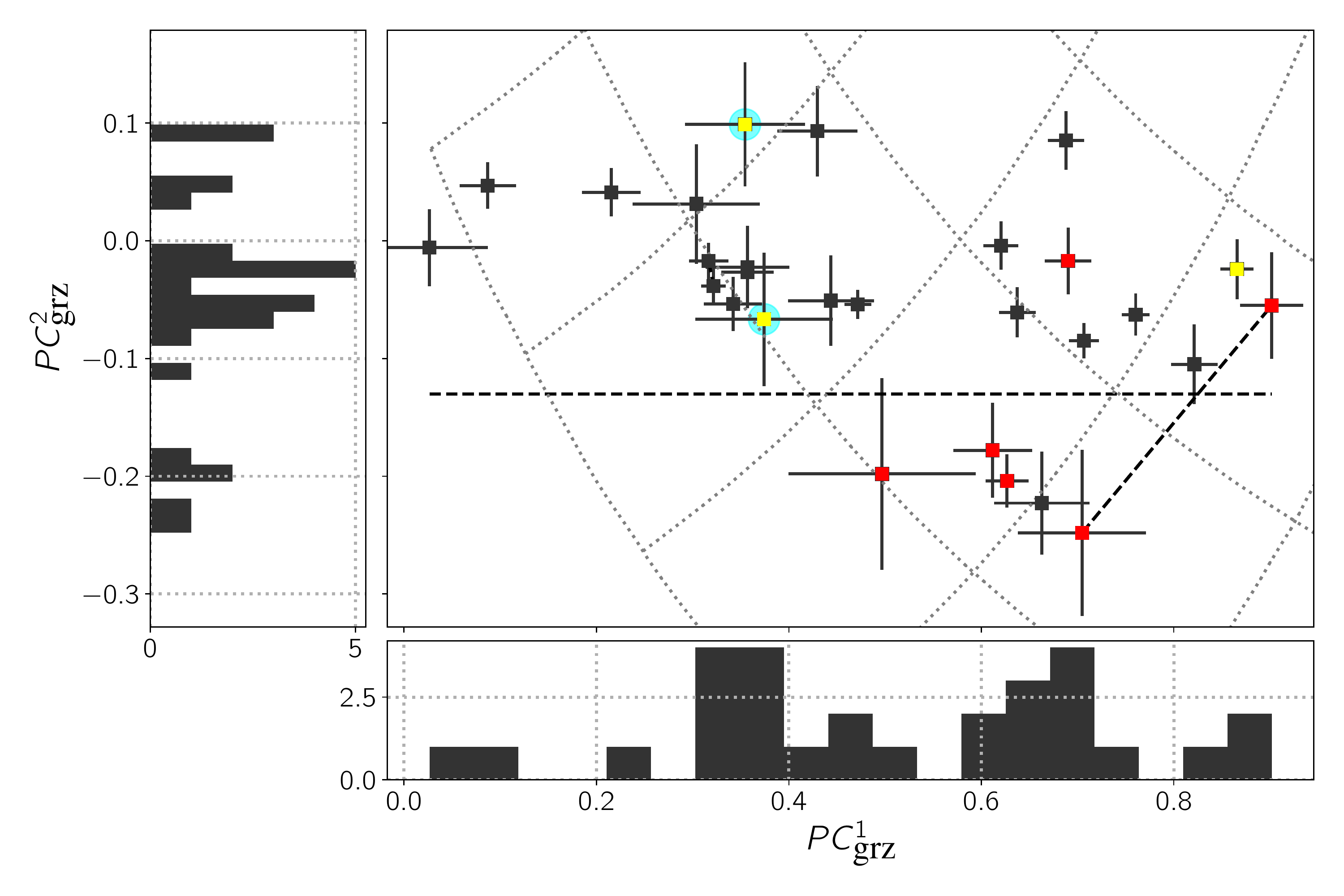}
\caption{\textbf{Left:} The (g-r) and (r-z) colours of Col-OSSOS targets. Repeat measurements for targets have been connected by dashed lines. \textbf{Right:} The projection values $PC^1$ and $PC^2$ of the (g-r) and (r-z) colours shown in the left panel. Red points are CCKBOs defined by their $i_{\textrm{free}}$, yellow and red points together are CCKBOs by the historical definition based on $i$, and black points are all those excited KBOs outside the CC region.  The horizontal line at $PC_\textrm{2}=-0.13$ marks the value below which all FaintIR objects are found. These mainly consist of CCKBOs. The colour scheme follows that of Figure~\ref{fig:grJ}. \label{fig:grz}}
\end{figure*}

The result that the majority of CCKBOs occupy a colour space of bluer NIR colours than similarly optically coloured excited KBOs is similar to the result found by \citet{Pike2017}. From Col-OSSOS data, they found that many CCKBOs have bluer (r-z) colours than their excited counterparts. We revisit that result here, and present (g-r) vs. (r-z) along with the reddening curve projection of those colours in Figure~\ref{fig:grz}. The colours we present here use a slightly enhanced reduction of the same data in \citet{Pike2017}. Unfortunately, a number of the colours reported in that paper were influenced by poor PSF star selection -- itself caused by the highly undersampled PSF in much of the raw imagery -- which resulted in poorly estimated aperture corrections. We have corrected this issue during a reanalysis of the data. Despite this error, the conclusions drawn in \citet{Pike2017} are not influenced: many CCKBOs occupy a unique region of (g-r) and (r-z) colour space. This is apparent from the distribution of $PC^2_{\textrm{grz}}$ values; all the dynamically excited KBOs for which we have (r-z) measurements have $PC^2_{\textrm{grz}}>-0.13$, and 5 of 10 CCKBO measurements fall below this value. 

Application of the AD statistic to the $PC^2_{\textrm{grz}}$ distribution demonstrates that the probability that the CC and non-CCKBO samples are drawn from the same parent distribution  is only 0.05\%. We draw the same conclusions as \citet{Pike2017}; like in the grJ space, the colours of CCKBOs are differently distributed than their excited counterparts in the grz space as well. Unlike in the grJ colour space, however, no statistically significant bifurcation in $PC^2_{\textrm{grz}}$ is apparent. This is likely a sample size effect as Col-OSSOS has only 26 objects (and two repeated) with (g-r) and (r-z) colours. Similarly, the lack of dynamically excited objects with $PC^2_\textrm{grz}<-0.13$ is probably just a sample size effect; from the grJ sample, we might expect one such excited object with these colours when we find zero.

We highlight that due to spectral behaviour differences between (r-z) and (r-J), the value of $PC^2_{\textrm{grJ}}=-0.13$ that separates BrightIR from FaintIR may not be the same for $PC^2_{\textrm{grz}}$. The small grz sample however, precludes searching for variations.

In the grz colour space, no observed objects are found with intermediate colours, (g-r)~$\sim0.8$. The DIP test, however, demonstrates that in projected grz the distribution of objects with $PC^2_{\textrm{grz}}>-0.13$ has a probability of unimodality of 73\%. Moreover, a number of objects are found with those (g-r) colours in the larger grJ sample, implying that the apparent lack of objects in that region is merely a result of the smaller grz sample, and not a real feature of the colour distribution. 

An intriguing result is found when one considers the (r-J) colours of those objects in the grz sample. All objects with $PC^1_{\textrm{grz}}>0.75$ and $PC^2_{\textrm{grz}}>-0.13$ have $PC^2_{\textrm{grJ}}<-0.13$.  It seems that just because an object is found to have $PC^2<-0.13$ in one NIR band, does not necessarily mean that it will have $PC^2<-0.13$ in a different NIR band. We will discuss this result within the context of a spectral model in Section~\ref{sec:model}.

We highlight the 5:2 resonant object, 2013 JK64 (OSSOS designation o3o11). This object exhibits significant variations in its (r-J) colour (see Figure~\ref{fig:grJ}). The $\sim0.5$ magnitude variation in is so extreme as to result in confident but different classification into BrightIR  at one visit, and FaintIR in the next. It is notable that JK64 does not exhibit intermediate colours between the BrightIR and FaintIR classes, but instead the measured values fall on either side of the $PC^2=-0.13$ value leading to an ambiguous classification for this interesting target. This is a similar result as that of CCKBO, 2013 UN15 (o3l63) which exhibits variable (r-J) and (r-z) colours, resulting in ambiguous classification from the (r-z) colour alone. We consider these targets along with KBOs that are known to be spectrally variable in the Discussion Section.

\begin{figure*}[h]
\epsscale{1.16}
\plottwo{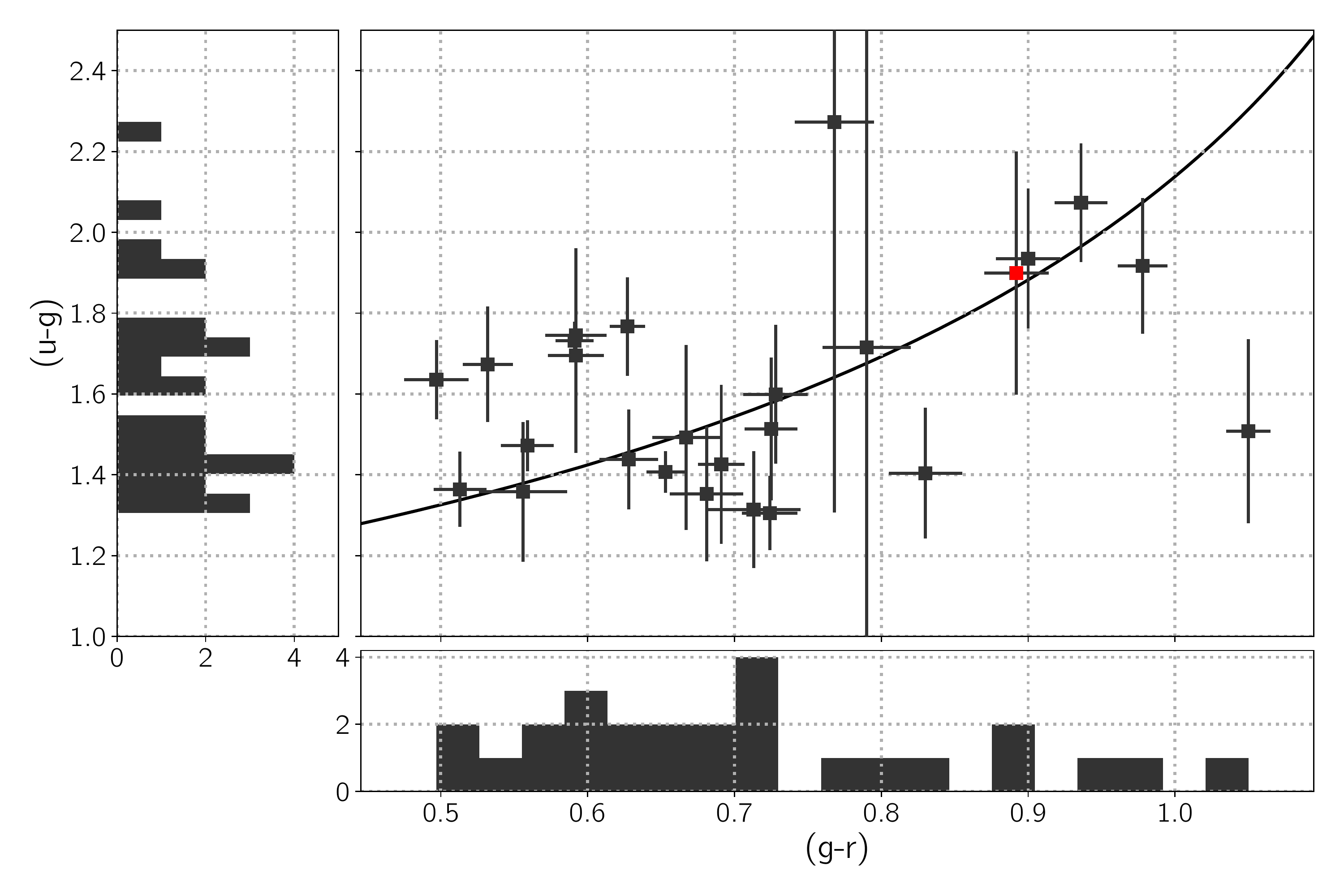}{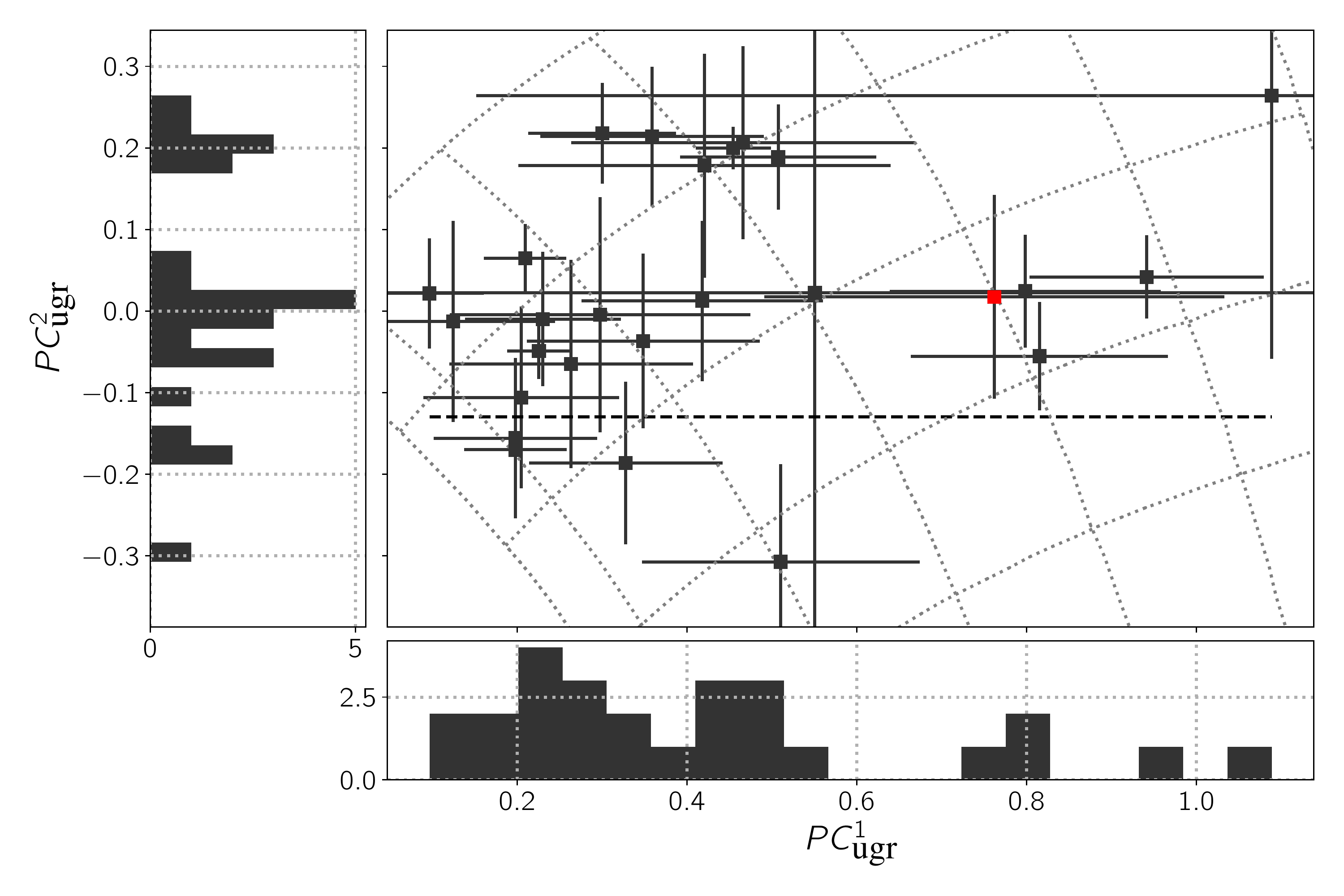}
\caption{\textbf{Left:}  The (u-g) and (g-r) colours of Col-OSSOS targets. \textbf{Right:} The projection values $PC^1$ and $PC^2$ of the (g-r) and (u-g) colours shown in the left panel. Cold classical CCKBOs are marked in red.  In these colours, CCKBOs cannot be differentiated from excited KBOs, to the SNR limits of the dataset. The colour scheme follows that of Figure~1; red points are CCKBOs defined by $i_{\textrm{free}}$, and black are KBOs outside the CC region. \label{fig:ugr}}
\end{figure*}

Finally, we consider the (u-g) colours sample, which we present alongside their projection in Figure~\ref{fig:ugr}. Unlike in the NIR bands, the single CCKBO in the (u-g) sample does not stand out, but rather, falls in the same range of (u-g) colours as the excited KBOs that possess similar (g-r) colours. This may be a result of the low SNR (u-g) colours compared to the (r-z) and (r-J), though it may be that the distribution of $PC^2$ is not bimodal in the NUV. 

For objects with (g-r)$<0.75$, the range of  (u-g) colours is compressed, with $1.25\lesssim(u-g)\lesssim1.8$ compared to the redder objects with $1.2\lesssim(u-g)\lesssim2.5$. This is most likely a result of the lower optical albedos of the less red KBOs \citep{Lacerda2014}. That is, because the albedos are lower, there simply is not as large a range of physically allowable (larger than zero) u-band albedos, while still having red (u-g) colours. This will be the topic of a future paper in which a compositional interpretation of the Col-OSSOS sample is presented. In summary, in the NUV-optical space, we find no evidence that the CCKBOs standout with a different colour distribution than the non-CCKBOs possessing similar optical colours.  

% in grz o3l63 crosses the classification line
% in 

\section{Published Colour Datasets} \label{sec:publishedcolours}

In the following section, we consider other optical and NIR colour datasets. In particular, we analyze how the KBO colour distributions in other filters appears in the reddening curve projection space. 

\subsection{H/WTSOSS}

We turn our attention to the observations of the Hubble/Wide Field Camera 3 Test of Surfaces of the Outer Solar System \citep[H/WTSOSS;][]{Fraser2012}.  Like Col-OSSOS, H/WTSOSS was an optical-NIR photometric survey of a large sample of KBOs, but of a target sample with unknown discovery and tracking biases and no limiting magnitude. Unlike in \citet{Fraser2012} we do not limit ourselves to only those data with colour uncertainty less than 0.1 magnitudes in (F606w-F814w) and (F814w-F139m), but consider the entire dataset in our projection analysis.

\begin{figure*}[h!]
\epsscale{1.16}
\plottwo{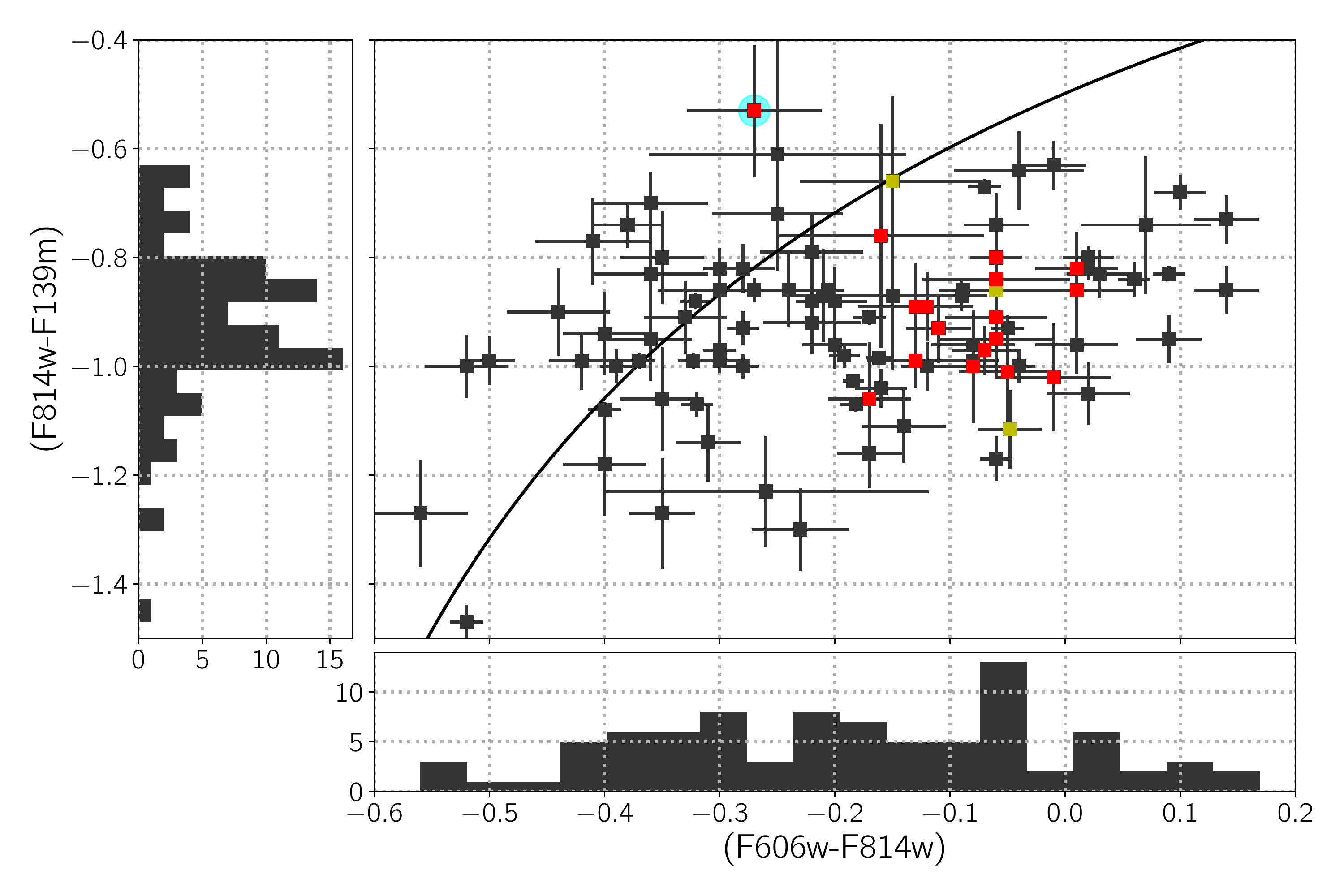}{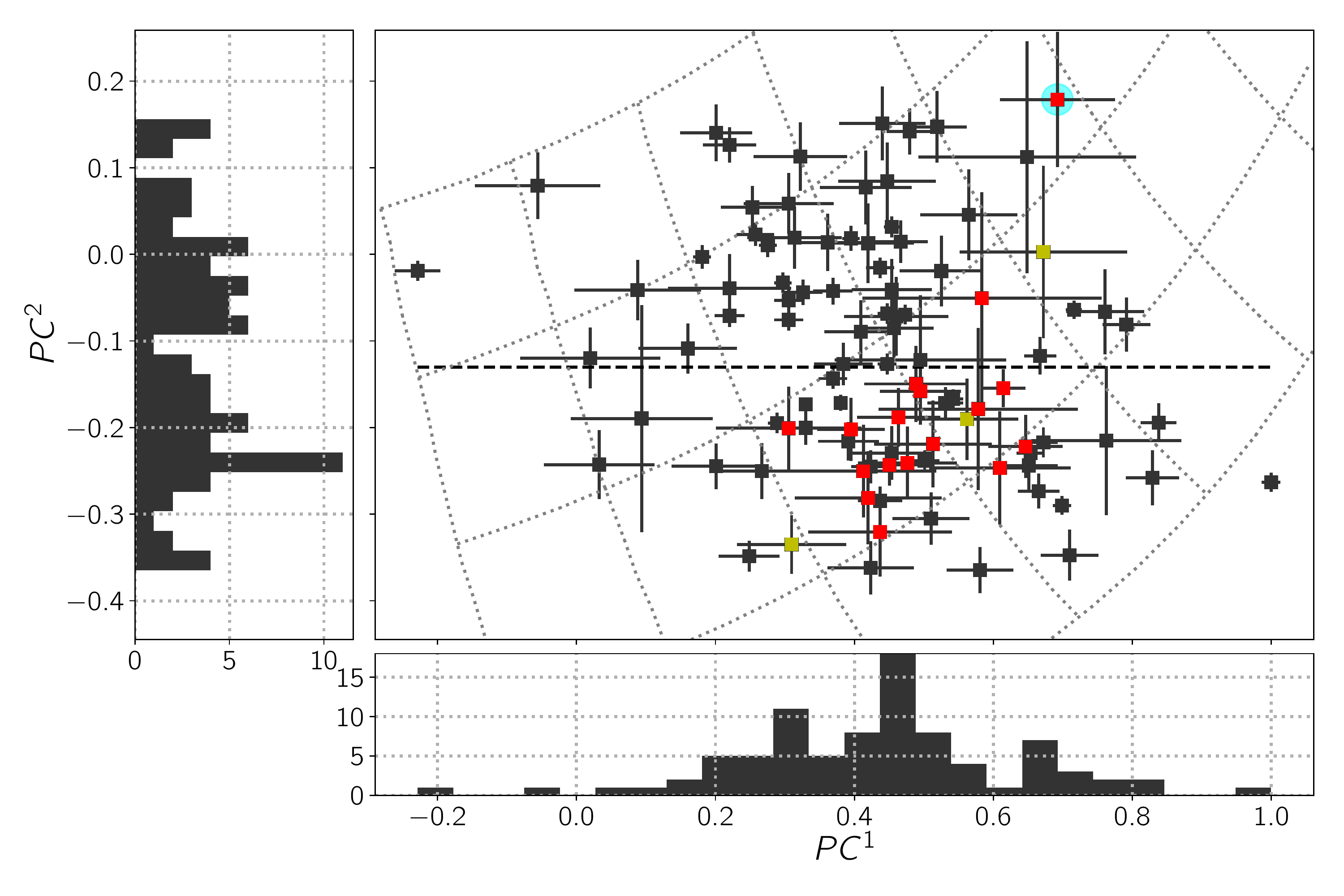}
\caption{\textbf{Left:}  The (F606w-F814w) and (F814w-F139m) colours of Kuiper Belt Objects and Centaurs reported in \citep{Fraser2012}. The reddening curve which is shown by the solid line. Colours are in the STMAG system. The left and bottom panels show histograms of the colours. \textbf{Right:} The projection values $PC^1$ and $PC^2$ of the (F606w-F814w) and (F814w-F139m) colours shown in the left panel. CCKBOs are marked in red.  The left and bottom panels show histograms of $PC^1$ and $PC^2$. The horizontal line at $PC^2=-0.13$ marks the value below which all FaintIR objects are found. These mainly consist of CCKBOs. The colour scheme follows that of Figure~\ref{fig:grJ}; red points are CCKBOs defined by $i_{\textrm{free}}$, yellow and red points are those based on $i$, and black are KBOs outside the CC region. The \emph{blue binary} 2003 HG57 is outlined with a cyan halo. All histograms show only objects that have uncertainties less than 0.1 magnitudes in both colours to avoid inclusion of uncertain measurements that may mask the presence of the bifurcation in $PC^2$. \label{fig:HWTSOSS}}
\end{figure*}

 We present a projection into PC colour space of the  (F606w-F814w) and (F814w-F139m) colours from the H/WTSOSS sample independent of the Col-OSSOS (g-r) and (r-J) in Figure~\ref{fig:HWTSOSS}. This sample reveals a similar bifurcation in $PC^2$ and at a similar value, $PC^2=-0.13$, as in the Col-OSSOS sample\footnote{For ease of reading, we omit the filter label subscripts when discussing the projection of the H/WTSOSS data.}. When considering only those objects with uncertainty of less than 0.1 magnitudes in each colour, we find an 8\% chance of a bifurcation in $PC^2$, as evaluated with the DIP test. We point out that this bifurcation is the same as the diagonal bifurcation in optical-NIR that has already been found in  the H/WTSOSS sample, albeit revealed with a much simpler analysis than the more complicated multi-dimensional clustering methods previously used \citep{Fraser2012}. 

Dynamically, the H/WTSOSS sample shows a very similar result as that of the Col-OSSOS sample. That is, objects with $PC^2>-0.13$ in the H/WTSOSS filters are predominantly dynamically excited objects:  only 3 of the 18 H/WTSOSS CCKBOs have $PC^2>-0.13$. Further, like the Col-OSSOS objects, the H/WTSOSS objects with $PC^2>-0.13$ span the full optical or $PC^1$ colour range.

In general, the H/WTSOSS and Col-OSSOS datasets present the same result:  in the optical-NIR colour space there is a bifurcation in $PC^2$ above and below a value of $\sim-0.13$, and CCKBOs are predominantly found with $PC^2<-0.13$, while most  excited KBOs have $PC^2>-0.13$. %We point out that the difference in the value of the $PC^1$ dividing line between the Col-OSSOS and H/WTSOSS samples is a result of the different wavelengths of the filters used in each survey. 
That the same systematic colour-dynamical divisions occur in both datasets is a demonstration that the apparent structures in projected colours are a robust feature of the KBO population, and not a consequence of filter or target selection. 

We note that no bifurcation in $PC^2$ was detected in the (F606w-F814w) and (F139m-F153m) colour space. That is to say, the two groups of objects found above and below $PC^2=-0.13$ fully overlap in the (F139m-F153m) colour. F153m samples the centre of the $1.5 \mbox{ $\mu$m}$ water-ice absorption band. The lack of bifurcation implies that the BrightIR and FaintIR objects share similar water-ice absorption depths, and differ only in the behaviour of their optical-gap absorptions which dominate at shorter wavelengths. 

\subsection{Optical Spectrophotometric Datasets}

The results we have presented thus far have demonstrated that in the NIR, the CCKBOs occupy a different region of colour space than do most dynamically excited KBOs. This is apparent at wavelengths, $\lambda \geq0.9 \mbox{ $\mu$m}$. We wish to test if such a bifurcation is present at shorter bands. To that end, we make use of the  \citet{Peixinho2015} optical colours dataset. This is a large and reliable multidimensional dataset of optical colours derived from Johnson BVRI photometry that was extracted from prior published optical colours. Importantly, they reject all colour measurements spanning timespans larger than 1.5~hours, in an attempt to minimize the influence of lightcurve effects on the reported colours. The main weakness of this colours sample is the lack of simultaneity in more than two bands. It is common that reported colours were gathered at different epochs. It should be noted however, that only a small fraction of KBOs have exhibited optical spectral variability \citep[see][and Table~2]{Fraser2015}, and as such, the Peixinho dataset is currently the best available for our purposes. 

In our analysis of the Peixinho dataset, we projected 4 different combinations of colours (B-V), (B-R), (V-R), (V-I), and (R-I). The most populous subset of that dataset is (V-R) and (R-I), the projection of which is presented in Figure~\ref{fig:Peixinho_VRI}. We also considered subsets with cuts in uncertainty (0.05 mags, 0.1 mags, etc.) and cuts in absolute magnitude. In all cases, only hints of bifurcation in $PC^2$ were found. This is most prominent in the colours involving I-band. No statistically significant bifurcation was found, however. The distribution of $PC^2$ appears to be unimodal when considering only optical colours. This is true regardless of dynamical population cuts. While the CCKBOs all possess red optical colours, and therefore have high $PC^1$ values, the do not differentiate from the excited objects in $PC^2$. It seems plausible that the FaintIR and BrightIR classes do differentiate in colours involving the I-band, though the two classes appear to be closer together in $PC^2$ compared to longer wavelengths, and so this signal is not detectable in the available colour datasets. Combining these results with those from Col-OSSOS, we conclude that the BrightIR and FaintIR classes as defined above, predominantly overlap in the ugr/BVRI colours space. That is, both classes can be found tracking the reddening curve, and exhibit nearly linear spectra through the optical range. Any spectral differences between the two classes only become readily apparent at wavelengths $\lambda\gtrsim 0.9 \mbox{ $\mu$m}$.

\section{Spectral Interpretation} \label{sec:model}

In this section, we consider the properties of KBO spectra that could account for the structures we have detected in the UV-optical-NIR colour distributions. Generally, there are five properties of the UV-optical-NIR colour space that we are interested in understanding:

\begin{enumerate}
\item The tendency to follow the reddening curve in the ugr/BVRI filter range
\item The bifurcation in the optical colour distributions (eg. (g-r) (B-R), etc.)
\item Deviations away from the reddening curve to less-red colours at NIR wavelengths ($\lambda \gtrsim 0.8 \mbox{ $\mu$m}$)
\item The separation of two groups in an optical-NIR colour space (e.g., grJ, grz, or the filters utilized by H/WTSOSS), separated at a $PC^2\sim-0.13$
\item The fact that not all objects consistently fall above or below $PC^2\sim-0.13$ across all of the band passes considered.
\end{enumerate}

In the most general sense, the spectra of KBOs can be described by red, linear optical spectra, that turnover to nearly linear NIR spectra with less-red slopes \citep{Cruikshank1998,Barucci2011,Gourgeot2015,Grundy2020,Seccull2021}. Observed exceptions to this rule include wavelength regions influenced by water-ice \citep{Barkume2008}, objects 2003 AZ84 and 2004 EW95 which appear to have spectra closer to C-type asteroids than KBOs \citep{Fornasier2004, Seccull2018}, and the spectra of the largest KBOs, which exhibit spectra dominated by select ices that are too volatile to survive on the surfaces of smaller KBOs \citep{Brown2011a,Brown2012}. To that end, we adopt a simple spectral model with which to interpret the colour distributions we present above. A model spectrum is defined by a linear optical spectrum and a linear NIR spectrum, with the two spectra meeting at some transition wavelength. Free parameters are the optical spectral slope, $s_{opt}$, the NIR spectral slope, $s_{ir}$, and a transition wavelength, $w$.

The distribution of (g-r) and (r-J) colours of Col-OSSOS targets shows a lot of similarity to the distribution of colours of the H/WTSOSS dataset. That is, both show evidence for two separate groups of objects occupying unique ranges of colour. Those classes also show a continuum of colours, with correlated optical and NIR colours. Inspired by the similarity between the two datasets, we turn to the mixing model put forth by \citet{Fraser2012} to model the colours and correlated optical-NIR colours of H/WTSOSS targets. In that model, the members of a compositional class are made up of a mixture of two main material components. The colours of a class then fall along a continuum of colour with the continuum's end-points matching the colours of the materials, with the specific colours of an object determined by the mixture of those two materials. \citet{Fraser2012} found that within that framework, the colours of the two materials that made up a class were consistent with one material being mainly silicate in nature (on the blue end of the class's continuum) and the other component being a water-rich organic material (on the red end of the continuum). 

\begin{figure*}[h]
\plotone{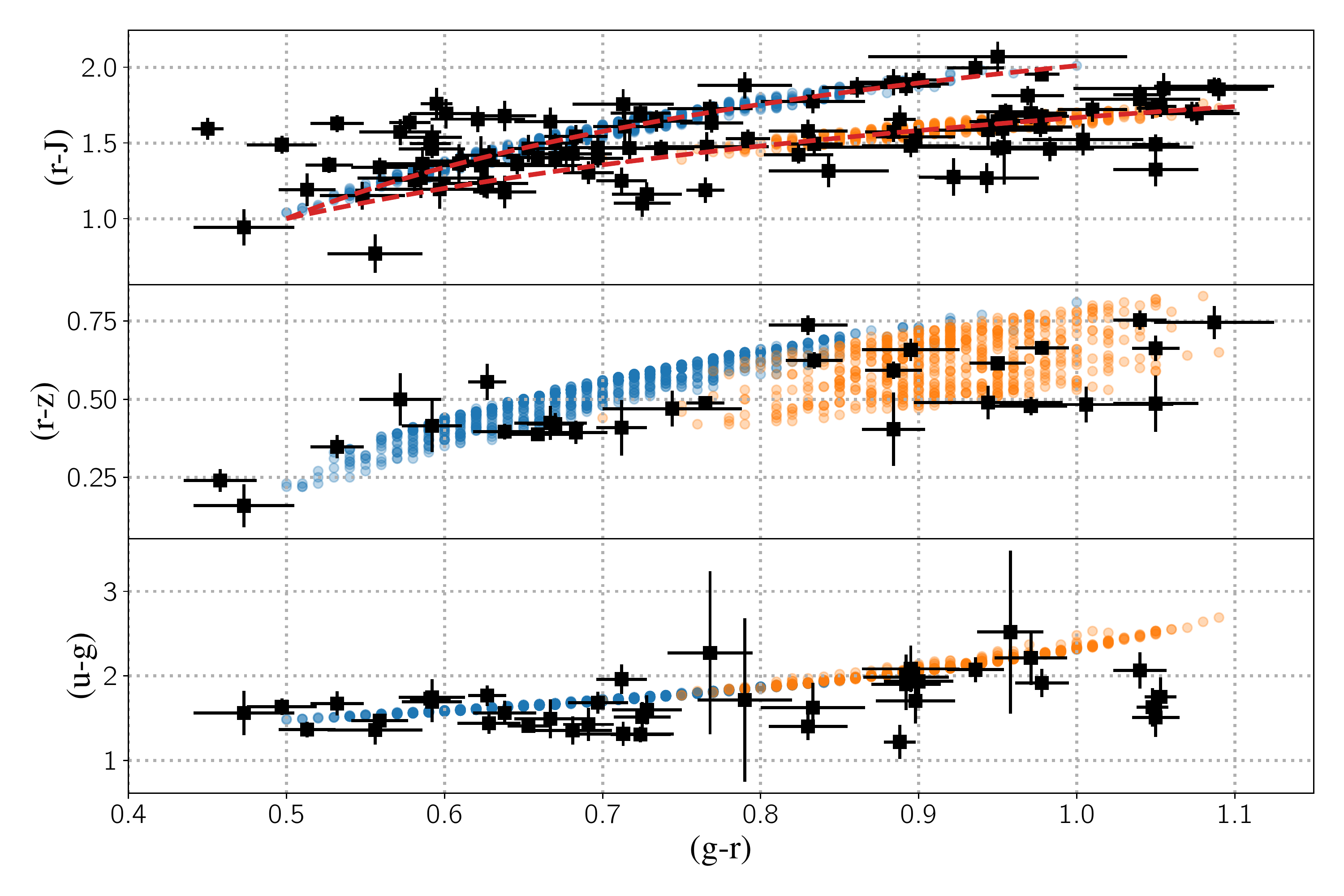}
\caption{ The Col-OSSOS dataset plotted with the mixing model representation (red dashed curves in the top panel). Simulated spectra were those randomly drawn that were consistent with the (g-r) and (r-J) colours of the mixing models. Other colours were evaluated from those spectra. The resultant colours for mixing model classes 1 and 2 are shown by the orange and blue points respectively. \label{fig:mixturemodel}}
\end{figure*}

Using the mixture model approach, we model the two colour groups detected in the Col-OSSOS grJ colour space with two separate classes  as seen in Figure~\ref{fig:mixturemodel}. These two mixture model curves have been chosen in a similar fashion to those proposed in \citet{Fraser2012}: both classes share the neutral-coloured material component, with each red material chosen to describe \emph{only} the features of the Col-OSSOS grJ distribution. These two mixture model curves are presented in Figure~\ref{fig:mixturemodel}. We emphasize that these mixture models are not a fit to observed colours, but merely are a representative model. Though we do point out that with only 6 parameters (two colours, (g-r) and (r-J), for each material component), 72 of the 110 observed (g-r) and (r-J) colours are $1-\sigma$ consistent with the chosen mixing models, and 91 of 110 are $2-\sigma$ consistent. We deem the chosen model as a reasonable representation of the grJ colour distribution.

From the two  mixture models, syntehtic spectra were sampled. This was done by randomly drawing model spectral parameters $(s_{opt}, s_{ir}, w)$, and calculating their (g-r) and (r-J) colours, collecting those that are within $0.04$\footnote{This value is chosen to approximate the slight apparent deviations away from non-linearity exhibited by KBOs in the optical, but is relatively arbitrary, as a range of similar values produce equally satisfactory results.} magnitudes of the mixing model curves. The only restriction was made on the NIR spectral slope, requiring $s_{ir}\geq0$, as few to no KBOs exhibit negative spectral slopes in the grzJ spectral range \citep[e.g.][]{Barucci2011}. In this way, we create representative model spectra that broadly reflect the known spectral behaviours of KBOs, and match the (g-r) and (r-J) colours of both surface classes in the Col-OSSOS dataset. The sample of synthetic KBO colours is shown in Figure~\ref{fig:mixturemodel}.

\begin{figure}[ht]
\plotone{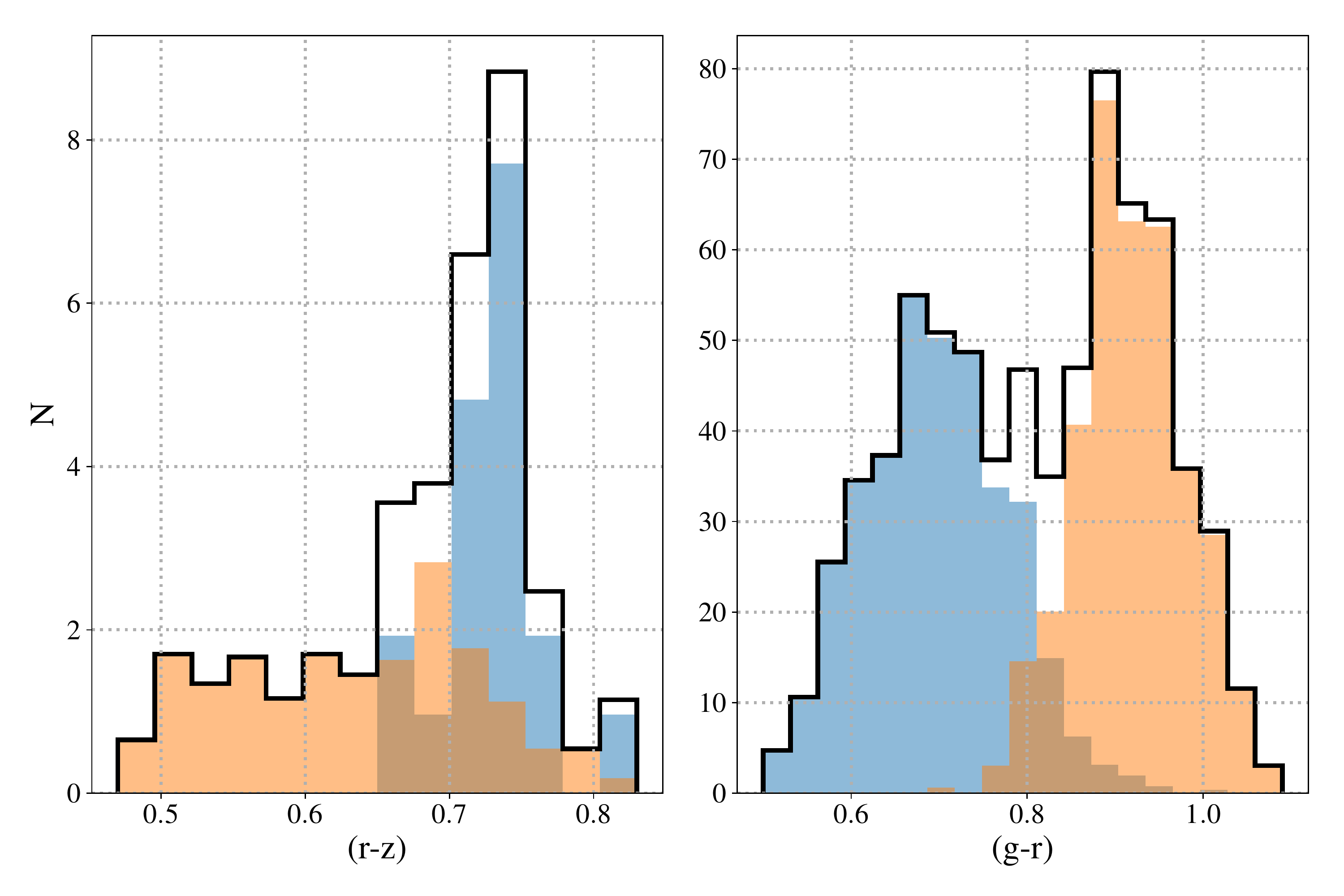}
\caption{ Histograms of (r-z) and (g-r) colour samples from our mixing model. Orange and blue points represent FaintIR and BrightIR objects, respectively, with the black curve showing the overall sample histograms. The histograms were weighted to have an equal number of objects in each class, which is approximately true of the observed sample. The absolute values in this figure are meaningless. The (r-z) histogram is only of optically red KBOs with (g-r)$>0.85$ which show a bimodality in (r-z) that is apparent in both the simulations and in the observations. Neither the simulated or observed (r-z) distribution is bimodal for objects with (g-r)$<0.85)$. \label{fig:model_histograms}}
\end{figure}

As the above modelling approach produces simple spectra, we can also draw colours using other filters, such as z, or VRI. The interesting result of the above modelling is apparent when we consider the predicted (r-z) colours of the model spectra. Specifically, we see that the BrightIR objects -- the objects with redder (r-J) colours -- also have redder (r-z) colours with a spread in (r-z) that falls along a relatively narrow strip. This is a result of the fact that to be consistent with the grJ colours of the redder class requires that $s_{ir}$ be only marginally bluer than $s_{opt}$. FaintIR objects -- the class with bluer (r-J) colours --  exhibit a larger spread of colours in (r-z). This is because of the degeneracy between spectral parameters $s_{ir}$ and $w$ which can result in a specific (r-J) colour from  a large spread of slope and transition wavelength values, thus resulting in a broad range in (r-z) for a single (r-J) value. Moreover, the (r-z) colours of this class tend to pile-up at values $(r-z)\sim0.55$. This occurs when the transition wavelength, $w$, is shorter than the blue edge of the z filter, $w\lesssim0.83\mbox{ $\mu$m}$. The BrightIR class however, has (r-z) values that pile-up near the reddening line. This has the effect of creating a bimodal (r-z) colour distribution for optically red $(g-r)\gtrsim0.85$ KBOs (see Figure~\ref{fig:model_histograms}). This is consistent with the  observed (r-z) colour distribution, which also appears bimodal for red KBOs  \citep{Pike2017}, and implies that the spectra of many CCKBOs have a transition wavelength, $w<0.83\mbox{ $\mu$m}$ \citep{Seccull2021}.

%A similar result is found when one considers the VRI colour space. Specifically,  a bimodal distribution of $PC^2$ is recreated, albeit with a more positive value of $PC^2$ separating the two classes, as was found with the real observations. Like in (r-z), this bimodality is purely the result of degeneracy between $w$ and $s_{ir}$ resulting in a broad range of (R-I) colours for the model FaintIR colours. Moreover, we draw a similar conclusion as above, in that many FaintIR objects must have a transition wavelength $w$ shortward of the I-band filter, with a central wavelength $\lambda\sim0.80\mbox{ $\mu$m}$.

In the framework of only two mixing models, the optical colour distributions of both compositional classes almost fully overlap one another. To account for the bimodal optical colour distribution therefore requires a non-uniform distribution of objects along each of the mixing model curves. That is, uniform sampling in the spectral model parameters, or uniform sampling along the model curves themselves results in a unimodal optical colour distribution. From the Col-OSSOS grzJ colour distribution (or the H/WTSOSS sample) however, it is clear that each class is not spread uniformly in colour space. Rather, BrightIR objects appear to cluster at $(g-r)\sim0.63$, or spectral slope values $s_{opt}=13 \%/100\mbox{ $n$m}$, while FaintIR objects cluster at redder optical colours, $(g-r)\sim0.86$ or $s_{opt}=30 \%/100\mbox{ $n$m}$. Inspired by this observation we adopt a trivial modelling approach. Specifically, when sampling synthetic KBO spectra parameters, we randomly sample $s_{opt}$ from two Gaussian curves, one for each class. We use median values of $s_{opt}=30$ and $13\%/100\mbox{ $n$m}$, with standard deviations of $4$ and $6\%/100\mbox{ $n$m}$, for the FaintIR and BrightIR classes, respectively. The BrightIR class requires a wider width to reflect the broader range of optical colours those objects occupy. $s_{ir}$ and $w$ were sampled uniformly. As seen in Figure~\ref{fig:model_histograms}, when sampled in this way, the model optical colour distribution (e.g. $(g-r)$, $(V-R)$, etc. ) is bimodal. Notably, by virtue of the relatively wide widths of the Gaussians from which $s$ were sampled, the resultant model colours still span the full optical-NIR range of both the BrightIR and FaintIR classes. 

We highlight that the double Gaussian sampling of $s$ is required only to broadly describe the distribution of sources along each mixing line, and hence the bimodal optical colour distribution. The bimodal (r-z) distribution for optically red KBOs is not a result of the Gaussian sampling, but rather, is entirely driven by the location of the wavelengths of the z-band filter, and the overlap between the two mixing model curves.  A bimodal distribution in (r-z) does result even when fictitious colours are sampled uniformly along the mixing model curves. 

To summarize, the four main model assumptions are:

\begin{itemize}
\item Two classes of KBO with the colours of each described by geometric mixing models, with the neutral-coloured end-member material being shared between both classes
\item Simple linear optical and NIR spectra with different slopes, and a transition wavelength
\item NIR spectra do not have negative spectral slopes
\item Sources are found spread across a mixing colour curve, but tend towards an optical spectral slope specific to each class.
\end{itemize}

\noindent
This model reproduces nearly all features of the distribution of optical-NIR colours of KBOs, though it is not without weakness. Specifically, the model breaks down at the shortest and longest wavelengths. That is, the model produces spectra that are too red in the (u-g) colour (Figure~\ref{fig:mixturemodel}). This argues for deviations away from linearity at wavelengths $\lambda\lesssim0.4 \mbox{ $\mu$m}$, or the approximate blue edges of the B and g-bands. Such deviations have been detected in some spectra, such as the spectrum of Pholus  which shows a slightly convex spectrum that is bluer  at $\lambda\sim0.4 \mbox{ $\mu$m}$ than at longer optical wavelengths \citep{Cruikshank1998}. Further weakness is shown for passbands beyond J. In particular, our  model results in objects from the  (r-J) red model compositional class that are too red in the Hubble passbands (F814w-F139m). To rectify this while preserving the (r-z) distribution requires the reddest NIR spectra to turnover to nearly neutral slopes at wavelengths beyond the J-band, $\lambda\gtrsim1.2 \mbox{ $\mu$m}$. An example of such a turnover can be seen in the spectrum of Arrokoth \citep{Grundy2020}.

We emphasize that the model we present here is not an attempt to reflect some of the quantitative properties of the observed or intrinsic colour distributions. For example, the model makes no robust attempt to match the distribution of colours along each mixing model curve, but merely adopts simple Gaussian distributions in optical spectral slope. Moreover, no attempt has been made to match the observed balance of objects between the two compositional classes. Other potential improvements include a more realistic spectral model that includes features such as an optical-NIR transition region and not just a single wavelength, consideration for water-ice absorption, and adjustments to account for a broader wavelength range where the assumption of linear NIR spectra breaks down. These and other considerations will be the topic of a future manuscript.

\section{Discussion} \label{sec:discussion}

%\section{Probability Hacking} \label{sec:p-hacking}
%
%In this section we discuss 

We have presented a new interpretation of KBO colours, in which two populations, which we label as BrightIR and FaintIR occupy different optical-NIR colour regions. Specifically, both classes span the full optical colour range exhibited by KBOs, but are separate in NIR colours: FaintIR is found at moderate  NIR colours while BrightIR has redder NIR colours just shy of the reddening curve. From the data spanning the NUV-optical-NIR range, a clean separation of the BrightIR and FaintIR classes  is not apparent in the ugr, or BVRI band-passses, as both classes span an overlapping range in those colours. The separation is only apparent for longer passes, z, J, F139m etc. This is supported by the results of the DIP test of unimodality. In the 8 different colour pairs we used in our search (3 from Col-OSSOS, 1 from H/WTSOSS, and 4 from the Peixinho dataset), only the projection of the colour pairs:  (g-r) and (r-J); and the (F606w-F814w) and (F814w-F139m) have  low probabilities of unimodality, at 0.0015\% and 8\% respectively. To test the overall significance of this result, we performed a numerical experiment whereby we simulated 8 separate uniform colour distributions of length matching those of the real datasets, and asked if across those 8 samples, if any set  had DIP-test probabilities less than each of the observed values. That is, one sample of the 8 with probability less than $0.0015\%$, and one with less than $8\%$. In 3,000 iterations of this experiment, only 26 met this threshold.  Therefore, we conclude that after accounting for the fact that bifurcations were searched for in multiple datasets, the detected bifurcation remains statistically significant. 

We do acknowledge that our analysis has an air of probability hacking. That is, the bifurcations in each of the colour sets we analyzed were only significant once the maximum measurement uncertainty of each sample was lowered below a specific value. We feel justified in this approach however, as the separations between the two groups detected in each sample were of order the uncertainty threshold below which the statistics become significant. Put another way, when the data were typically more uncertain than the apparent separation of the two underlying groups, the presence of the FaintIR and BrightIR classes was masked. We emphasize that the presence of two groups was detected in three entirely unrelated colour datasets, and across many different unrelated optical and NIR colours. Despite this weakness in our analysis, the detection of the BrightIR and FaintIR classes is statistically significant, and robust. 

% filter sets gru, grz, grJ, BVRI, VRI, BRI, BVI, 6-8-1.3

 The work we present here is not the first time a bifurcation of KBOs into two populations has been published. It has long been apparent that the optical colour distribution of small ($H\gtrsim6$) KBOs is bifurcated \citep[see][for recent examples]{Fraser2012, Peixinho2012, Peixinho2015, Tegler2016, Marsset2019}. This can also be seen in the (g-r) colours of Col-OSSOS (see Figure~\ref{fig:grJ}). The bifurcation in the optical colour distribution has been interpreted as evidence for the presence of two distinct populations of KBOs, separated with optical spectral slopes smaller and larger than $s\sim20\%/100 \mbox{ $n$m}$. These populations are now colloquially referred to as \emph{Less Red} (LR) and \emph{Very Red} (VR). \citet{Tegler2016} pointed out that the LR and VR colour distribution tails must slightly overlap, hence why published significance of bifurcation values rarely exceeds 2-$\sigma$, even for colours samples counting in the hundreds. The H/WTSOSS sample, and the data presented in \citet{Schwamb2019} demonstrated that a cleaner bifurcation is apparent in optical-NIR colour space. Generally, past works have concluded that the LR population spans a spectral slope range $0\lesssim s\lesssim25\%/100 \mbox{ $n$m}$, and  $20\lesssim s\lesssim50\%/100 \mbox{ $n$m}$ for the VR population.

We and other teams have argued in the past for three spectro-dynamical classes. That is, alongside the LR, and VR KBOs,  the CCKBOs make a third class \citep{Fraser2012, Pike2017}. Here, we argue that this interpretation is incorrect. Rather, we have found evidence for only two spectral classes that significantly overlap in optical colours. The bimodality in optical colour of excited KBOs and Centaurs is a result of the distribution of each spectral class along their mixing lines, with a preference for one class to be found at more neutral optical colours than the other. Though we emphasize that both classes have members with optical colours $0.7<(g-r)<1.0$.  As the BrightIR and FaintIR class have an overlap that spans much of the optical colour range of KBOs, a statistically robust bifurcation will never be found with optical colours alone. This subtle transformation of the modern interpretation of the KBO colours distribution has cosmogonic implications that are less than subtle, and will be discussed below.

We find no evidence of any further bifurcation beyond the two classes. This is true of the Col-OSSOS, H/WTSOSS, and Peixinho samples, and subsets thereof. At most, only two groups are apparent in all of the colour datasets we have considered, which span the UV-optical-NIR range. And notably, the bulk properties of all observed colours across all colours datasets can be modelled with only two classes.

The classification scheme we present here is a two class system, and is different than the three (or more) taxon systems previously put forth. In most cases, the CCKBOs were assigned their own taxon. This was motivated by their distinct physical properties including binarity \citep{Noll2008}, size distribution \citep{Fraser2014a},  high albedos \citep{Brucker2009}, and unique colours \citep{Pike2017} compared to the excited KBOs. The optical bimodality of the excited sample was then interpreted as evidence for at least two additional taxons. We have shown here that once the true shape of the FaintIR class is revealed, no additional bifurcation in the optical-NIR colour distribution of the excited KBOs is apparent. Rather, we find evidence for only two main taxa.

Our assertion of the presence of only two classes for the bulk of small KBOs comes with two predictions about KBOs that must hold true:

\begin{itemize}
\item The albedos of VR KBOs that belong to the BrightIR class will be as equally low as the LR objects
\item The size distribution of excited KBOs from the FaintIR class will match the size distribution of red cold classical KBOs, and inversely, BrightIR objects in the cold classical range will have a size distribution that matches that of dynamically excited BrightIR objects.
\end{itemize}

The former prediction arises from the observation that LR objects (less red optical colours, $s\lesssim 20 \%/100 \mbox{ $n$m}$) typically have lower albedos \citep{Lacerda2014,Fraser2012}. It seems that CCKBOs have systematically higher albedos than at least the LR KBOs, with most having albedos $\rho \gtrsim 0.1$ \citep{Brucker2009,Lacerda2014}. Considering our two group classification then, it follows that all FaintIR objects will possess equally high albedos. VR objects (red optical spectral slopes, $s\gtrsim 20 \%/100 \mbox{ $n$m}$)  with high albedos can be found in all excited classes, consistent with this idea. Similarly, the LR objects have low albedos $\rho\lesssim0.1$, with only a few exceptions. It follows that all BrightIR KBOs will have lower albedos. The observed albedos are compatible with this prediction, as optically red KBOs with low-albedos have been detected in the excited KBO populations. Alternatively,  it is possible that optically red BrightIR bodies may have albedos like those of FaintIR objects, though this would imply that BrightIR objects have a large range of albedos alongside the large range of optical colours.

The second prediction arises from the observation that the low-inclination component of the Kuiper Belt possesses a different size distribution than the excited bodies \citep{Bernstein2004,Fraser2009,Fuentes2009,Petit2011,Fraser2014a}. While the low and high-inclination components are mixtures of bodies from BrightIR and FaintIR objects, the CCKBOs (or the low-i component) are predominantly FaintIR members. This implies that the steep size distribution of the low-i component is a reflection of the size distribution of FaintIR bodies, and that BrightIR objects have a shallower size distribution for bodies larger than the break diameter $D\gtrsim100$~km. The prediction then is that FaintIR objects found on excited orbits will have a steep size distribution like that of the CCKBOs. The contrary possibility, that of differing size distributions for different dynamically separated classes of FaintIR object, would be very hard to rectify in our current understanding of the formation of the Kuiper Belt, and would challenge the two group classification scheme we advocate for here.

\subsection{Spectral Shapes of BrightIR and FaintIR Objects}

\begin{figure}[h]
\plotone{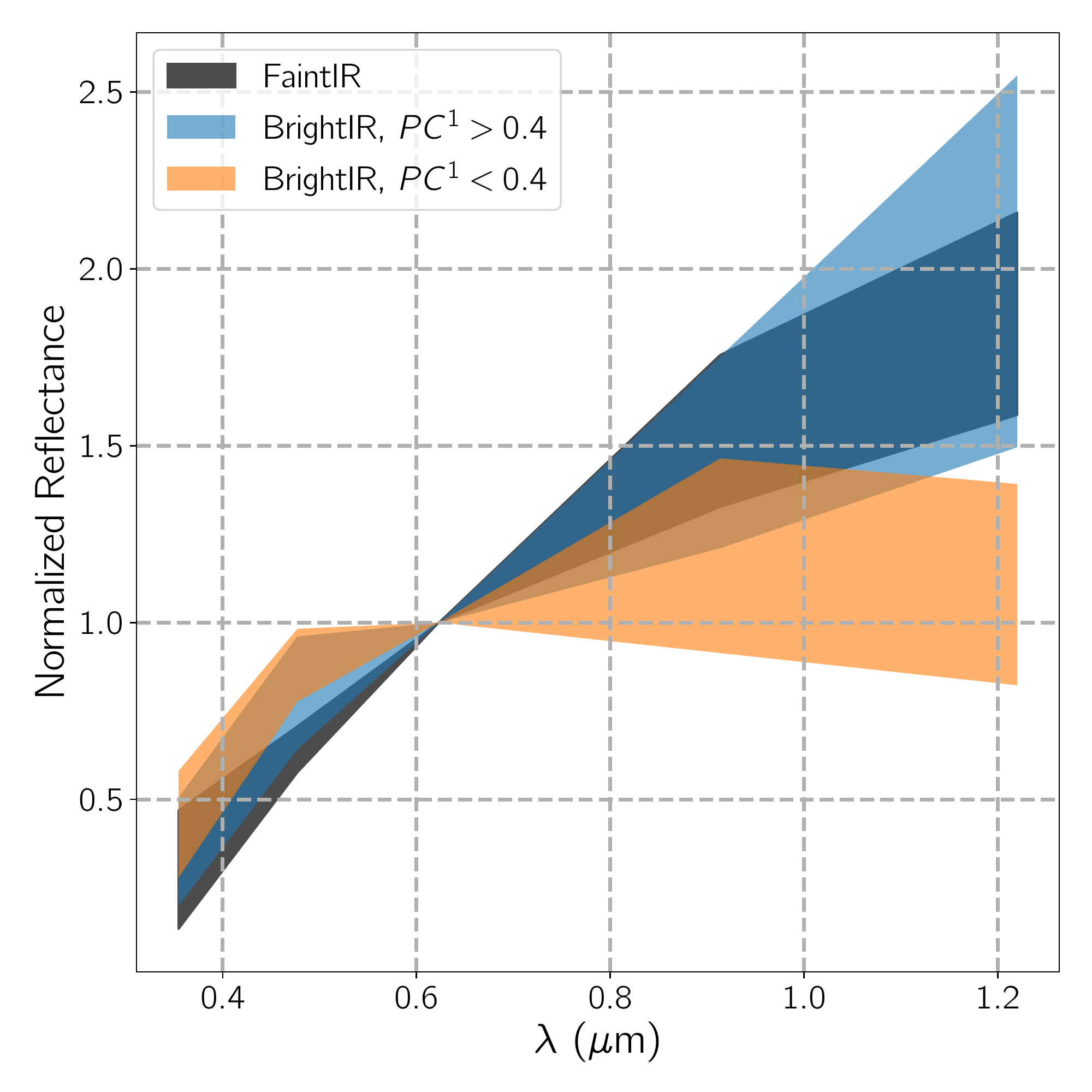}
\caption{The range of reflectivities for Col-OSSOS targets for BrightIR and FaintIR objects. The BrightIR objects have been divided into those objects with $PC^1$ less than and greater than $0.4$ for presentation purposes only; no evidence for any further subdivisions beyond BrightIR and FaintIR has been detected. Relative reflectance values were derived from the observed ugrzJ  colours and adopting Solar colours (u-g)=1.43, (g-r)=0.44, (r-z)=0.18, and (r-J)=0.98 (see Appendix C). Spectra are normalized to unity in the r-band. Uncertainties in the observed colours were not considered when generating the range of spectra, only the nominal measured values.    \label{fig:spectra}}
\end{figure}

The classification scheme we present here results in two classes with distinct colour distributions, and therefore distinct spectral shapes. These differences can be summarized as lower NIR reflectance for FaintIR objects compared to BrightIR objects with the same optical colours. Though we highlight that there are BrightIR objects with equally low NIR reflectance as FaintIR bodies, but only with bluer optical colours. This is visually demonstrated in Figure~\ref{fig:spectra}, where we plot relative reflectances of Col-OSSOS targets. These reflectances are derived from their observed mean ugrzJ photometry. It can be seen that FaintIR objects typically have lower relative reflectances for wavelengths $\lambda\gtrsim 0.9 \mbox{ $\mu$m}$ compared to BrightIR objects with $PC^1_{\textrm{grJ}}>0.4$. The BrightIR objects with $PC^1_{\textrm{grJ}}<0.4$ have NIR reflectances  fully overlap that of the FaintIR objects. 

\citet{Pike2017} speculated that the differences between BrightIR and FaintIR objects were due to an absorption feature that influences the z-band reflectance of CCKBOs that is not present to the same extent on excited KBOs. Our finding that the bifurcation is apparent at even longer wavelengths disfavours the absorption band hypothesis, as the purported absorption band must span $0.9\lesssim \lambda \lesssim 1.4 \mbox{ $\mu$m}$ so as to account for the separation of the two classes at these wavelengths. Such an absorption is extremely wide compared to most solid-state absorptions, and thus, this hypothesis seems unlikely.

The high quality spectra of CCKBOs reported by \citet{Seccull2021} support our interpretation. That is, they find no evidence for a specific absorption feature influencing the $\gtrsim0.9 \mbox{ $\mu$m}$ region. The same is true of the reflectance spectrum of 2014 MU69 Arrokoth. This small CCKBO (mean diameter $\sim40~km$) has colours that reveal it to be a FaintIR member, and does not exhibit any notable absorptions in the $0.9 \lesssim \lambda \lesssim 1.5 \mbox{ $\mu$m}$ range \citep{Grundy2020}. Rather, like most small and excited KBOs, CCKBOs exhibit spectra that can be characterized as near-linear in the optical, that roll over to shallower slopes in the NIR. It happens that FaintIR objects, which dominate the CCKBOs, exhibit a roll-over at shorter wavelengths than their excited counterparts \citep[e.g.,][]{Barucci2011}. Like the excited KBOs, the full spectral behaviour of CCKBOs from $0.3\lesssim \lambda \lesssim 1.4 \mbox{ $\mu$m}$ appears to be dominated almost solely by the optical-gap feature of organic rich materials.

It seems that the apparent spectral differences between the BrightIR and FaintIR classes are driven by differences in the overall shapes of their optical-gap absorptions. The optical-gap is a broad feature that is itself a sum of overlapping absorptions driven by the varied organic materials present on the surfaces of a body. It follows then that organic materials present on KBOs differ between the two classes. This is likely caused by many factors, including but not limited to the budget of non-organic contaminants within the molecular structure of the organic materials \citep[e.g.,][]{Izawa2014}, the duration of short-lived volatiles on their surfaces \citep{Brown2011a}, and the amount of irradiation driven de-hydrogenation that each class has experienced \citep{Brunetto2006}. Specific compositional interpretations will require observations at wavelengths $0.1\lesssim \lambda \lesssim 0.4 \mbox{ $\mu$m}$, and at $\lambda\gtrsim 3 \mbox{ $\mu$m}$ where many organic materials exhibit diagnostic absorption features.

\subsection{The Orbital Distributions and Origins of FaintIR and BrightIR Objects}

%\textbf{ADJUST TO REFLECT THE inc free cold classicals}

\begin{figure}[h]
\plotone{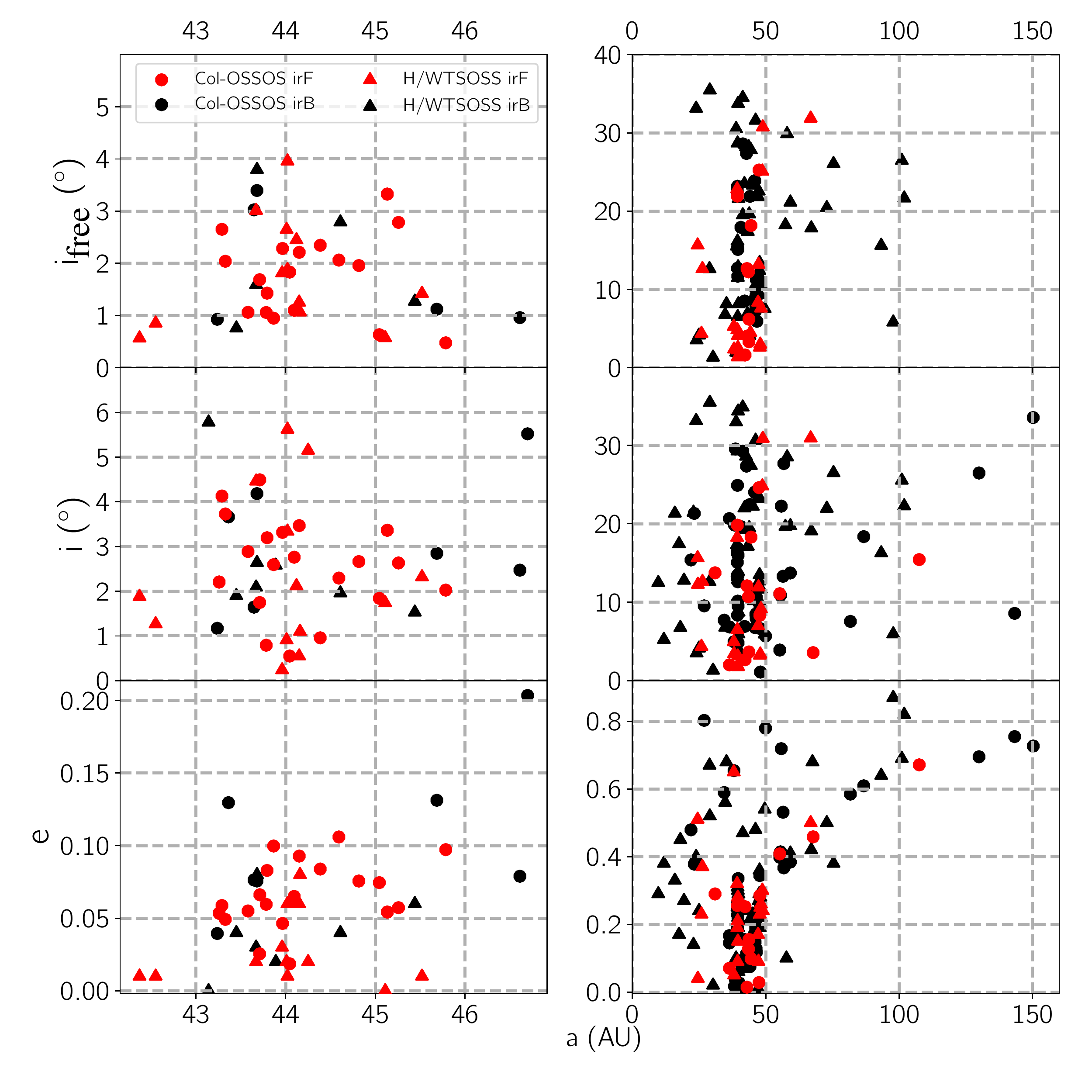}
\caption{Free inclination, inclination, and eccentricity vs. semi-major axes of the objects from Col-OSSOS and H/WTSOSS. Those two surveys are the only two from which reliable classification into BrightIR and FaintIR can be made. The left column shows the cold classical region, while the right column shows the broader extended Kuiper belt.  \label{fig:elements}}
\end{figure}

We now turn our attention to the orbital element distributions of the FaintIR and BrightIR objects. In Figure~\ref{fig:elements} we present the orbital elements of both classes. No significant difference is detected in the eccentricity or inclination distributions of FaintIR and BrightIR CCKBOs. This is also true of the eccentricity distribution of the excited objects in both classes. Interestingly, the AD test suggests that the probability that the semi-major axis distributions of the BrightIR and FaintIR CCKBOs share the same parent distribution, is only 5\%. This is true of both the H/WTSOSS and Col-OSSOS datasets. We caution though that the former has unknown observational biases, and the latter remains incomplete. Therefore, this result while intriguing, should be considered tentative at best. 

The only immediately apparent appreciable difference in orbital element distributions rests within the inclination distributions of the excited objects. Specifically, the highest inclination objects are predominantly found in the BrightIR class. \citet{Marsset2019} demonstrated that the inclination distributions of the VR and LR KBO populations were statistically different throughout the observed excited TNO region. That is, a distinct decrease in density of VR objects occurs for inclinations greater than $i\sim21^{\circ}$. We find a similar result here, but rephrase that result in terms of FaintIR and BrightIR. Only 1 of the 12 objects with $i\geq21^{\circ}$ are FaintIR members. This is a similar ratio to the ratio of VR to LR objects in the Marsset sample: of the 44 objects that can be reliably classed as either VR or LR, only 5 are VR. While many of the VR objects are also FaintIR members, the correspondence is not 1-to-1. In the Col-OSSOS (g-r) and (r-J) sample, 9 of the 44  VR objects ((g-r)$>0.78$) actually belong to the BrightIR class. We draw the reader's attention to Marsset et al. (in prep; to be updated at redline stage) where the colour-inclination correlation is revisited in the Bright/FaintIR taxonomy we present here. We also point out Pike et al. (in prep; to be updated at redline stage) who present an analysis of the distribution of BrightIR and FaintIR bodies in the Col-OSSOS resonant populations.

The preponderance of FaintIR objects found in the cold classical region (by either definition), and their reduced occurrence  at high inclinations argues for different dynamical origins of these two populations. It is generally accepted that the origins of the chemo-dynamical classes of object are primordial, being the result of compositional differences in the protoplanetesimal disk from which KBOs originated. This interpretation is favoured over the possibility of the different classes forming as a result of some evolutionary process that occurred after disk dispersal and emplacement in the Kuiper Belt. The evidence against the latter idea is numerous, but rests mainly on the fact that both BrightIR and FaintIR objects are found in various ratios across all dynamical classes and regions of the Kuiper Belt. Thus, it is hard to construct an evolutionary mechanism that selectively alters the colours of only some KBOs, that otherwise have experienced the same dynamical, thermal, and collisional histories as dynamically nearby bodies. This selective mechanism must also preserve the nearly identical colours of components in binary systems \citep{Benecchi2009}. Rather, it seems most likely that objects in the protoplanetesimal belt earned their spectral types before dispersal, probably as a result of processes that alter the compositional makeups of planetesimal bodies according to their locales \citep[e.g.][]{Brown2011a}. 

The above reasoning has led to recent efforts to use the observed distribution of compositional classes to infer properties of the protoplanetesimal disk, given a dynamical model \citep[see][for a review]{Gladman2021}. From the properties of a subset of BrightIR bodies found in the cold classical region -- the so-called \emph{blue binaries}, \citet{Fraser2017} inferred that some BrightIR objects must have originated only a handful of AU interior to the current cold classical region where the majority of FaintIR objects are found \citep[also see][]{Fraser2021}. From a simple exponential disk model, \citet{Schwamb2019} used the ratio of LR and VR excited bodies to infer that the LR bodies formed approximately between $\sim30$ and $\sim39$~AU, though this was made under the assumption that KBOs consist of three separate classes. From this work we can now appreciate that many of the red dynamically excited KBOs are in fact just the reddest members of the FaintIR class, and therefore must have formed alongside the rest of the FaintIR bodies.

To date the most comprehensive analysis of KBO formation locations is the work of \citet{Nesvorny2020}. They consider only two classes of object, the LR and VR objects, with CCKBOs being considered members of the latter. From the ratios of LR to VR objects found in various orbital classes, they conclude that the LR objects must have formed interior 40~AU. Such a formation location for the LR objects naturally explains why that class has higher inclinations than the VR class, as the latter, by virtue of their formation locations, experienced weaker excitation during dispersal. 

Considering the results we present here, one cannot simply apply a cut in optical colour alone to properly divide a sample of KBOs into their intrinsic classes. From the Col-OSSOS grJ sample, 48 or, 52\% of the 91 objects with unambiguous BrightIR/FaintIR classification would belong to the LR class ((g-r)$<0.78$). These are the KBOs that are considered to have formed in regions closer than they currently are. Of the 91 objects, 58, or 63\% belong to the BrightIR class, suggesting that many more objects formed closer to the Sun than would be implied by a simple cut in optical colour. The difference is even more stark when considering only the  64 excited bodies; 44 would be classed as LR but 51 belong to the BrightIR class. Therefore, when using only optical colour alone to classify objects, one underestimates the fraction of excited objects that formed at closer regions by $\sim15\%$. While this correction may seem small, the observed fraction of excited objects that belong to the BrightIR class is $0.79\pm0.06$, which is $2-\sigma$ discrepant from the  equivalent value inferred when considering the LR/VR classification. 

We highlight that a few objects with FaintIR colour measurements are found on highly excited orbits. Namely, 2004 PB112 and 2013 JK64 belong to the 27:4 and 5:2 mean-motion resonance respectively, and 2014 UA225 is found on a detached orbit. Intriguingly, these three objects have relatively low inclinations, $i\leq15^\circ$ (UA225 has $i=3.57^\circ$). It seems likely that the dynamical pathways each of these objects experienced to place them on their current orbits will provide significant constraint on the formation locations of FaintIR bodies, and of the migration history of Neptune.  

We also see a few BrightIR objects amongst the CCKBOs. Many of these bodies are found near some of the prominent mean-motion resonances. \citet{Huang2022preprint} demonstrated that objects in mean-motion resonances can experience large inclination variations. We speculate that some of these BrightIR CCKBOs were originally loosely captured resonant bodies that experienced a downward diffusion to low-i before being dropped from the resonance. 

In the scenario discussed above, in which BrightIR bodies formed closer to the Sun than the FaintIR bodies, the formation of the \emph{blue binaries} remains a challenge.
These bodies tend to have inflated $i_\textrm{free}$ compared to most other low inclination classical KBOs \citep{Fraser2021}. Some \emph{blue binaries}, are members of the CCKBOs based on $i_\textrm{free}$. Take for example 2003 HG57, which has $i_\textrm{free}=2.1^\circ$. In principle the slightly inflated $i_\textrm{free}$ values of the \emph{blue binaries} are consistent with the idea that these objects have been pushed out to the cold classical region. Though, as pointed out in \citet{Nesvorny2022preprint} when considering detailed migration histories, the push-out scenario would still produce more single objects than binary systems, and so the origins of these peculiar objects remain unclear. It seems clear that like the excited FaintIR bodies, the \emph{blue binaries} are bound to provide some significant cosmogonic insights. 

We point out that objects 2013 UN15, 2013 JK64, and 2013 SA100 have highly variable surface colours, with variations much larger than the uncertainties in their colours. Similarly large colour variations have been detected on objects 1998 SM165, 2005 TV189, and 2001 PK47 \citep{Fraser2015}. All 6 of these targets are unambiguously dynamically excited, with inclinations $i>8^\circ$. Within the dynamically excited sample, the prevalence of highly spectrally variable targets is $\sim5\%$, and so the lack of detection of similar variability in the Col-OSSOS CCKBO sample is not particularly meaningful.  Of these objects, JK64 and PK47, both have ambiguous BrightIR/FaintIR classifications, with each exhibiting statistically significantly different colours with class consistency that changes from epoch to epoch. These two objects are of particular interest as understanding the full extent of their spectral variability may provide insight regarding the true nature of the difference between BrightIR and FaintIR surfaces.

\section{Conclusions} \label{sec:conclusions}

Here we have presented the full colours dataset of the Colours of the Outer Solar System Origins Survey. This includes a sample of 98 objects with (g-r) and (r-J) colours, a number of which that have also received near-simultaneous (u-g) and (r-z) measurements. Using a technique which projects a two dimensional colour space onto the reddening curve, we find a bifurcation in the observed (g-r) and (r-J) colours, into two spectral classes. The BrightIR class closely follows the reddening curve, while the FaintIR class has lower reflectivity at NIR wavelengths compared to BrightIR objects of similar optical colour. The presence of this bifurcation is confirmed with the H/WTSOSS dataset, the (g-r) and (r-z) colours from Col-OSSOS, and the optical colours from \citet{Peixinho2015}. Available optical colour datasets reveal that this bifurcation is not as apparent at wavelengths $\lambda \lesssim 0.8 \mbox{ $\mu$m}$, resulting in unimodal optical colour distributions, to the precision of available datasets. 

We present a simple model that can fully account for all of the observed structures in the optical-NIR colour space. A KBO spectrum is modelled with a simple linear optical and linear NIR spectrum, that intersect at some transition wavelength. Parameter values are chosen to match two mixing model curves which follow the main trends seen in the the grJ colour space, and account for nearly 80\% of the observed colours. In this way, the model reproduces:

\begin{itemize}
\item The separation of two groups in an optical-NIR colour space
\item The bifurcation in the optical colour distributions
\item Deviations away from the reddening curve to less-red colours at NIR wavelengths ($\lambda \gtrsim 0.85 \mbox{ $\mu$m}$)
\item The fact that not all objects fall above or below $PC^2\sim-0.13$ across all band passes considered.
\end{itemize}

\noindent The bimodal optical colour distribution requires that objects are not uniformly distributed along the mixing model curves, but rather cluster around optical spectral slopes of $s=13$ and $30 \%/100 \mbox{ $n$m}$, for the BrightIR and FaintIR classes, respectively.

In summary, we find that all the major properties of the UV-optical-NIR colour distribution can be accounted for with a model consisting of only two compositional classes. This model predicts that the size distribution of dynamically excited FaintIR objects will match that of the red cold classical KBOs. And conversely, the size distribution of BrightIR objects in the cold classical region will match that of the dynamically excited BrightIR objects. Testing such a prediction will be possible with the completion of the (g-r) and (r-J) colour measurements of the outstanding ST and E-block Col-OSSOS targets.

We also argue that when making use of KBO population statistics to constrain cosmogonic models, one cannot simply apply a selection in optical colour to separate the objects into the two compositional classes. Such a cut will overestimate the fraction of objects that formed inside the transition distance that hypothetically divided the two classes. Rather a selection in optical-NIR colour space is required. Cosmogonic inference based on population statistics drawn from optical colours alone should be considered with an appropriate amount of caution.

\pagebreak

\begin{deluxetable*}{cccccccl}. 
\tabletypesize{\tiny}
\tablecaption{Summary of OSSOS Survey Fields\label{tab:fields}}
\tablewidth{0pt}
\tablehead{
\colhead{Field Label} & \colhead{Area} & \colhead{R.A.} & \colhead{Dec.} & \colhead{Ec. Lon.} & \colhead{Ec. Lat.} & \colhead{N(r$<23.6$)} & \colhead{Complete} \\
\colhead{} & \colhead{ ($\circ^2$)} & \colhead{($\circ$)} & \colhead{($\circ$)} & \colhead{($\circ$)} & \colhead{($\circ$)} & \colhead{} & \colhead{} 
}
\startdata
E & 21 & 14:15:28.89 & -12:32:28.5 & 215:50:40.3 & +00:59:02.5 & 32 & N \\
O & 21 & 15:58:01.35 & -12:19:54.2 & 239:57:37.5 & +07:59:03.1 & 13 & Y\tablenotemark{*}\\
H & 21 & 01:35:14.39 & +13:28:25.3 & 26:58:37.1 & +03:18:09.0 & 22 & Y \\
L & 20 & 00:52:55.81 & +03:43:49.1 & 13:37:29.5 & -01:47:10.0 & 18 & Y \\
S & 10.827 & 00:30:08.35 & +06:00:09.5 & 09:17:17.9 & +02:31:33.3 & 11 & N \\
T & 10.827 & 00:35:08.35 & +04:02:04.5 & 09:39:27.8 & +00:13:35.4 & 6 & N \\
\enddata
\tablecomments{Listed coordinates are the central coordinates of each field, and in the J2000 epoch. For further details, see \citet{Bannister2018}. \tablenotetext{*}{ --- The centaur, 2013 JC64 - o3o01 has drifted into the galactic plane, and is not retrievable in ground-based data. O-block should only be considered complete for orbits with semi-major axes beyond Neptune.} }
\end{deluxetable*}

\begin{deluxetable*}{llllcccccc}. 
\tabletypesize{\tiny}
\tablecaption{Observations\label{tab:observations}}
\tablewidth{0pt}
\tablehead{
\multicolumn3c{Designation} \\
\colhead{MPC} & \colhead{OSSOS} & \colhead{Header} & \colhead{Exposure} & \colhead{Filter} & \colhead{MJD} & \colhead{Gemini Mag.} & \colhead{Zeropoint} &\colhead{PS Mag.} & \colhead{Exp. Time (s)}  
}
\startdata
2014 UE225 & o4h50 & Col3N01 & N20140825S0353.fits & r$_{G0303}$ & 56894.511300 & $22.852 \pm 0.019$ & $28.254 \pm 0.007$ & $22.899 \pm 0.019$ & 200 \\
2014 UE225 & o4h50 & Col3N01 & N20140825S0376.fits & r$_{G0303}$ & 56894.562150 & $22.867 \pm 0.019$ & $28.259 \pm 0.007$ & $22.915 \pm 0.019$ & 200 \\
2014 UE225 & o4h50 & Col3N01 & N20140825S0354.fits & g$_{G0301}$ & 56894.514650 & $23.873 \pm 0.030$ & $28.178 \pm 0.009$ & $23.832 \pm 0.030$ & 200 \\
2014 UE225 & o4h50 & Col3N01 & N20140825S0355.fits & g$_{G0301}$ & 56894.517910 & $23.831 \pm 0.028$ & $28.163 \pm 0.008$ & $23.790 \pm 0.028$ & 200 \\
2014 UE225 & o4h50 & Col3N01 & N20140825S0375.fits & g$_{G0301}$ & 56894.558820 & $23.819 \pm 0.028$ & $28.175 \pm 0.009$ & $23.778 \pm 0.028$ & 200 \\
2014 UE225 & o4h50 & Col3N01 & Col3N01\_0.fits & J & 56894.540820 & $20.967 \pm 0.086$ & $23.888 \pm 0.020$ & -- & 720 \\
2014 UE225 & o4h50 & Col3N01 & Col3N01\_1.fits & J & 56894.550450 & $21.227 \pm 0.100$ & $23.888 \pm 0.020$ & -- & 840 \\
\enddata
\tablecomments{Example observations for 2014 UE225. A full list of observations is available in the online version.}
\end{deluxetable*}

\startlongtable
\begin{deluxetable*}{llccccccccccccccc}. % example column alignments. h - hidden. D - aligned on decimal
\tabletypesize{\tiny}
\tablecaption{Col-OSSOS Colours\label{tab:colours}}
\tablewidth{0pt}
\tablehead{
\multicolumn2c{Object} & \colhead{(g-r)} & \colhead{(g-r)} & \colhead{$\Delta$(g-r)}  
& \colhead{(r-z)} & \colhead{(r-z)} & \colhead{$\Delta$(r-z)} 
& \colhead{(r-J)} & \colhead{(r-J)} & \colhead{$\Delta$(r-J)} 
& \colhead{(u-g)} & \colhead{(u-g)} & \colhead{$\Delta$(u-g)} 
& \colhead{Bracket} & \colhead{Second} & \colhead{Simult.}\\
\colhead{OSSOS} & \colhead{MPC} & \colhead{PS} & \colhead{SDSS} & \colhead{}
& \colhead{PS} & \colhead{SDSS} & \colhead{} 
& \colhead{PS} & \colhead{SDSS} & \colhead{} 
& \colhead{PS} & \colhead{SDSS} & \colhead{} 
& \colhead{r} & \colhead{Half} & \colhead{u}
}
%\decimalcolnumbers
\startdata
Sun & -- & -- & 0.453 & -- & -- & 0.091 & -- & -- & 1.020 & -- & -- & 1.430 & -- & - & T & - \\
o3e01 & 2002 GG166 & 0.503 & 0.591 & 0.013 & -- & -- & -- & 1.493 & 1.496 & 0.050 & 1.823 & 1.732 & 0.047 & T & T & F \\
o3e02 & 2013 GH137 & 0.609 & 0.713 & 0.032 & -- & -- & -- & 1.753 & 1.757 & 0.097 & 1.423 & 1.314 & 0.145 & T & T & T \\
o3e04 & 2013 GJ137 & 0.512 & 0.601 & 0.023 & -- & -- & -- & 1.690 & 1.693 & 0.097 & -- & -- & -- & T & T & - \\
o3e04 & 2013 GJ137 & 0.529 & 0.621 & 0.030 & -- & -- & -- & 1.654 & 1.657 & 0.086 & -- & -- & -- & T & T & - \\
o3e04 & 2013 GJ137 & 0.663 & 0.776 & 0.082 & -- & -- & -- & -- & -- & -- & -- & -- & -- & F & F & - \\
o3e05 & 2013 GW136 & 0.618 & 0.724 & 0.019 & -- & -- & -- & 1.692 & 1.697 & 0.065 & 1.415 & 1.305 & 0.092 & T & T & F \\
o3e09 & 2013 GY136 & 0.435 & 0.512 & 0.010 & 0.385 & 0.394 & 0.012 & -- & -- & -- & -- & -- & -- & T & T & - \\
o3e09 & 2013 GY136 & 0.436 & 0.513 & 0.018 & -- & -- & -- & 1.189 & 1.191 & 0.109 & 1.443 & 1.364 & 0.093 & T & T & T \\
o3e11 & 2013 GZ136 & 0.609 & 0.714 & 0.020 & -- & -- & -- & 1.604 & 1.608 & 0.122 & -- & -- & -- & T & T & - \\
o3e16 & 2013 GS137 & 0.865 & 1.010 & 0.022 & -- & -- & -- & 1.713 & 1.720 & 0.080 & -- & -- & -- & T & T & - \\
o3e19 & 2013 GR136 & 0.612 & 0.717 & 0.026 & -- & -- & -- & 1.462 & 1.466 & 0.102 & -- & -- & -- & T & T & - \\
o3e20PD & 2001 FK185 & 0.713 & 0.833 & 0.033 & -- & -- & -- & 1.768 & 1.773 & 0.077 & 1.750 & 1.624 & 0.293 & T & T & T \\
o3e21 & 2013 GQ137 & 0.763 & 0.892 & 0.022 & -- & -- & -- & 1.868 & 1.874 & 0.062 & 2.034 & 1.899 & 0.301 & T & T & T \\
o3e22 & 2013 GN137 & 0.902 & 1.053 & 0.010 & -- & -- & -- & 1.734 & 1.742 & 0.070 & 1.912 & 1.753 & 0.230 & F & T & T \\
o3e23PD & 2001 FO185 & 0.717 & 0.838 & 0.080 & -- & -- & -- & -- & -- & -- & -- & -- & -- & F & F & - \\
o3e23PD & 2001 FO185 & 0.737 & 0.861 & 0.022 & -- & -- & -- & 1.861 & 1.866 & 0.068 & -- & -- & -- & T & T & - \\
o3e23PD & 2001 FO185 & 0.840 & 0.981 & 0.071 & -- & -- & -- & -- & -- & -- & -- & -- & -- & F & F & - \\
o3e27PD & 2004 EU95 & 0.830 & 0.969 & 0.023 & -- & -- & -- & 1.806 & 1.813 & 0.064 & -- & -- & -- & T & T & - \\
o3e28 & 2013 GX137 & 0.842 & 0.983 & 0.028 & -- & -- & -- & 1.452 & 1.460 & 0.081 & -- & -- & -- & T & T & - \\
o3e29 & 2013 GO137 & 0.657 & 0.768 & 0.027 & -- & -- & -- & 1.723 & 1.727 & 0.059 & 2.389 & 2.273 & 0.966 & T & T & T \\
o3e30PD & 2013 EM149 & 0.820 & 0.958 & 0.021 & -- & -- & -- & 1.644 & 1.651 & 0.060 & 2.662 & 2.518 & 0.966 & T & T & T \\
o3e31 & 2013 GT137 & 0.891 & 1.040 & 0.038 & -- & -- & -- & 1.780 & 1.788 & 0.086 & -- & -- & -- & T & T & - \\
o3e34PD & 2013 GF138 & 0.919 & 1.073 & 0.026 & -- & -- & -- & 1.700 & 1.708 & 0.065 & -- & -- & -- & T & T & - \\
o3e35 & 2013 GP137 & 0.807 & 0.943 & 0.033 & -- & -- & -- & 1.261 & 1.268 & 0.100 & -- & -- & -- & T & T & - \\
o3e37PD & 2004 HJ79 & 0.816 & 0.953 & 0.020 & -- & -- & -- & 1.588 & 1.595 & 0.068 & -- & -- & -- & T & T & - \\
o3e39 & 2013 GP136 & 0.657 & 0.769 & 0.020 & -- & -- & -- & 1.628 & 1.633 & 0.066 & -- & -- & -- & T & T & - \\
o3e43 & 2013 GV137 & 0.789 & 0.922 & 0.022 & -- & -- & -- & 1.269 & 1.275 & 0.124 & -- & -- & -- & T & T & - \\
o3e43 & 2013 GV137 & 0.810 & 0.947 & 0.060 & -- & -- & -- & -- & -- & -- & -- & -- & -- & F & F & - \\
o3e44 & 2013 GG138 & 0.933 & 1.090 & 0.031 & -- & -- & -- & 1.845 & 1.854 & 0.075 & -- & -- & -- & T & T & - \\
o3e45 & 2013 GQ136 & 0.862 & 1.006 & 0.055 & 0.446 & 0.483 & 0.057 & -- & -- & -- & -- & -- & -- & F & F & - \\
o3e45 & 2013 GQ136 & 0.923 & 1.078 & 0.025 & -- & -- & -- & -- & -- & -- & -- & -- & -- & T & T & - \\
o3e49 & 2013 HR156 & 0.498 & 0.586 & 0.027 & -- & -- & -- & 1.360 & 1.363 & 0.112 & -- & -- & -- & T & T & - \\
o3e51 & 2013 GM137 & 0.508 & 0.597 & 0.041 & -- & -- & -- & 1.191 & 1.194 & 0.129 & -- & -- & -- & T & T & - \\
o3e54 & 2013 GW137 & 0.813 & 0.949 & 0.023 & -- & -- & -- & -- & -- & -- & -- & -- & -- & T & T & - \\
o3e55 & 2013 GX136 & 0.627 & 0.734 & 0.023 & -- & -- & -- & 1.641 & 1.645 & 0.071 & -- & -- & -- & T & T & - \\
o3l01 & 2013 UR15 & 0.569 & 0.667 & 0.023 & -- & -- & -- & 1.637 & 1.641 & 0.093 & 1.594 & 1.492 & 0.229 & T & T & T \\
o3l05 & 2013 UJ17 & 0.582 & 0.682 & 0.036 & -- & -- & -- & -- & -- & -- & -- & -- & -- & T & T & - \\
o3l06PD & 2001 QF331 & 0.710 & 0.830 & 0.025 & -- & -- & -- & 1.574 & 1.579 & 0.074 & 1.530 & 1.404 & 0.162 & T & T & T \\
o3l09 & 2013 US15 & 0.900 & 1.050 & 0.015 & -- & -- & -- & 1.482 & 1.491 & 0.065 & 1.667 & 1.508 & 0.228 & T & T & T \\
o3l13PD & 2003 SR317 & 0.551 & 0.646 & 0.012 & -- & -- & -- & 1.356 & 1.359 & 0.065 & -- & -- & -- & T & T & - \\
o3l15 & 2013 SZ99 & 0.504 & 0.592 & 0.019 & -- & -- & -- & 1.535 & 1.538 & 0.083 & 1.785 & 1.695 & 0.241 & T & T & T \\
o3l18 & 2010 RE188 & 0.577 & 0.677 & 0.015 & -- & -- & -- & 1.426 & 1.429 & 0.079 & -- & -- & -- & T & T & - \\
o3l32 & 2013 SP99 & 0.837 & 0.977 & 0.020 & -- & -- & -- & 1.598 & 1.606 & 0.067 & -- & -- & -- & T & T & - \\
o3l39 & 2016 BP81 & 0.486 & 0.572 & 0.026 & -- & -- & -- & 1.570 & 1.572 & 0.086 & -- & -- & -- & T & T & - \\
o3l43 & 2013 UL15 & 0.766 & 0.895 & 0.031 & -- & -- & -- & 1.476 & 1.482 & 0.076 & 2.219 & 2.083 & 0.276 & T & T & T \\
o3l46 & 2013 UP15 & 0.756 & 0.884 & 0.020 & -- & -- & -- & 1.896 & 1.902 & 0.087 & -- & -- & -- & T & T & - \\
o3l50 & 2013 UO15 & 0.817 & 0.955 & 0.019 & -- & -- & -- & 1.698 & 1.705 & 0.055 & -- & -- & -- & T & T & - \\
o3l57 & 2013 UM15 & 0.898 & 1.048 & 0.010 & -- & -- & -- & 1.721 & 1.729 & 0.068 & 1.787 & 1.628 & 0.228 & F & T & F \\
o3l60 & 2006 QF181 & 0.769 & 0.898 & 0.025 & -- & -- & -- & 1.533 & 1.539 & 0.072 & 1.839 & 1.704 & 0.270 & T & T & F \\
o3l63 & 2013 UN15 & 0.900 & 1.050 & 0.027 & 0.446 & 0.486 & 0.090 & 1.316 & 1.324 & 0.110 & -- & -- & -- & T & T & - \\
o3l63 & 2013 UN15 & 0.931 & 1.087 & 0.038 & 0.704 & 0.746 & 0.053 & 1.866 & 1.875 & 0.057 & -- & -- & -- & T & T & - \\
o3l63 & 2013 UN15 & 0.935 & 1.091 & 0.032 & -- & -- & -- & -- & -- & -- & -- & -- & -- & T & T & - \\
o3l69 & 2013 UX18 & 0.760 & 0.888 & 0.010 & -- & -- & -- & 1.649 & 1.655 & 0.093 & 1.353 & 1.219 & 0.200 & F & T & T \\
o3l76 & 2013 SQ99 & 0.831 & 0.971 & 0.023 & 0.444 & 0.478 & 0.030 & 1.689 & 1.696 & 0.072 & 2.359 & 2.212 & 0.320 & T & T & T \\
o3l77 & 2013 UQ15 & 0.401 & 0.473 & 0.032 & -- & -- & -- & 0.940 & 0.942 & 0.120 & 1.634 & 1.560 & 0.263 & T & T & F \\
o3l79 & 2013 SA100 & 0.544 & 0.638 & 0.020 & 0.381 & 0.396 & 0.025 & 1.675 & 1.678 & 0.100 & 1.659 & 1.562 & 0.144 & T & T & F \\
o3l79 & 2013 SA100 & 0.562 & 0.659 & 0.020 & 0.372 & 0.388 & 0.015 & 1.401 & 1.405 & 0.054 & -- & -- & -- & T & T & - \\
o3o09 & 2013 JB65 & 0.619 & 0.725 & 0.018 & -- & -- & -- & 1.097 & 1.101 & 0.090 & 1.623 & 1.513 & 0.177 & T & T & T \\
o3o11 & 2013 JK64 & 0.770 & 0.900 & 0.022 & -- & -- & -- & 1.911 & 1.917 & 0.061 & 2.071 & 1.935 & 0.173 & T & T & T \\
o3o11 & 2013 JK64 & 0.904 & 1.055 & 0.057 & -- & -- & -- & 1.853 & 1.861 & 0.101 & -- & -- & -- & F & F & - \\
o3o14 & 2013 JO64 & 0.466 & 0.548 & 0.027 & -- & -- & -- & 1.149 & 1.152 & 0.092 & -- & -- & -- & T & T & - \\
o3o15 & 2013 JD65 & 0.675 & 0.790 & 0.030 & -- & -- & -- & 1.877 & 1.881 & 0.087 & 1.835 & 1.715 & 0.966 & T & T & T \\
o3o18 & 2013 JE64 & 0.592 & 0.694 & 0.032 & 0.493 & 0.512 & 0.035 & -- & -- & -- & -- & -- & -- & T & T & - \\
o3o18 & 2013 JE64 & 0.816 & 0.954 & 0.120 & -- & -- & -- & 1.465 & 1.472 & 0.247 & -- & -- & -- & F & T & - \\
o3o20PD & 2007 JF43 & 0.838 & 0.978 & 0.011 & -- & -- & -- & 1.946 & 1.953 & 0.048 & -- & -- & -- & T & T & - \\
o3o21 & 2013 JR65 & 0.380 & 0.450 & 0.010 & -- & -- & -- & 1.592 & 1.593 & 0.073 & -- & -- & -- & F & T & - \\
o3o21 & 2013 JR65 & 0.485 & 0.570 & 0.035 & 0.429 & 0.440 & 0.043 & -- & -- & -- & -- & -- & -- & T & T & - \\
o3o27 & 2013 JJ65 & 0.922 & 1.076 & 0.027 & -- & -- & -- & 1.686 & 1.695 & 0.075 & -- & -- & -- & T & T & - \\
o3o28 & 2013 JN65 & 0.492 & 0.578 & 0.013 & -- & -- & -- & 1.631 & 1.634 & 0.062 & -- & -- & -- & T & T & - \\
o3o29 & 2013 JL64 & 0.581 & 0.681 & 0.025 & -- & -- & -- & 1.424 & 1.428 & 0.132 & 1.457 & 1.353 & 0.167 & T & T & T \\
o3o34 & 2013 JH64 & 0.595 & 0.697 & 0.019 & -- & -- & -- & 1.466 & 1.470 & 0.079 & 1.788 & 1.682 & 0.130 & T & T & T \\
o3o51 & 2013 JX67 & 0.595 & 0.697 & 0.016 & -- & -- & -- & 1.395 & 1.399 & 0.062 & -- & -- & -- & T & T & - \\
o4h01 & 2014 UJ225 & 0.535 & 0.627 & 0.012 & -- & -- & -- & 1.235 & 1.238 & 0.105 & 1.863 & 1.767 & 0.122 & T & T & T \\
o4h03 & 2014 UQ229 & 0.801 & 0.936 & 0.018 & -- & -- & -- & 1.989 & 1.996 & 0.062 & 2.215 & 2.073 & 0.147 & T & T & T \\
o4h05 & 2014 UX229 & 0.557 & 0.653 & 0.013 & -- & -- & -- & 1.460 & 1.463 & 0.090 & 1.506 & 1.407 & 0.052 & T & T & T \\
o4h07 & 2010 TJ182 & 0.476 & 0.559 & 0.018 & -- & -- & -- & 1.338 & 1.340 & 0.064 & 1.558 & 1.472 & 0.063 & T & T & T \\
o4h09 & 2014 UV228 & 0.504 & 0.592 & 0.021 & -- & -- & -- & 1.458 & 1.461 & 0.060 & 1.836 & 1.745 & 0.216 & T & T & T \\
o4h11 & 2014 UO229 & 0.622 & 0.728 & 0.022 & -- & -- & -- & 1.158 & 1.162 & 0.078 & 1.710 & 1.599 & 0.172 & T & T & T \\
o4h13 & 2014 UD229 & 0.590 & 0.691 & 0.016 & -- & -- & -- & 1.300 & 1.304 & 0.076 & 1.532 & 1.426 & 0.197 & T & T & T \\
o4h14 & 2014 US229 & 0.535 & 0.628 & 0.020 & -- & -- & -- & 1.415 & 1.418 & 0.068 & 1.534 & 1.438 & 0.124 & T & T & T \\
o4h18 & 2014 UX228 & 0.422 & 0.497 & 0.022 & -- & -- & -- & 1.486 & 1.488 & 0.060 & 1.713 & 1.635 & 0.098 & T & T & T \\
o4h18 & 2014 UX228 & 0.469 & 0.552 & 0.017 & 0.513 & 0.523 & 0.024 & -- & -- & -- & -- & -- & -- & T & T & - \\
o4h19 & 2014 UK225 & 0.837 & 0.978 & 0.017 & 0.628 & 0.664 & 0.022 & 1.670 & 1.677 & 0.055 & 2.064 & 1.917 & 0.168 & T & T & T \\
o4h20 & 2014 UL225 & 0.388 & 0.458 & 0.023 & 0.235 & 0.240 & 0.036 & -- & -- & -- & -- & -- & -- & T & T & - \\
o4h20 & 2014 UL225 & 0.473 & 0.556 & 0.030 & -- & -- & -- & 0.767 & 0.769 & 0.128 & 1.444 & 1.358 & 0.173 & T & T & F \\
o4h29 & 2014 UH225 & 0.452 & 0.532 & 0.017 & 0.338 & 0.348 & 0.037 & 1.626 & 1.628 & 0.056 & 1.756 & 1.673 & 0.143 & T & T & T \\
o4h31 & 2014 UM225 & 0.677 & 0.792 & 0.014 & -- & -- & -- & 1.524 & 1.529 & 0.061 & -- & -- & -- & T & T & - \\
o4h39 & 2007 TC434 & 0.572 & 0.670 & 0.015 & -- & -- & -- & 1.495 & 1.499 & 0.063 & -- & -- & -- & T & T & - \\
o4h45 & 2014 UD225 & 0.608 & 0.712 & 0.016 & -- & -- & -- & 1.247 & 1.251 & 0.087 & 2.069 & 1.961 & 0.173 & T & T & F \\
o4h48 & 2001 RY143 & 0.764 & 0.892 & 0.027 & -- & -- & -- & 1.884 & 1.890 & 0.074 & 2.119 & 1.984 & 0.267 & T & T & T \\
o4h50 & 2014 UE225 & 0.891 & 1.040 & 0.017 & -- & -- & -- & 1.809 & 1.817 & 0.067 & 2.220 & 2.063 & 0.215 & T & T & T \\
o4h67PD & 2006 QP180 & 0.814 & 0.950 & 0.082 & -- & -- & -- & 2.062 & 2.069 & 0.099 & -- & -- & -- & T & T & - \\
o4h69PD & 1995 QY9 & 0.629 & 0.737 & 0.021 & -- & -- & -- & 1.460 & 1.464 & 0.057 & -- & -- & -- & T & T & - \\
o4h70 & 2014 UF228 & 0.520 & 0.611 & 0.024 & -- & -- & -- & 1.377 & 1.380 & 0.072 & -- & -- & -- & T & T & - \\
o4h75 & 2014 UN228 & 0.497 & 0.585 & 0.040 & -- & -- & -- & 1.264 & 1.267 & 0.130 & -- & -- & -- & T & T & - \\
o4h75 & 2014 UN228 & 0.506 & 0.595 & 0.010 & -- & -- & -- & 1.757 & 1.760 & 0.103 & -- & -- & -- & F & T & - \\
o4h75 & 2014 UN228 & 0.531 & 0.623 & 0.064 & -- & -- & -- & 1.348 & 1.351 & 0.194 & -- & -- & -- & T & T & - \\
o4h76PD & 2001 RX143 & 0.721 & 0.843 & 0.038 & -- & -- & -- & 1.309 & 1.315 & 0.109 & -- & -- & -- & T & T & - \\
o5s01 & 2015 RK277 & 0.447 & 0.527 & 0.015 & -- & -- & -- & 1.352 & 1.355 & 0.052 & -- & -- & -- & T & T & - \\
o5s05 & 2015 RV245 & 0.572 & 0.670 & 0.018 & 0.404 & 0.421 & 0.035 & 1.398 & 1.402 & 0.088 & -- & -- & -- & T & T & - \\
o5s06 & 2015 RW245 & 0.583 & 0.683 & 0.020 & 0.377 & 0.394 & 0.037 & 1.539 & 1.543 & 0.075 & -- & -- & -- & T & T & - \\
o5s07 & 2015 RU277 & 0.494 & 0.581 & 0.023 & -- & -- & -- & 1.251 & 1.254 & 0.094 & -- & -- & -- & T & T & - \\
o5s07 & 2015 RU277 & 0.533 & 0.625 & 0.028 & -- & -- & -- & 1.229 & 1.232 & 0.093 & -- & -- & -- & T & T & - \\
o5s16 & 2004 PB112 & 0.636 & 0.744 & 0.044 & 0.449 & 0.470 & 0.057 & -- & -- & -- & -- & -- & -- & F & F & - \\
o5s16 & 2004 PB112 & 0.705 & 0.824 & 0.022 & -- & -- & -- & 1.417 & 1.422 & 0.059 & -- & -- & -- & T & T & - \\
o5s32 & 2015 RJ277 & 0.544 & 0.638 & 0.020 & -- & -- & -- & 1.172 & 1.175 & 0.108 & -- & -- & -- & T & T & - \\
o5s36 & 2015 RB281 & 0.654 & 0.766 & 0.039 & -- & -- & -- & 1.471 & 1.475 & 0.095 & -- & -- & -- & T & T & - \\
o5s36 & 2015 RB281 & 0.860 & 1.004 & 0.038 & -- & -- & -- & 1.514 & 1.522 & 0.104 & -- & -- & -- & T & T & - \\
o5s45 & 2015 RG277 & 0.814 & 0.950 & 0.018 & -- & -- & -- & 1.635 & 1.642 & 0.061 & -- & -- & -- & T & T & - \\
o5s52 & 2015 RU278 & 0.518 & 0.609 & 0.030 & -- & -- & -- & 1.361 & 1.364 & 0.126 & -- & -- & -- & T & T & - \\
o5s68 & 2015 RR245 & 0.654 & 0.765 & 0.012 & 0.466 & 0.488 & 0.018 & 1.183 & 1.188 & 0.083 & -- & -- & -- & T & T & - \\
o5t04 & 2015 RU245 & 0.757 & 0.884 & 0.018 & 0.563 & 0.592 & 0.028 & 1.565 & 1.572 & 0.066 & -- & -- & -- & T & T & - \\
o5t09 & 2014 UA225 & 0.814 & 0.950 & 0.018 & 0.581 & 0.615 & 0.019 & 1.458 & 1.464 & 0.060 & -- & -- & -- & T & T & - \\
o5t11PD & 2001 QE298 & 0.714 & 0.834 & 0.018 & 0.598 & 0.624 & 0.027 & 1.488 & 1.494 & 0.063 & -- & -- & -- & T & T & - \\
o5t31 & 2015 RT245 & 0.808 & 0.944 & 0.047 & 0.456 & 0.489 & 0.054 & 1.578 & 1.585 & 0.122 & -- & -- & -- & F & F & - \\
\enddata
\tablecomments{Colours are reported for u, g, r, and z observations in both the PS and SDSS filter systems. Each separate epoch for a target is reported. Magnitudes are in AB magnitudes for all but the J-band which is in the Vega magnitude system. Each row is marked according to if the sequence used in deriving that colour had bracketing r-band observations, if that sequence was missing one half of the GMOS observations, and if the reported u-band observations were acquired simultaneously with the Gemini observations. OSSOS and Minor Planet Center (MPC) designations are shown.}
\end{deluxetable*}

\startlongtable
\begin{deluxetable*}{llcccc}. % example column alignments. h - hidden. D - aligned on decimal
\tabletypesize{\tiny}
\tablecaption{Subaru Colours\label{tab:subaru_colours}}
\tablewidth{0pt}
\tablehead{
\multicolumn2c{Object} 
& \colhead{(r-i)} &  \colhead{$\Delta$(r-i)} 
& \colhead{(r-z)} &  \colhead{$\Delta$(r-z)} \\
\colhead{OSSOS} & \colhead{MPC} 
& \colhead{SDSS} & \colhead{}
& \colhead{SDSS} & \colhead{}
}
%\decimalcolnumbers
\startdata
o4h50 & 2014 UE225 & 0.468 & 0.024 & 0.753 & 0.031\\
o4h01 & 2014 UJ225 &   --  &  -- & 0.555 & 0.058 \\
o4h45 & 2014 UD225 &   --  &  -- & 0.409 & 0.089 \\
o4h31 & 2014 UM225 &   --  &  -- & 0.360 & 0.100 \\
o3l39 & 2016 BP81 &   --  &  -- & 0.499 & 0.084 \\
o3l77 & 2013 UQ15 &   -- &   -- &  0.159 &  0.069 \\
o3l63 & 2013 UN15 &   -- &   -- & 0.595 & 0.077 \\
o3l09 & 2013 US15 &   -- &   -- & 0.663 & 0.041 \\
o3l01 & 2013 UR15 & 0.392 & 0.035 & 0.424 & 0.055 \\
o3l43 & 2013 UL15 & 0.414 & 0.027 &  0.658 & 0.036 \\
o3l46 & 2013 UP15 &  --  &  -- & 0.404 &  0.117 \\
o3l06PD & 2001 QF331 & 0.469 & 0.027 &  0.737 &  0.032 \\
o3l15 & 2013 SZ99 &   --  &  -- & 0.415 &  0.084 \\
o3l79 & 2013 SA100 &   -- &   -- & 0.500 & 0.048 \\
o3l76 & 2013 SQ99 &   -- &   -- & 0.738 &  0.071 
\enddata
\tablecomments{Updated colours of Col-OSSOS targets acquired at Subaru in the 2014B semester during simultaneous observations with the Gemini telescope. For more details, see .}
\end{deluxetable*}

\startlongtable
\begin{deluxetable*}{lccccccccc}. % example column alignments. h - hidden. D - aligned on decimal
\tabletypesize{\tiny}
\tablecaption{PC Values\label{tab:colours}}
\tablewidth{0pt}
\tablehead{
\multicolumn2c{Object} & 
\colhead{$PC^1_{grz}$} & \colhead{$\Delta PC^1_{grz}$} & 
\colhead{$PC^2_{grz}$}  & \colhead{$\Delta PC^2_{grz}$} & 
\colhead{$PC^1_{grJ}$} & \colhead{$\Delta PC^1_{grJ}$} & 
\colhead{$PC^2_{grJ}$}  & \colhead{$\Delta PC^2_{grJ}$} 
}
%\decimalcolnumbers
\startdata
Sun & -- & 0.0 & -- & 0.0 & -- & 0.0 & -- & 0.0 & -- \\
o4h50 & 2014 UE225 & -- & -- & -- & -- & 1.067 & 0.060 & -0.192 & 0.060 \\
o4h01 & 2014 UJ225 & -- & -- & -- & -- & 0.298 & 0.105 & -0.093 & 0.105 \\
o4h45 & 2014 UD225 & -- & -- & -- & -- & 0.337 & 0.087 & -0.169 & 0.087 \\
o4h20 & 2014 UL225 & -- & -- & -- & -- & -0.188 & 0.129 & -0.137 & 0.129 \\
o3l43 & 2013 UL15 & -- & -- & -- & -- & 0.653 & 0.078 & -0.251 & 0.078 \\
o3l18 & 2010 RE188 & -- & -- & -- & -- & 0.501 & 0.076 & -0.074 & 0.076 \\
o3l39 & 2016 BP81 & -- & -- & -- & -- & 0.591 & 0.077 & 0.079 & 0.077 \\
o3l77 & 2013 UQ15 & -- & -- & -- & -- & -0.035 & 0.120 & -0.022 & 0.120 \\
o3l63 & 2013 UN15 & 0.705 & 0.067 & -0.248 & 0.071 & 0.557 & 0.128 & -0.459 & 0.128 \\
o3l57 & 2013 UM15 & -- & -- & -- & -- & 0.991 & 0.064 & -0.253 & 0.064 \\
o3l09 & 2013 US15 & -- & -- & -- & -- & 0.748 & 0.072 & -0.384 & 0.072 \\
o3l01 & 2013 UR15 & -- & -- & -- & -- & 0.691 & 0.082 & 0.025 & 0.082 \\
o3l69 & 2013 UX18 & -- & -- & -- & -- & 0.817 & 0.087 & -0.162 & 0.087 \\
o3l46 & 2013 UP15 & -- & -- & -- & -- & 1.031 & 0.069 & -0.017 & 0.069 \\
o3l06PD & 2001 QF331 & -- & -- & -- & -- & 0.715 & 0.070 & -0.149 & 0.070 \\
o3l32 & 2013 SP99 & -- & -- & -- & -- & 0.823 & 0.067 & -0.263 & 0.067 \\
o3l13PD & 2003 SR317 & -- & -- & -- & -- & 0.425 & 0.062 & -0.071 & 0.062 \\
o3l15 & 2013 SZ99 & -- & -- & -- & -- & 0.569 & 0.075 & 0.047 & 0.075 \\
o3l79 & 2013 SA100 & 0.317 & 0.020 & -0.017 & 0.015 & 0.712 & 0.086 & 0.067 & 0.086 \\
o3l76 & 2013 SQ99 & 0.627 & 0.022 & -0.204 & 0.023 & 0.909 & 0.068 & -0.209 & 0.068 \\
o3l60 & 2006 QF181 & -- & -- & -- & -- & 0.712 & 0.074 & -0.228 & 0.074 \\
o3l79 & 2013 SA100 & 0.322 & 0.013 & -0.039 & 0.016 & 0.472 & 0.051 & -0.067 & 0.051 \\
o4h29 & 2014 UH225 & 0.216 & 0.030 & 0.041 & 0.021 & 0.624 & 0.047 & 0.139 & 0.047 \\
o4h19 & 2014 UK225 & 0.760 & 0.014 & -0.063 & 0.018 & 0.894 & 0.052 & -0.226 & 0.052 \\
o3l63 & 2013 UN15 & 0.902 & 0.033 & -0.055 & 0.045 & 1.151 & 0.049 & -0.188 & 0.049 \\
o4h31 & 2014 UM225 & -- & -- & -- & -- & 0.646 & 0.059 & -0.138 & 0.059 \\
o3l50 & 2013 UO15 & -- & -- & -- & -- & 0.906 & 0.051 & -0.191 & 0.051 \\
o4h39 & 2007 TC434 & -- & -- & -- & -- & 0.565 & 0.060 & -0.040 & 0.060 \\
o4h03 & 2014 UQ229 & -- & -- & -- & -- & 1.137 & 0.046 & 0.003 & 0.046 \\
o4h05 & 2014 UX229 & -- & -- & -- & -- & 0.526 & 0.084 & -0.039 & 0.084 \\
o4h07 & 2010 TJ182 & -- & -- & -- & -- & 0.375 & 0.060 & 0.002 & 0.060 \\
o4h09 & 2014 UV228 & -- & -- & -- & -- & 0.497 & 0.055 & 0.018 & 0.055 \\
o4h11 & 2014 UO229 & -- & -- & -- & -- & 0.251 & 0.079 & -0.212 & 0.079 \\
o4h13 & 2014 UD229 & -- & -- & -- & -- & 0.384 & 0.076 & -0.132 & 0.076 \\
o4h14 & 2014 US229 & -- & -- & -- & -- & 0.472 & 0.065 & -0.033 & 0.065 \\
o4h18 & 2014 UX228 & -- & -- & -- & -- & 0.489 & 0.054 & 0.115 & 0.054 \\
o4h48 & 2001 RY143 & -- & -- & -- & -- & 1.026 & 0.060 & -0.031 & 0.060 \\
o3e01 & 2002 GG166 & -- & -- & -- & -- & 0.529 & 0.046 & 0.032 & 0.046 \\
o3e05 & 2013 GW136 & -- & -- & -- & -- & 0.770 & 0.056 & 0.001 & 0.056 \\
o3e09 & 2013 GY136 & -- & -- & -- & -- & 0.219 & 0.106 & 0.000 & 0.106 \\
o3e22 & 2013 GN137 & -- & -- & -- & -- & 1.007 & 0.066 & -0.249 & 0.066 \\
o3e29 & 2013 GO137 & -- & -- & -- & -- & 0.817 & 0.052 & -0.022 & 0.052 \\
o3e30PD & 2013 EM149 & -- & -- & -- & -- & 0.856 & 0.059 & -0.223 & 0.059 \\
o3e39 & 2013 GP136 & -- & -- & -- & -- & 0.735 & 0.059 & -0.070 & 0.059 \\
o3e44 & 2013 GG138 & -- & -- & -- & -- & 1.137 & 0.065 & -0.205 & 0.065 \\
o3o11 & 2013 JK64 & -- & -- & -- & -- & 1.114 & 0.086 & -0.174 & 0.086 \\
o3o14 & 2013 JO64 & -- & -- & -- & -- & 0.192 & 0.090 & -0.042 & 0.090 \\
o3o15 & 2013 JD65 & -- & -- & -- & -- & 0.959 & 0.070 & 0.045 & 0.070 \\
o3o20PD & 2007 JF43 & -- & -- & -- & -- & 1.134 & 0.037 & -0.055 & 0.037 \\
o3o27 & 2013 JJ65 & -- & -- & -- & -- & 0.980 & 0.074 & -0.296 & 0.074 \\
o3o34 & 2013 JH64 & -- & -- & -- & -- & 0.549 & 0.076 & -0.077 & 0.076 \\
o3e02 & 2013 GH137 & -- & -- & -- & -- & 0.814 & 0.081 & 0.041 & 0.081 \\
o3e04 & 2013 GJ137 & -- & -- & -- & -- & -- & -- & -- & -- \\
o3e16 & 2013 GS137 & -- & -- & -- & -- & 0.956 & 0.075 & -0.228 & 0.075 \\
o3e20PD & 2001 FK185 & -- & -- & -- & -- & 0.894 & 0.066 & -0.052 & 0.066 \\
o3e21 & 2013 GQ137 & -- & -- & -- & -- & 1.012 & 0.050 & -0.041 & 0.050 \\
o3e23PD & 2001 FO185 & -- & -- & -- & -- & -- & -- & -- & -- \\
o3e23PD & 2001 FO185 & -- & -- & -- & -- & -- & -- & -- & -- \\
o3e35 & 2013 GP137 & -- & -- & -- & -- & 0.447 & 0.110 & -0.381 & 0.110 \\
o3e37PD & 2004 HJ79 & -- & -- & -- & -- & 0.799 & 0.068 & -0.248 & 0.068 \\
o3e43 & 2013 GV137 & -- & -- & -- & -- & -- & -- & -- & -- \\
o3e55 & 2013 GX136 & -- & -- & -- & -- & 0.728 & 0.063 & -0.033 & 0.063 \\
o3o09 & 2013 JB65 & -- & -- & -- & -- & 0.192 & 0.093 & -0.226 & 0.093 \\
o3o11 & 2013 JK64 & -- & -- & -- & -- & 1.054 & 0.048 & -0.020 & 0.048 \\
o3o18 & 2013 JE64 & -- & -- & -- & -- & 0.671 & 0.257 & -0.309 & 0.257 \\
o3o21 & 2013 JR65 & -- & -- & -- & -- & 0.561 & 0.062 & 0.199 & 0.062 \\
o3o28 & 2013 JN65 & -- & -- & -- & -- & 0.646 & 0.053 & 0.100 & 0.053 \\
o3o29 & 2013 JL64 & -- & -- & -- & -- & 0.501 & 0.125 & -0.078 & 0.125 \\
o5t09 & 2014 UA225 & 0.707 & 0.016 & -0.085 & 0.015 & 0.660 & 0.065 & -0.309 & 0.065 \\
o5s68 & 2015 RR245 & 0.472 & 0.014 & -0.054 & 0.012 & 0.293 & 0.086 & -0.239 & 0.086 \\
o5s06 & 2015 RW245 & 0.342 & 0.030 & -0.054 & 0.023 & 0.610 & 0.070 & -0.034 & 0.070 \\
o5t04 & 2015 RU245 & 0.637 & 0.019 & -0.061 & 0.021 & 0.735 & 0.064 & -0.200 & 0.064 \\
o5t11PD & 2001 QE298 & 0.620 & 0.018 & -0.004 & 0.020 & 0.632 & 0.063 & -0.191 & 0.063 \\
o5t31 & 2015 RT245 & 0.612 & 0.041 & -0.178 & 0.040 & 0.783 & 0.122 & -0.246 & 0.122 \\
o5s05 & 2015 RV245 & 0.357 & 0.027 & -0.027 & 0.022 & 0.472 & 0.084 & -0.078 & 0.084 \\
o5s16 & 2004 PB112 & 0.444 & 0.045 & -0.051 & 0.038 & -- & -- & -- & -- \\
o5s45 & 2015 RG277 & -- & -- & -- & -- & 0.841 & 0.058 & -0.221 & 0.058 \\
o5s01 & 2015 RK277 & -- & -- & -- & -- & 0.380 & 0.049 & 0.040 & 0.049 \\
o5s32 & 2015 RJ277 & -- & -- & -- & -- & 0.241 & 0.109 & -0.122 & 0.109 \\
o3l63 & 2013 UN15 & -- & -- & -- & -- & -- & -- & -- & -- \\
o3l05 & 2013 UJ17 & -- & -- & -- & -- & -- & -- & -- & -- \\
o3e09 & 2013 GY136 & 0.240 & 0.009 & 0.084 & 0.008 & -- & -- & -- & -- \\
o4h18 & 2014 UX228 & 0.362 & 0.017 & 0.129 & 0.015 & -- & -- & -- & -- \\
o3e11 & 2013 GZ136 & -- & -- & -- & -- & 0.685 & 0.110 & -0.033 & 0.110 \\
o3e19 & 2013 GR136 & -- & -- & -- & -- & 0.553 & 0.097 & -0.097 & 0.097 \\
o3e23PD & 2001 FO185 & -- & -- & -- & -- & 0.989 & 0.055 & -0.021 & 0.055 \\
o3e27PD & 2004 EU95 & -- & -- & -- & -- & 1.012 & 0.055 & -0.139 & 0.055 \\
o3e28 & 2013 GX137 & -- & -- & -- & -- & 0.674 & 0.087 & -0.340 & 0.087 \\
o3e31 & 2013 GT137 & -- & -- & -- & -- & 1.041 & 0.078 & -0.210 & 0.078 \\
o3e34PD & 2013 GF138 & -- & -- & -- & -- & 0.989 & 0.063 & -0.285 & 0.063 \\
o3e43 & 2013 GV137 & -- & -- & -- & -- & 0.442 & 0.134 & -0.359 & 0.134 \\
o3e45 & 2013 GQ136 & 0.663 & 0.050 & -0.223 & 0.044 & -- & -- & -- & -- \\
o3e49 & 2013 HR156 & -- & -- & -- & -- & 0.407 & 0.106 & -0.013 & 0.106 \\
o3e45 & 2013 GQ136 & -- & -- & -- & -- & -- & -- & -- & -- \\
o3e51 & 2013 GM137 & -- & -- & -- & -- & 0.246 & 0.127 & -0.077 & 0.127 \\
o3e54 & 2013 GW137 & -- & -- & -- & -- & -- & -- & -- & -- \\
o5s07 & 2015 RU277 & -- & -- & -- & -- & 0.298 & 0.090 & -0.044 & 0.090 \\
o5s07 & 2015 RU277 & -- & -- & -- & -- & 0.293 & 0.094 & -0.093 & 0.094 \\
o5s16 & 2004 PB112 & -- & -- & -- & -- & 0.557 & 0.060 & -0.212 & 0.060 \\
o5s36 & 2015 RB281 & -- & -- & -- & -- & 0.584 & 0.093 & -0.138 & 0.093 \\
o5s36 & 2015 RB281 & -- & -- & -- & -- & 0.754 & 0.110 & -0.329 & 0.110 \\
o3e04 & 2013 GJ137 & -- & -- & -- & -- & 0.685 & 0.074 & 0.072 & 0.074 \\
o3e04 & 2013 GJ137 & -- & -- & -- & -- & 0.708 & 0.081 & 0.107 & 0.081 \\
o3o51 & 2013 JX67 & -- & -- & -- & -- & 0.481 & 0.061 & -0.104 & 0.061 \\
o3o21 & 2013 JR65 & 0.309 & 0.032 & 0.064 & 0.028 & -- & -- & -- & -- \\
o3o18 & 2013 JE64 & 0.441 & 0.026 & 0.015 & 0.024 & -- & -- & -- & -- \\
o4h67PD & 2006 QP180 & -- & -- & -- & -- & 1.199 & 0.069 & 0.045 & 0.069 \\
o4h69PD & 1995 QY9 & -- & -- & -- & -- & 0.561 & 0.054 & -0.116 & 0.054 \\
o4h70 & 2014 UF228 & -- & -- & -- & -- & 0.429 & 0.068 & -0.031 & 0.068 \\
o4h75 & 2014 UN228 & -- & -- & -- & -- & 0.313 & 0.126 & -0.044 & 0.126 \\
o4h75 & 2014 UN228 & -- & -- & -- & -- & 0.407 & 0.184 & -0.052 & 0.184 \\
o4h75 & 2014 UN228 & -- & -- & -- & -- & 0.761 & 0.084 & 0.144 & 0.084 \\
o4h76PD & 2001 RX143 & -- & -- & -- & -- & 0.455 & 0.116 & -0.271 & 0.116 \\
o4h20 & 2014 UL225 & 0.087 & 0.029 & 0.047 & 0.020 & -- & -- & -- & -- \\
o5s52 & 2015 RU278 & -- & -- & -- & -- & 0.416 & 0.120 & -0.035 & 0.120 \\
\enddata
\tablecomments{Projection values for the grz and grJ colour spaces, calculated from the colours in the SDSS system that are tabulated in Table~3. \label{tab:projextras}}
\end{deluxetable*}

\pagebreak

\begin{acknowledgements}

The authors wish to recognize and acknowledge the very significant cultural role and reverence that the summit of Maunakea has always had within the indigenous Hawaiian community.  We are most fortunate to have the opportunity to conduct observations from this mountain.

This research is based on observations from the Large and Long Program GN-2014B-LP-1, GN-2015A-LP-1, GN-2015B-LP-1, GN-2016A-LP-1, GN-2016B-LP-1,  GN-2017A-LP-1, GN-2018A-Q-118, GN-2018A-Q-223, and GN-2020B-Q-127 obtained at the international Gemini Observatory, a program of NSF’s NOIRLab, which is managed by the Association of Universities for Research in Astronomy (AURA) under a cooperative agreement with the National Science Foundation. on behalf of the Gemini Observatory partnership: the National Science Foundation (United States), National Research Council (Canada), Agencia Nacional de Investigaci\'{o}n y Desarrollo (Chile), Ministerio de Ciencia, Tecnolog\'{i}a e Innovaci\'{o}n (Argentina), Minist\'{e}rio da Ci\^{e}ncia, Tecnologia, Inova\c{c}\~{o}es e Comunica\c{c}\~{o}es (Brazil), and Korea Astronomy and Space Science Institute (Republic of Korea).

This research is also based on observations obtained with MegaPrime/MegaCam, a joint project of CFHT and CEA/DAPNIA, at the Canada-France-Hawaii Telescope (CFHT) which is operated by the National Research Council (NRC) of Canada, the Institut National des Science de l'Univers of the Centre National de la Recherche Scientifique (CNRS) of France, and the University of Hawaii. The observations at the Canada-France-Hawaii Telescope were performed with care and respect from the summit of Maunakea which is a significant cultural and historic site.

Finally, this research is based in part on data collected at the Subaru Telescope, which is operated by the National Astronomical Observatory of Japan.

This research used the facilities of the Canadian Astronomy Data Centre operated by the National Research Council of Canada with the support of the Canadian Space Agency. This work also made use of the Gemini Observatory Archive, NASA’s Astrophysics Data System Bibliographic Services, and the JPL HORIZONS web interface (\url{https://ssd.jpl.nasa.gov/horizons.cgi}).

This work made use of the the Pan-STARRS1 Surveys (PS1) and the PS1 public science archive, which have been made possible through contributions by the Institute for Astronomy, the University of Hawaii, the Pan-STARRS Project Office, the Max-Planck Society and its participating institutes, the Max Planck Institute for Astronomy, Heidelberg and the Max Planck Institute for Extraterrestrial Physics, Garching, The Johns Hopkins University, Durham University, the University of Edinburgh, the Queen's University Belfast, the Harvard-Smithsonian Center for Astrophysics, the Las Cumbres Observatory Global Telescope Network Incorporated, the National Central University of Taiwan, the Space Telescope Science Institute, the National Aeronautics and Space Administration under Grant No. NNX08AR22G issued through the Planetary Science Division of the NASA Science Mission Directorate, the National Science Foundation Grant No. AST-1238877, the University of Maryland, Eotvos Lorand University (ELTE), the Los Alamos National Laboratory, and the Gordon and Betty Moore Foundation.

REP acknowledges funding from NASA Emerging Worlds grant 80NSSC21K0376.
\end{acknowledgements}

\vspace{5mm}
\facilities{Gemini-North(GMOS, NIRI), CFHT(MegaCam), Subaru (Suprime-cam)}

\software{numpy \citep{Harris2020}, matplotlib \citep{Hunter2007}, TRailed Imaging in Python, \citep{Fraser2016}, astropy \citep{Astropy2013, Astropy2018}, 
          nirlin\footnote{\url{https://staff.gemini.edu/~astephens/niri/nirlin/nirlin.py}}, 
          SExtractor \citep{hihi}, SExtractor in Python \citep{Barbary2016}, synphot \citep{Lim2013}
          }

\appendix
\restartappendixnumbering
\section{Photometry of Highly Azimuthally Variable Point Sources \label{sec:Jcals}}

Here we present a new technique developed to accurately measure the photometry of the stellar calibration imagery acquired as part of every NIR observation. These sequences were a series of 10 frames acquired with integrations of a few to $\sim6$~s, with dithers of a few tens of arcseconds between each frame. An issue with inconsistent zeropoints was found when studying the calibration frames in depth. This has been attributed to the highly variable shape of the stellar images acquired during these sequences. When measuring zeropoints from these images, the standard method of measuring of measuring flux in a large aperture, $r=4$~FWHM -- was adopted. The FWHM itself was determined from the radially averaged stellar profile. Such a large aperture radius is usually enough to encompass $\sim99\%$ of a point source flux, and provide stable absolute calibrations. It was found however, that the resultant zeropoints would vary by a couple tenths of a magnitude, a factor of $\sim5$ more than expected. This variation disappeared however, when unusually large apertures were considered, hinting that oddly, some large, and variable fraction of the stellar flux was being excluded from the normally large apertures. 

\begin{figure}[h]
\plotone{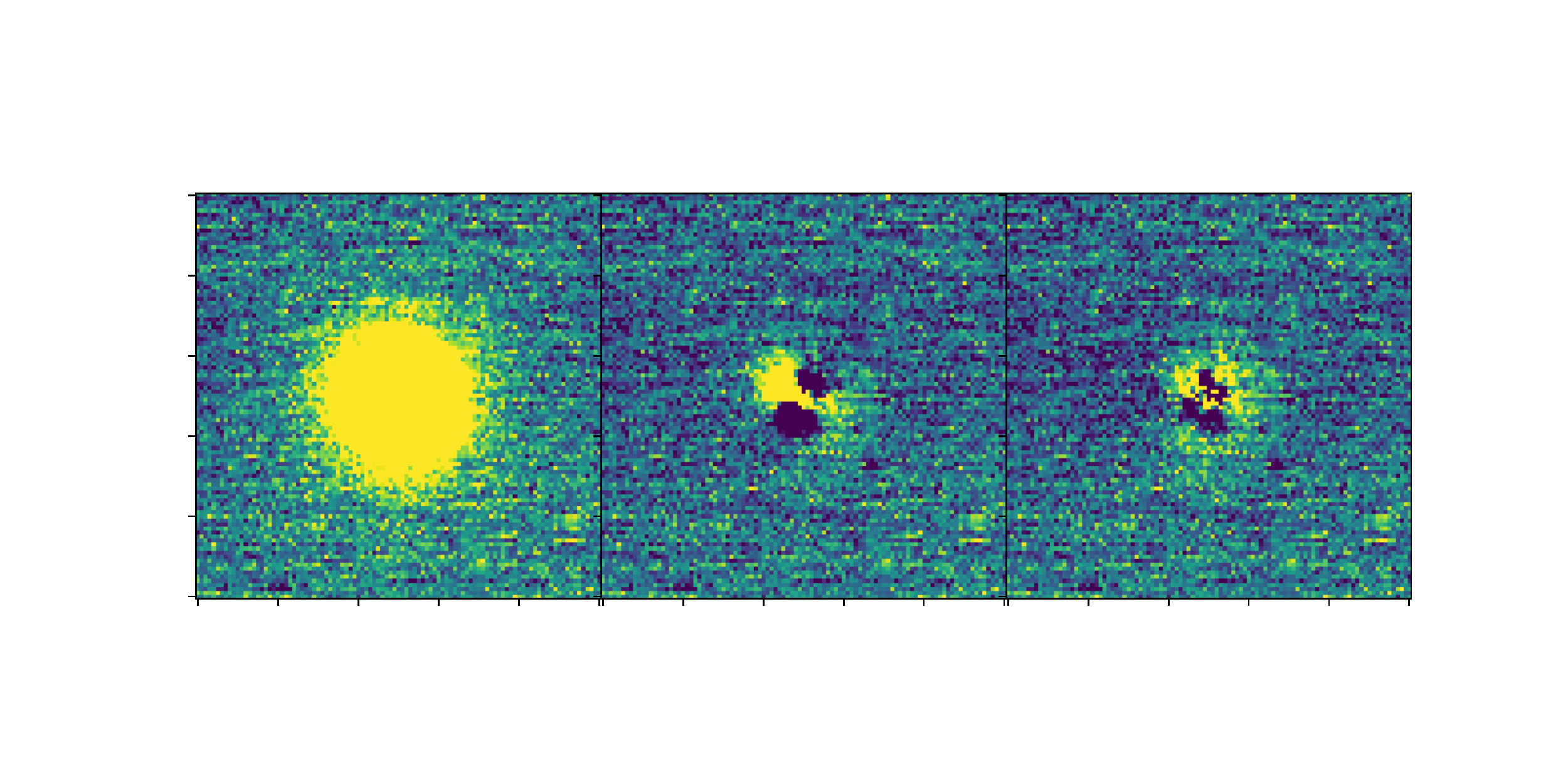}
\caption{ Cutouts around a typical NIRI calibration star. \textbf{Left:} original image. \textbf{Middle:} Subtraction of an average moffat profile. \textbf{Right:} Subtraction of the azimuthally variable moffat profile. The residual flux in the wings are improved through use of the azimuthally variable profile.  \label{fig:subtracted}}
\end{figure}

The solution to the highly variable zeropoints was discovered when examining the residuals of subtracting the radially averaged moffat profile from the stellar images. See Figure~\ref{fig:subtracted}. Large asymmetries containing substantial portions of the stellar flux were present, including a significant portion of flux in the profile wings. These asymmetries inspired the new technique. The stellar image is broken up into $N$ equal sized wedges, that pivot around the source centroid. In each wedge, a moffat profile is fit to the pixels in that wedge.  Linear interpolation is used to generate functions dependent on the azimuthal angle $\theta$ that replace the usual moffat parameter, $\alpha = \alpha\left(\theta\right)$ with a constant $\beta$. We adopted $N=6$ as a good compromise between risks of overfitting, and quality of the modelling of the profile wings. 

At the cost of potentially worsening the subtraction residuals near the core of the profile, the azimuthally variable profile resulted in much better modelling of the profile wings. We found that by adopting an aperture radius $r=4$~FWHM where the FWHM is the maximum determined from each wedge, the resultant zeropoint scatter was substantially reduced, with a scatter of order $\sim0.1$ magnitudes. The adopted zeropoint was taken as the median of zeropoints of a set of calibration frames, after rejecting the two most discrepant points. Accordingly, we adopt a zeropoint uncertainty of 0.03 magnitudes for each sequence.

While successful at reducing calibration uncertainties to acceptable levels, the above method still leaves significant room for improvement; from poissonian consideration alone, sub-1\% calibration quality should be achievable. We speculate that the calibration issues we (and likely many other groups) experienced is due to temporal variations in the telescope image shape that do not average out sufficiently in short exposures. The obvious solution is to use significantly longer exposure times for the calibration frames, that in itself requires a set of fainter sources than those typically used. Further efforts will be required to determine appropriate solutions. We would like to raise a point of caution in considering published science results that make use of the typical type of calibration images like we gathered, but with standard photometry calibration techniques.

\section{Colour Terms between the Gemini, Pan-STAARS, and SDSS Systems \label{sec:c-terms}}
\begin{figure}[h]
\plotone{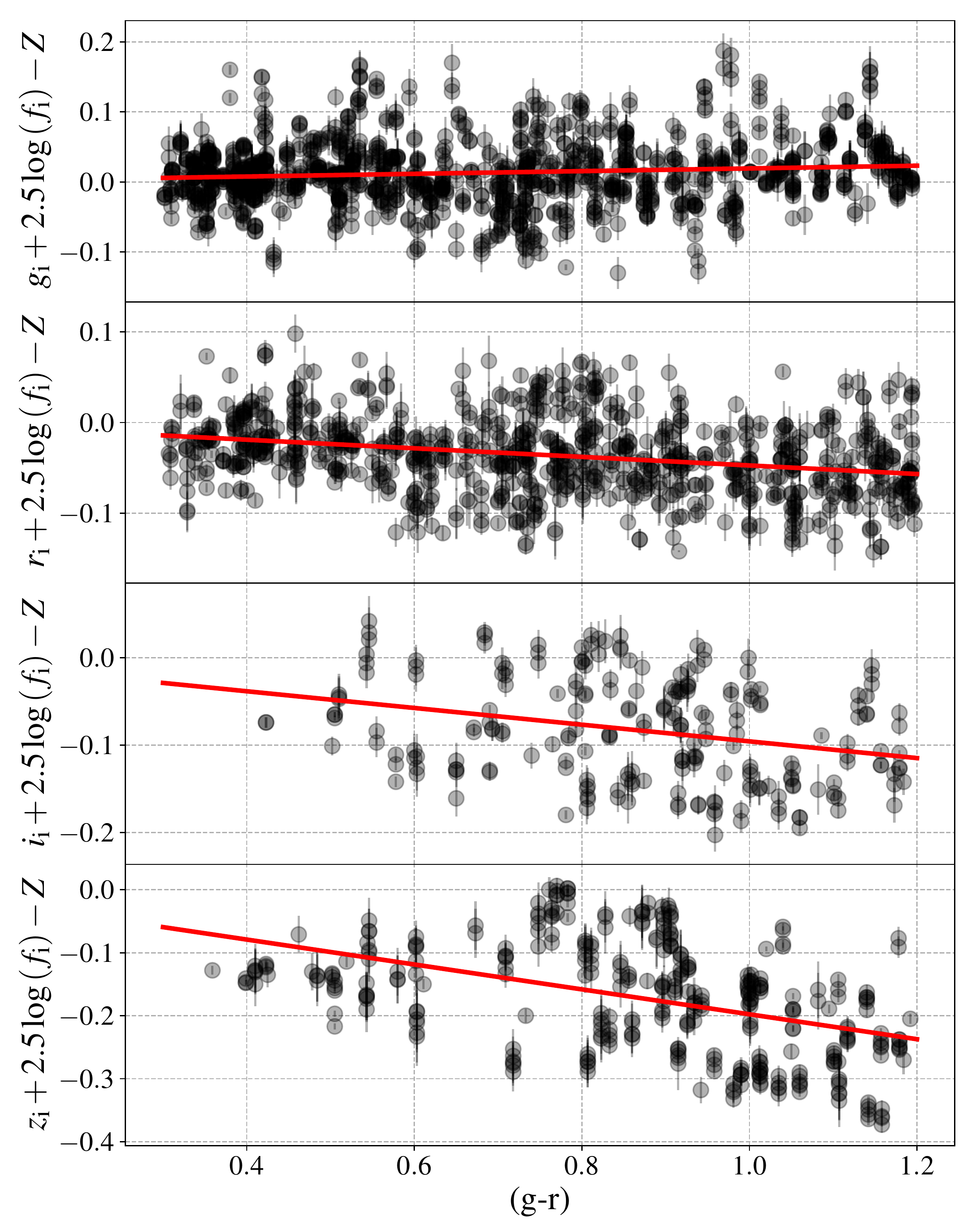}
\caption{GMOS Instrumental magnitudes vs. (g-r) colours of stars in the Pan-STAARS v1 dataset. Red lines show the linear colour terms derived from those measurements.  \label{fig:colourterms}}
\end{figure}

Colour terms between the Gemini and Pan-STAARS griz filter sets were calculated in the same way as presented in \citep{Schwamb2019}. Stars were selected as those sources in the Pan-STAARS release 1 dataset with difference between Kron and aperture magnitudes of less than 0.05 magnitudes, and  magnitudes $17<r<22.5$, $g<22.5$, $z>17.0$ and colours $0.3<(g-r)<1.2$.  Linear colour terms were fit to all measurements of catalog sources across all GMOS data acquired prior to publication of this manuscript. Terms were calculated separately for the E2V and Hamamatsu detectors. The difference between instrumental magnitudes and (g-r) colours for the catalog stars in the images acquired with the Hamamatsu detectors are shown in Figure~\ref{fig:colourterms}. The resultant linear colour terms are:

\begin{eqnarray}
g_\textrm{G} & = & g_{\textrm{PS}} + 0.019(\pm 0.005)\times(g-r)_{\textrm{PS}} \\
r_\textrm{G} & = & r_{\textrm{PS}} - 0.047(\pm 0.004)\times(g-r)_{\textrm{PS}} \\
i_\textrm{G} & = & i_{\textrm{PS}} - 0.198(\pm 0.016)\times(g-r)_{\textrm{PS}} \\
z_\textrm{G} & = & z_{\textrm{PS}} - 0.096(\pm 0.012)\times(g-r)_{\textrm{PS}} 
\end{eqnarray}

\noindent for the Hamamatsu GMOS detectors, and

\begin{eqnarray}
g_\textrm{G} & = & g_{\textrm{PS}} +0.046(\pm 0.003)\times(g-r)_{\textrm{PS}} \\
r_\textrm{G} & = & r_{\textrm{PS}} - 0.053(\pm 0.002)\times(g-r)_{\textrm{PS}} \\
i_\textrm{G} & = & i_{\textrm{PS}} - 0.12(\pm 0.03)\times(g-r)_{\textrm{PS}} \\
z_\textrm{G} & = & z_{\textrm{PS}} - 0.118(\pm 0.007)\times(g-r)_{\textrm{PS}} 
\end{eqnarray}

\noindent for the E2V GMOS detectors.

\section{The Reddening Curve Projection}

\begin{figure}[h]
\plotone{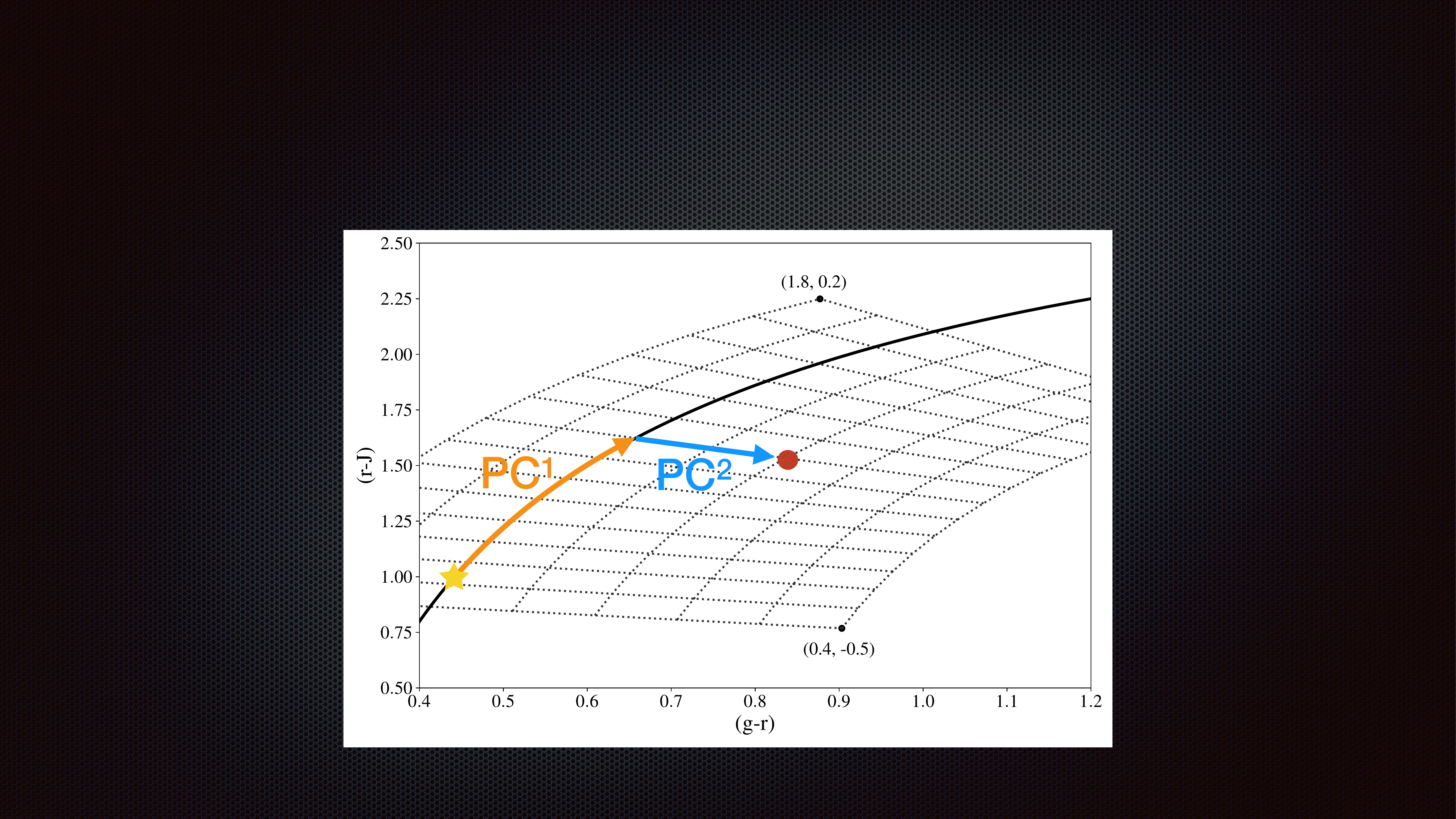}
\caption{Visual demonstration of the reddening curve projection. The dashed grid shows curves of constant $PC^1$ and $PC^2$ in intervals of 0.1. Objects with positive and negative values of $PC^2$ have convex and concave spectra, and are found above and below the reddening curve, respectively. The point values show the extremal projection values.
   \label{fig:projection_instruction}}
\end{figure}

Here we provide a visual demonstration of the reddening curve (Fig.~\ref{fig:projection_instruction}). The $PC^1$ and $PC^2$ basis vectors each take on the instantaneous tangential and perpendicular vectors of the reddening line, respectively, with projection values increasing with increasing red colours. The software written to perform this non-linear projection is available at \url{https://github.com/fraserw/projection} (Will be available at time of submission).

In this work, we adopt Solar Colours of (u-g)=1.43, (g-r)=0.44, (r-z)=0.18, (r-J)=0.98, (F606w-F814w)=-0.47, (F814w-F139m)=-1.24, (B-V)=0.66, (V-R)=0.36, and (R-I)=0.35. The bandpasses u, g, r, and z, are AB magnitudes in the SDSS system. The J bandpass is specifically the MaunaKea Observatories J-band filter\footnote{\url{http://www.gemini.edu/sciops/instruments/niri/filters/}} in Vega magnitudes. The F606w, F814w, and F139m filters are references to the Hubble Space Telescope STMAG ``flight" system \citep{Fraser2012}. We adopt Vega magnitudes for the B, V, R, and I bandpasses. For B, V, R, I, J, and the HST STMAG system, Solar colours were computed using the \emph{synphot} package, with the specific bandpasses of each filter, and the \emph{sun\_reference\_001.fits} data file provided as part of the \emph{synphot data files}.

\bibliography{astroelsart}{}

\begin{thebibliography}{}
\expandafter\ifx\csname natexlab\endcsname\relax\def\natexlab#1{#1}\fi
\providecommand{\url}[1]{\href{#1}{#1}}
\providecommand{\dodoi}[1]{doi:~\href{http://doi.org/#1}{\nolinkurl{#1}}}
\providecommand{\doeprint}[1]{\href{http://ascl.net/#1}{\nolinkurl{http://ascl.net/#1}}}
\providecommand{\doarXiv}[1]{\href{https://arxiv.org/abs/#1}{\nolinkurl{https://arxiv.org/abs/#1}}}

\bibitem[{{Astropy Collaboration} {et~al.}(2013){Astropy Collaboration},
  {Robitaille}, {Tollerud}, {Greenfield}, {Droettboom}, {Bray}, {Aldcroft},
  {Davis}, {Ginsburg}, {Price-Whelan}, {Kerzendorf}, {Conley}, {Crighton},
  {Barbary}, {Muna}, {Ferguson}, {Grollier}, {Parikh}, {Nair}, {Unther},
  {Deil}, {Woillez}, {Conseil}, {Kramer}, {Turner}, {Singer}, {Fox}, {Weaver},
  {Zabalza}, {Edwards}, {Azalee Bostroem}, {Burke}, {Casey}, {Crawford},
  {Dencheva}, {Ely}, {Jenness}, {Labrie}, {Lim}, {Pierfederici}, {Pontzen},
  {Ptak}, {Refsdal}, {Servillat}, \& {Streicher}}]{Astropy2013}
{Astropy Collaboration}, {Robitaille}, T.~P., {Tollerud}, E.~J., {et~al.} 2013,
  \aap, 558, A33, \dodoi{10.1051/0004-6361/201322068}

\bibitem[{{Astropy Collaboration} {et~al.}(2018){Astropy Collaboration},
  {Price-Whelan}, {Sip{\H{o}}cz}, {G{\"u}nther}, {Lim}, {Crawford}, {Conseil},
  {Shupe}, {Craig}, {Dencheva}, {Ginsburg}, {VanderPlas}, {Bradley},
  {P{\'e}rez-Su{\'a}rez}, {de Val-Borro}, {Aldcroft}, {Cruz}, {Robitaille},
  {Tollerud}, {Ardelean}, {Babej}, {Bach}, {Bachetti}, {Bakanov}, {Bamford},
  {Barentsen}, {Barmby}, {Baumbach}, {Berry}, {Biscani}, {Boquien}, {Bostroem},
  {Bouma}, {Brammer}, {Bray}, {Breytenbach}, {Buddelmeijer}, {Burke},
  {Calderone}, {Cano Rodr{\'\i}guez}, {Cara}, {Cardoso}, {Cheedella}, {Copin},
  {Corrales}, {Crichton}, {D'Avella}, {Deil}, {Depagne}, {Dietrich}, {Donath},
  {Droettboom}, {Earl}, {Erben}, {Fabbro}, {Ferreira}, {Finethy}, {Fox},
  {Garrison}, {Gibbons}, {Goldstein}, {Gommers}, {Greco}, {Greenfield},
  {Groener}, {Grollier}, {Hagen}, {Hirst}, {Homeier}, {Horton}, {Hosseinzadeh},
  {Hu}, {Hunkeler}, {Ivezi{\'c}}, {Jain}, {Jenness}, {Kanarek}, {Kendrew},
  {Kern}, {Kerzendorf}, {Khvalko}, {King}, {Kirkby}, {Kulkarni}, {Kumar},
  {Lee}, {Lenz}, {Littlefair}, {Ma}, {Macleod}, {Mastropietro}, {McCully},
  {Montagnac}, {Morris}, {Mueller}, {Mumford}, {Muna}, {Murphy}, {Nelson},
  {Nguyen}, {Ninan}, {N{\"o}the}, {Ogaz}, {Oh}, {Parejko}, {Parley}, {Pascual},
  {Patil}, {Patil}, {Plunkett}, {Prochaska}, {Rastogi}, {Reddy Janga},
  {Sabater}, {Sakurikar}, {Seifert}, {Sherbert}, {Sherwood-Taylor}, {Shih},
  {Sick}, {Silbiger}, {Singanamalla}, {Singer}, {Sladen}, {Sooley},
  {Sornarajah}, {Streicher}, {Teuben}, {Thomas}, {Tremblay}, {Turner},
  {Terr{\'o}n}, {van Kerkwijk}, {de la Vega}, {Watkins}, {Weaver}, {Whitmore},
  {Woillez}, {Zabalza}, \& {Astropy Contributors}}]{Astropy2018}
{Astropy Collaboration}, {Price-Whelan}, A.~M., {Sip{\H{o}}cz}, B.~M., {et~al.}
  2018, \aj, 156, 123, \dodoi{10.3847/1538-3881/aabc4f}

\bibitem[{{Bannister} {et~al.}(2018){Bannister}, {Gladman}, {Kavelaars},
  {Petit}, {Volk}, {Chen}, {Alexand ersen}, {Gwyn}, {Schwamb}, {Ashton},
  {Benecchi}, {Cabral}, {Dawson}, {Delsanti}, {Fraser}, {Granvik},
  {Greenstreet}, {Guilbert-Lepoutre}, {Ip}, {Jakubik}, {Jones}, {Kaib},
  {Lacerda}, {Van Laerhoven}, {Lawler}, {Lehner}, {Lin}, {Lykawka}, {Marsset},
  {Murray-Clay}, {Pike}, {Rousselot}, {Shankman}, {Thirouin}, {Vernazza}, \&
  {Wang}}]{Bannister2018}
{Bannister}, M.~T., {Gladman}, B.~J., {Kavelaars}, J.~J., {et~al.} 2018, \apjs,
  236, 18, \dodoi{10.3847/1538-4365/aab77a}

\bibitem[{Barbary(2016)}]{Barbary2016}
Barbary, K. 2016, Journal of Open Source Software, 1, 58,
  \dodoi{10.21105/joss.00058}

\bibitem[{{Barkume} {et~al.}(2008){Barkume}, {Brown}, \&
  {Schaller}}]{Barkume2008}
{Barkume}, K.~M., {Brown}, M.~E., \& {Schaller}, E.~L. 2008, \aj, 135, 55,
  \dodoi{10.1088/0004-6256/135/1/55}

\bibitem[{{Barucci} {et~al.}(2011){Barucci}, {Alvarez-Candal}, {Merlin},
  {Belskaya}, {de Bergh}, {Perna}, {DeMeo}, \& {Fornasier}}]{Barucci2011}
{Barucci}, M.~A., {Alvarez-Candal}, A., {Merlin}, F., {et~al.} 2011, \icarus,
  214, 297, \dodoi{10.1016/j.icarus.2011.04.019}

\bibitem[{{Benecchi} {et~al.}(2009){Benecchi}, {Noll}, {Grundy}, {Buie},
  {Stephens}, \& {Levison}}]{Benecchi2009}
{Benecchi}, S.~D., {Noll}, K.~S., {Grundy}, W.~M., {et~al.} 2009, \icarus, 200,
  292, \dodoi{10.1016/j.icarus.2008.10.025}

\bibitem[{{Bernstein} {et~al.}(2004){Bernstein}, {Trilling}, {Allen}, {Brown},
  {Holman}, \& {Malhotra}}]{Bernstein2004}
{Bernstein}, G.~M., {Trilling}, D.~E., {Allen}, R.~L., {et~al.} 2004, AJ, 128,
  1364, \dodoi{10.1086/422919}

\bibitem[{{Bertin} \& {Arnouts}(1996)}]{hihi}
{Bertin}, E., \& {Arnouts}, S. 1996, \aaps, 117, 393

\bibitem[{{Boulade} {et~al.}(2003){Boulade}, {Charlot}, {Abbon}, {Aune},
  {Borgeaud}, {Carton}, {Carty}, {Da Costa}, {Deschamps}, {Desforge},
  {Eppell{\'e}}, {Gallais}, {Gosset}, {Granelli}, {Gros}, {de Kat}, {Loiseau},
  {Ritou}, {Rouss{\'e}}, {Starzynski}, {Vignal}, \& {Vigroux}}]{Boulade2003}
{Boulade}, O., {Charlot}, X., {Abbon}, P., {et~al.} 2003, in Society of
  Photo-Optical Instrumentation Engineers (SPIE) Conference Series, Vol. 4841,
  Instrument Design and Performance for Optical/Infrared Ground-based
  Telescopes, ed. M.~{Iye} \& A.~F.~M. {Moorwood}, 72--81,
  \dodoi{10.1117/12.459890}

\bibitem[{{Brown}(2001)}]{Brown2001}
{Brown}, M.~E. 2001, AJ, 121, 2804, \dodoi{10.1086/320391}

\bibitem[{{Brown} {et~al.}(2011){Brown}, {Schaller}, \& {Fraser}}]{Brown2011a}
{Brown}, M.~E., {Schaller}, E.~L., \& {Fraser}, W.~C. 2011, \apjl, 739, L60,
  \dodoi{10.1088/2041-8205/739/2/L60}

\bibitem[{{Brown} {et~al.}(2012){Brown}, {Schaller}, \& {Fraser}}]{Brown2012}
---. 2012, \aj, 143, 146, \dodoi{10.1088/0004-6256/143/6/146}

\bibitem[{{Brucker} {et~al.}(2009){Brucker}, {Grundy}, {Stansberry}, {Spencer},
  {Sheppard}, {Chiang}, \& {Buie}}]{Brucker2009}
{Brucker}, M.~J., {Grundy}, W.~M., {Stansberry}, J.~A., {et~al.} 2009, Icarus,
  201, 284, \dodoi{10.1016/j.icarus.2008.12.040}

\bibitem[{{Brunetto} {et~al.}(2006){Brunetto}, {Barucci}, {Dotto}, \&
  {Strazzulla}}]{Brunetto2006}
{Brunetto}, R., {Barucci}, M.~A., {Dotto}, E., \& {Strazzulla}, G. 2006, \apj,
  644, 646, \dodoi{10.1086/503359}

\bibitem[{{Cruikshank} {et~al.}(1998){Cruikshank}, {Roush}, {Bartholomew},
  {Geballe}, {Pendleton}, {White}, {Bell}, {Davies}, {Owen}, {de Bergh},
  {Tholen}, {Bernstein}, {Brown}, {Tryka}, \& {Dalle Ore}}]{Cruikshank1998}
{Cruikshank}, D.~P., {Roush}, T.~L., {Bartholomew}, M.~J., {et~al.} 1998,
  \icarus, 135, 389, \dodoi{10.1006/icar.1998.5997}

\bibitem[{{Dalle Ore} {et~al.}(2013){Dalle Ore}, {Dalle Ore}, {Roush},
  {Cruikshank}, {Emery}, {Pinilla-Alonso}, \& {Marzo}}]{DalleOre2013}
{Dalle Ore}, C.~M., {Dalle Ore}, L.~V., {Roush}, T.~L., {et~al.} 2013, \icarus,
  222, 307, \dodoi{10.1016/j.icarus.2012.11.015}

\bibitem[{{Delsanti} {et~al.}(2006){Delsanti}, {Peixinho}, {Boehnhardt},
  {Barucci}, {Merlin}, {Doressoundiram}, \& {Davies}}]{Delsanti2006}
{Delsanti}, A., {Peixinho}, N., {Boehnhardt}, H., {et~al.} 2006, \aj, 131,
  1851, \dodoi{10.1086/499402}

\bibitem[{{Elliot} {et~al.}(2005){Elliot}, {Kern}, {Clancy}, {Gulbis},
  {Millis}, {Buie}, {Wasserman}, {Chiang}, {Jordan}, {Trilling}, \&
  {Meech}}]{Elliot2005}
{Elliot}, J.~L., {Kern}, S.~D., {Clancy}, K.~B., {et~al.} 2005, AJ, 129, 1117,
  \dodoi{10.1086/427395}

\bibitem[{{Emery} {et~al.}(2011){Emery}, {Burr}, \& {Cruikshank}}]{Emery2011}
{Emery}, J.~P., {Burr}, D.~M., \& {Cruikshank}, D.~P. 2011, \aj, 141, 25,
  \dodoi{10.1088/0004-6256/141/1/25}

\bibitem[{{Fornasier} {et~al.}(2004){Fornasier}, {Doressoundiram}, {Tozzi},
  {Barucci}, {Boehnhardt}, {de Bergh}, {Delsanti}, {Davies}, \&
  {Dotto}}]{Fornasier2004}
{Fornasier}, S., {Doressoundiram}, A., {Tozzi}, G.~P., {et~al.} 2004, \aap,
  421, 353, \dodoi{10.1051/0004-6361:20041221}

\bibitem[{{Fornasier} {et~al.}(2009){Fornasier}, {Barucci}, {de Bergh},
  {Alvarez-Candal}, {DeMeo}, {Merlin}, {Perna}, {Guilbert}, {Delsanti},
  {Dotto}, \& {Doressoundiram}}]{Fornasier2009}
{Fornasier}, S., {Barucci}, M.~A., {de Bergh}, C., {et~al.} 2009, \aap, 508,
  457, \dodoi{10.1051/0004-6361/200912582}

\bibitem[{{Fraser} {et~al.}(2016){Fraser}, {Alexandersen}, {Schwamb},
  {Marsset}, {Pike}, {Kavelaars}, {Bannister}, {Benecchi}, \&
  {Delsanti}}]{Fraser2016}
{Fraser}, W., {Alexandersen}, M., {Schwamb}, M.~E., {et~al.} 2016, \aj, 151,
  158, \dodoi{10.3847/0004-6256/151/6/158}

\bibitem[{{Fraser} \& {Brown}(2012)}]{Fraser2012}
{Fraser}, W.~C., \& {Brown}, M.~E. 2012, \apj, 749, 33,
  \dodoi{10.1088/0004-637X/749/1/33}

\bibitem[{{Fraser} {et~al.}(2015){Fraser}, {Brown}, \& {Glass}}]{Fraser2015}
{Fraser}, W.~C., {Brown}, M.~E., \& {Glass}, F. 2015, \apj, 804, 31,
  \dodoi{10.1088/0004-637X/804/1/31}

\bibitem[{{Fraser} {et~al.}(2014){Fraser}, {Brown}, {Morbidelli}, {Parker}, \&
  {Batygin}}]{Fraser2014a}
{Fraser}, W.~C., {Brown}, M.~E., {Morbidelli}, A., {Parker}, A., \& {Batygin},
  K. 2014, \apj, 782, 100, \dodoi{10.1088/0004-637X/782/2/100}

\bibitem[{{Fraser} \& {Kavelaars}(2009)}]{Fraser2009}
{Fraser}, W.~C., \& {Kavelaars}, J.~J. 2009, \aj, 137, 72,
  \dodoi{10.1088/0004-6256/137/1/72}

\bibitem[{{Fraser} {et~al.}(2017){Fraser}, {Bannister}, {Pike}, {Marsset},
  {Schwamb}, {Kavelaars}, {Lacerda}, {Nesvorn{\'y}}, {Volk}, {Delsanti},
  {Benecchi}, {Lehner}, {Noll}, {Gladman}, {Petit}, {Gwyn}, {Chen}, {Wang},
  {Alexandersen}, {Burdullis}, {Sheppard}, \& {Trujillo}}]{Fraser2017}
{Fraser}, W.~C., {Bannister}, M.~T., {Pike}, R.~E., {et~al.} 2017, Nature
  Astronomy, 1, 0088, \dodoi{10.1038/s41550-017-0088}

\bibitem[{{Fraser} {et~al.}(2021){Fraser}, {Benecchi}, {Kavelaars}, {Marsset},
  {Pike}, {Bannister}, {Schwamb}, {Volk}, {Nesvorny}, {Alexandersen}, {Chen},
  {Gwyn}, {Lehner}, \& {Wang}}]{Fraser2021}
{Fraser}, W.~C., {Benecchi}, S.~D., {Kavelaars}, J., {et~al.} 2021, arXiv
  e-prints, arXiv:2104.00028.
\newblock \doarXiv{2104.00028}

\bibitem[{{Fuentes} {et~al.}(2009){Fuentes}, {George}, \&
  {Holman}}]{Fuentes2009}
{Fuentes}, C.~I., {George}, M.~R., \& {Holman}, M.~J. 2009, \apj, 696, 91,
  \dodoi{10.1088/0004-637X/696/1/91}

\bibitem[{{Fukugita} {et~al.}(1996){Fukugita}, {Ichikawa}, {Gunn}, {Doi},
  {Shimasaku}, \& {Schneider}}]{Fukugita1996}
{Fukugita}, M., {Ichikawa}, T., {Gunn}, J.~E., {et~al.} 1996, \aj, 111, 1748,
  \dodoi{10.1086/117915}

\bibitem[{{Gladman} \& {Volk}(2021)}]{Gladman2021}
{Gladman}, B., \& {Volk}, K. 2021, \araa, 59,
  \dodoi{10.1146/annurev-astro-120920-010005}

\bibitem[{{Gourgeot} {et~al.}(2015){Gourgeot}, {Barucci}, {Alvarez-Candal},
  {Merlin}, {Perna}, \& {Lazzaro}}]{Gourgeot2015}
{Gourgeot}, F., {Barucci}, M.~A., {Alvarez-Candal}, A., {et~al.} 2015, \aap,
  582, A13, \dodoi{10.1051/0004-6361/201526014}

\bibitem[{{Grundy} {et~al.}(2020){Grundy}, {Bird}, {Britt}, {Cook},
  {Cruikshank}, {Howett}, {Krijt}, {Linscott}, {Olkin}, {Parker}, {Protopapa},
  {Ruaud}, {Umurhan}, {Young}, {Dalle Ore}, {Kavelaars}, {Keane}, {Pendleton},
  {Porter}, {Scipioni}, {Spencer}, {Stern}, {Verbiscer}, {Weaver}, {Binzel},
  {Buie}, {Buratti}, {Cheng}, {Earle}, {Elliott}, {Gabasova}, {Gladstone},
  {Hill}, {Horanyi}, {Jennings}, {Lunsford}, {McComas}, {McKinnon}, {McNutt},
  {Moore}, {Parker}, {Quirico}, {Reuter}, {Schenk}, {Schmitt}, {Showalter},
  {Singer}, {Weigle}, \& {Zangari}}]{Grundy2020}
{Grundy}, W.~M., {Bird}, M.~K., {Britt}, D.~T., {et~al.} 2020, Science, 367,
  aay3705, \dodoi{10.1126/science.aay3705}

\bibitem[{{Gwyn}(2008)}]{Gwyn2008}
{Gwyn}, S.~D.~J. 2008, \pasp, 120, 212, \dodoi{10.1086/526794}

\bibitem[{Harris {et~al.}(2020)Harris, Millman, van~der Walt, Gommers,
  Virtanen, Cournapeau, Wieser, Taylor, Berg, Smith, Kern, Picus, Hoyer, van
  Kerkwijk, Brett, Haldane, del R{\'{i}}o, Wiebe, Peterson,
  G{\'{e}}rard-Marchant, Sheppard, Reddy, Weckesser, Abbasi, Gohlke, \&
  Oliphant}]{Harris2020}
Harris, C.~R., Millman, K.~J., van~der Walt, S.~J., {et~al.} 2020, Nature, 585,
  357, \dodoi{10.1038/s41586-020-2649-2}

\bibitem[{{Hartigan} \& {Hartigan}(1985)}]{Hartigan1985}
{Hartigan}, J.~A., \& {Hartigan}, P.~M. 1985, Annals of Statistics, 13, 70

\bibitem[{{Hodapp} {et~al.}(2003){Hodapp}, {Jensen}, {Irwin}, {Yamada},
  {Chung}, {Fletcher}, {Robertson}, {Hora}, {Simons}, {Mays}, {Nolan}, {Bec},
  {Merrill}, \& {Fowler}}]{Hodapp2003}
{Hodapp}, K.~W., {Jensen}, J.~B., {Irwin}, E.~M., {et~al.} 2003, \pasp, 115,
  1388, \dodoi{10.1086/379669}

\bibitem[{{Hook} {et~al.}(2004){Hook}, {J{\o}rgensen}, {Allington-Smith},
  {Davies}, {Metcalfe}, {Murowinski}, \& {Crampton}}]{Hook2004}
{Hook}, I.~M., {J{\o}rgensen}, I., {Allington-Smith}, J.~R., {et~al.} 2004,
  \pasp, 116, 425, \dodoi{10.1086/383624}

\bibitem[{{Huang} {et~al.}(2022){Huang}, {Gladman}, \&
  {Volk}}]{Huang2022preprint}
{Huang}, Y., {Gladman}, B., \& {Volk}, K. 2022, arXiv e-prints,
  arXiv:2202.09045.
\newblock \doarXiv{2202.09045}

\bibitem[{Hunter(2007)}]{Hunter2007}
Hunter, J.~D. 2007, Computing in Science \& Engineering, 9, 90,
  \dodoi{10.1109/MCSE.2007.55}

\bibitem[{{Ivezi{\'c}} {et~al.}(2001){Ivezi{\'c}}, {Tabachnik}, {Rafikov},
  {Lupton}, {Quinn}, {Hammergren}, {Eyer}, {Chu}, {Armstrong}, {Fan},
  {Finlator}, {Geballe}, {Gunn}, {Hennessy}, {Knapp}, {Leggett}, {Munn},
  {Pier}, {Rockosi}, {Schneider}, {Strauss}, {Yanny}, {Brinkmann}, {Csabai},
  {Hindsley}, {Kent}, {Lamb}, {Margon}, {McKay}, {Smith}, {Waddel}, {York}, \&
  {SDSS Collaboration}}]{Ivezic2001}
{Ivezi{\'c}}, {\v{Z}}., {Tabachnik}, S., {Rafikov}, R., {et~al.} 2001, \aj,
  122, 2749, \dodoi{10.1086/323452}

\bibitem[{{Izawa} {et~al.}(2014){Izawa}, {Applin}, {Norman}, \&
  {Cloutis}}]{Izawa2014}
{Izawa}, M.~R.~M., {Applin}, D.~M., {Norman}, L., \& {Cloutis}, E.~A. 2014,
  \icarus, 237, 159, \dodoi{10.1016/j.icarus.2014.04.033}

\bibitem[{{Lacerda} {et~al.}(2014){Lacerda}, {Fornasier}, {Lellouch}, {Kiss},
  {Vilenius}, {Santos-Sanz}, {Rengel}, {M{\"u}ller}, {Stansberry}, {Duffard},
  {Delsanti}, \& {Guilbert-Lepoutre}}]{Lacerda2014}
{Lacerda}, P., {Fornasier}, S., {Lellouch}, E., {et~al.} 2014, \apjl, 793, L2,
  \dodoi{10.1088/2041-8205/793/1/L2}

\bibitem[{{Marsset} {et~al.}(2019){Marsset}, {Fraser}, {Pike}, {Bannister},
  {Schwamb}, {Volk}, {Kavelaars}, {Alexandersen}, {Chen}, {Gladman}, {Gwyn},
  {Lehner}, {Peixinho}, {Petit}, \& {Wang}}]{Marsset2019}
{Marsset}, M., {Fraser}, W.~C., {Pike}, R.~E., {et~al.} 2019, \aj, 157, 94,
  \dodoi{10.3847/1538-3881/aaf72e}

\bibitem[{{Marsset} {et~al.}(2020){Marsset}, {Fraser}, {Bannister}, {Schwamb},
  {Pike}, {Benecchi}, {Kavelaars}, {Alexandersen}, {Chen}, {Gladman}, {Gwyn},
  {Petit}, \& {Volk}}]{Marsset2020}
{Marsset}, M., {Fraser}, W.~C., {Bannister}, M.~T., {et~al.} 2020, The
  Planetary Science Journal, 1, 16, \dodoi{10.3847/PSJ/ab8cc0}

\bibitem[{{Miyazaki} {et~al.}(2002){Miyazaki}, {Komiyama}, {Sekiguchi},
  {Okamura}, {Doi}, {Furusawa}, {Hamabe}, {Imi}, {Kimura}, {Nakata}, {Okada},
  {Ouchi}, {Shimasaku}, {Yagi}, \& {Yasuda}}]{Miyazaki2002}
{Miyazaki}, S., {Komiyama}, Y., {Sekiguchi}, M., {et~al.} 2002, \pasj, 54, 833

\bibitem[{{Nesvorny} {et~al.}(2022){Nesvorny}, {Vokrouhlicky}, \&
  {Fraser}}]{Nesvorny2022preprint}
{Nesvorny}, D., {Vokrouhlicky}, D., \& {Fraser}, W.~C. 2022, arXiv e-prints,
  arXiv:2201.02747.
\newblock \doarXiv{2201.02747}

\bibitem[{{Nesvorn{\'y}} {et~al.}(2020){Nesvorn{\'y}}, {Vokrouhlick{\'y}},
  {Alexandersen}, {Bannister}, {Buchanan}, {Chen}, {Gladman}, {Gwyn},
  {Kavelaars}, {Petit}, {Schwamb}, \& {Volk}}]{Nesvorny2020}
{Nesvorn{\'y}}, D., {Vokrouhlick{\'y}}, D., {Alexandersen}, M., {et~al.} 2020,
  \aj, 160, 46, \dodoi{10.3847/1538-3881/ab98fb}

\bibitem[{{Noll} {et~al.}(2008){Noll}, {Grundy}, {Chiang}, {Margot}, \&
  {Kern}}]{Noll2008}
{Noll}, K.~S., {Grundy}, W.~M., {Chiang}, E.~I., {Margot}, J.-L., \& {Kern},
  S.~D. 2008, {Binaries in the Kuiper Belt} (The Solar System Beyond Neptune),
  345--363

\bibitem[{{Padmanabhan} {et~al.}(2008){Padmanabhan}, {Schlegel}, {Finkbeiner},
  {Barentine}, {Blanton}, {Brewington}, {Gunn}, {Harvanek}, {Hogg},
  {Ivezi{\'c}}, {Johnston}, {Kent}, {Kleinman}, {Knapp}, {Krzesinski}, {Long},
  {Neilsen}, {Nitta}, {Loomis}, {Lupton}, {Roweis}, {Snedden}, {Strauss}, \&
  {Tucker}}]{Padmanabhan2008}
{Padmanabhan}, N., {Schlegel}, D.~J., {Finkbeiner}, D.~P., {et~al.} 2008, \apj,
  674, 1217, \dodoi{10.1086/524677}

\bibitem[{{Peixinho} {et~al.}(2015){Peixinho}, {Delsanti}, \&
  {Doressoundiram}}]{Peixinho2015}
{Peixinho}, N., {Delsanti}, A., \& {Doressoundiram}, A. 2015, \aap, 577, A35,
  \dodoi{10.1051/0004-6361/201425436}

\bibitem[{{Peixinho} {et~al.}(2012){Peixinho}, {Delsanti}, {Guilbert-Lepoutre},
  {Gafeira}, \& {Lacerda}}]{Peixinho2012}
{Peixinho}, N., {Delsanti}, A., {Guilbert-Lepoutre}, A., {Gafeira}, R., \&
  {Lacerda}, P. 2012, ArXiv e-prints.
\newblock \doarXiv{1206.3153}

\bibitem[{{Petit} {et~al.}(2011){Petit}, {Kavelaars}, {Gladman}, {Jones},
  {Parker}, {Van Laerhoven}, {Nicholson}, {Mars}, {Rousselot}, {Mousis},
  {Marsden}, {Bieryla}, {Taylor}, {Ashby}, {Benavidez}, {Campo Bagatin}, \&
  {Bernabeu}}]{Petit2011}
{Petit}, J.-M., {Kavelaars}, J.~J., {Gladman}, B.~J., {et~al.} 2011, \aj, 142,
  131, \dodoi{10.1088/0004-6256/142/4/131}

\bibitem[{{Pike} {et~al.}(2017){Pike}, {Fraser}, {Schwamb}, {Kavelaars},
  {Marsset}, {Bannister}, {Lehner}, {Wang}, {Alexandersen}, {Chen}, {Gladman},
  {Gwyn}, {Petit}, \& {Volk}}]{Pike2017}
{Pike}, R.~E., {Fraser}, W.~C., {Schwamb}, M.~E., {et~al.} 2017, \aj, 154, 101,
  \dodoi{10.3847/1538-3881/aa83b1}

\bibitem[{{Schwamb} {et~al.}(2019){Schwamb}, {Fraser}, {Bannister}, {Marsset},
  {Pike}, {Kavelaars}, {Benecchi}, {Lehner}, {Wang}, {Thirouin}, {Delsanti},
  {Peixinho}, {Volk}, {Alexandersen}, {Chen}, {Gladman}, {Gwyn}, \&
  {Petit}}]{Schwamb2019}
{Schwamb}, M.~E., {Fraser}, W.~C., {Bannister}, M.~T., {et~al.} 2019, \apjs,
  243, 12, \dodoi{10.3847/1538-4365/ab2194}

\bibitem[{{Seccull} {et~al.}(2021){Seccull}, {Fraser}, \&
  {Puzia}}]{Seccull2021}
{Seccull}, T., {Fraser}, W.~C., \& {Puzia}, T.~H. 2021, arXiv e-prints,
  arXiv:2102.05076.
\newblock \doarXiv{2102.05076}

\bibitem[{{Seccull} {et~al.}(2018){Seccull}, {Fraser}, {Puzia}, {Brown}, \&
  {Sch{\"o}nebeck}}]{Seccull2018}
{Seccull}, T., {Fraser}, W.~C., {Puzia}, T.~H., {Brown}, M.~E., \&
  {Sch{\"o}nebeck}, F. 2018, \apjl, 855, L26, \dodoi{10.3847/2041-8213/aab3dc}

\bibitem[{{Snodgrass} {et~al.}(2010){Snodgrass}, {Carry}, {Dumas}, \&
  {Hainaut}}]{Snodgrass2010}
{Snodgrass}, C., {Carry}, B., {Dumas}, C., \& {Hainaut}, O. 2010, \aap, 511,
  A72+, \dodoi{10.1051/0004-6361/200913031}

\bibitem[{{STScI Development Team}(2013)}]{Lim2013}
{STScI Development Team}. 2013, {pysynphot: Synthetic photometry software
  package}, Astrophysics Source Code Library, record ascl:1303.023.
\newblock \doeprint{1303.023}

\bibitem[{{Szab{\'o}} {et~al.}(2007){Szab{\'o}}, {Ivezi{\'c}}, {Juri{\'c}}, \&
  {Lupton}}]{Szabo2007}
{Szab{\'o}}, G.~M., {Ivezi{\'c}}, {\v Z}., {Juri{\'c}}, M., \& {Lupton}, R.
  2007, \mnras, 377, 1393, \dodoi{10.1111/j.1365-2966.2007.11687.x}

\bibitem[{{Tegler} \& {Romanishin}(2003)}]{Tegler2003b}
{Tegler}, S.~C., \& {Romanishin}, W. 2003, \icarus, 161, 181,
  \dodoi{10.1016/S0019-1035(02)00021-0}

\bibitem[{{Tegler} {et~al.}(2016){Tegler}, {Romanishin}, {Consolmagno}, \&
  {J.}}]{Tegler2016}
{Tegler}, S.~C., {Romanishin}, W., {Consolmagno}, G.~J., \& {J.}, S. 2016, \aj,
  152, 210, \dodoi{10.3847/0004-6256/152/6/210}

\bibitem[{{Tonry} {et~al.}(2012){Tonry}, {Stubbs}, {Lykke}, {Doherty},
  {Shivvers}, {Burgett}, {Chambers}, {Hodapp}, {Kaiser}, {Kudritzki},
  {Magnier}, {Morgan}, {Price}, \& {Wainscoat}}]{Tonry2012}
{Tonry}, J.~L., {Stubbs}, C.~W., {Lykke}, K.~R., {et~al.} 2012, \apj, 750, 99,
  \dodoi{10.1088/0004-637X/750/2/99}

\bibitem[{{Trujillo} {et~al.}(2011){Trujillo}, {Sheppard}, \&
  {Schaller}}]{Trujillo2011}
{Trujillo}, C.~A., {Sheppard}, S.~S., \& {Schaller}, E.~L. 2011, \apj, 730,
  105, \dodoi{10.1088/0004-637X/730/2/105}

\bibitem[{{Van Laerhoven} {et~al.}(2019){Van Laerhoven}, {Gladman}, {Volk},
  {Kavelaars}, {Petit}, {Bannister}, {Alexandersen}, {Chen}, \&
  {Gwyn}}]{vanLaerhoven2019}
{Van Laerhoven}, C., {Gladman}, B., {Volk}, K., {et~al.} 2019, \aj, 158, 49,
  \dodoi{10.3847/1538-3881/ab24e1}

\bibitem[{{Volk} \& {Malhotra}(2017)}]{Volk2017}
{Volk}, K., \& {Malhotra}, R. 2017, \aj, 154, 62,
  \dodoi{10.3847/1538-3881/aa79ff}

\end{thebibliography}
\bibliographystyle{aasjournal}

\end{document}